%% file: Signal_Processing_part2.tex
\documentclass[a4paper,11pt]{article}
\pdfoutput=1 

\usepackage{jinstpub} 

\usepackage{color}
\usepackage[normalem]{ulem}
\usepackage{lineno}
\usepackage{graphicx}
\usepackage{subcaption}
\usepackage{mwe}
\usepackage{amsmath}
\usepackage{amssymb}
\usepackage{siunitx}
\usepackage{afterpage}
\usepackage{newtxmath,newtxtext}

\usepackage{helvet}
\def\Put(#1,#2)#3{\leavevmode\makebox(0,0){\put(#1,#2){#3}}}

\newlength{\figwidth}
\newlength{\fighalfwidth}
\collaboration{MicroBooNE Collaboration}

\title{\boldmath \center \LARGE Ionization Electron Signal Processing \\
  in Single Phase LArTPCs \\
  II. Data/Simulation Comparison \\
  and Performance in MicroBooNE}

\input{uBauthorlistSP2}

\emailAdd{microboone\_info@fnal.gov}

\abstract{
  The single-phase liquid argon time projection chamber (LArTPC) provides a large amount
  of detailed information in the form of fine-grained drifted ionization charge from
  particle traces. To fully utilize this information, the deposited charge must be
  accurately extracted from the raw digitized waveforms via a robust signal processing chain.
  Enabled by the ultra-low noise levels associated with cryogenic electronics
  in the MicroBooNE detector, the precise extraction of 
  ionization charge from the induction wire planes in a single-phase LArTPC
  is qualitatively demonstrated on MicroBooNE data with event display images, and
  quantitatively demonstrated via waveform-level and track-level metrics.  Improved performance
  of induction plane calorimetry is demonstrated through the agreement of
  extracted ionization charge measurements across different wire planes for various event
  topologies.  In addition to the comprehensive waveform-level
  comparison of data and simulation, a calibration of the cryogenic electronics response
  is presented and solutions to various MicroBooNE-specific TPC
  issues are discussed.  This work presents an important improvement in LArTPC signal
  processing, the foundation of reconstruction and therefore physics analyses in MicroBooNE.}

\keywords{MicroBooNE, LArTPC, Signal Processing, Deconvolution, Calorimetry}

\begin{document}
\maketitle
\flushbottom

\pagebreak

\section{Introduction} \label{sec:introduction}

\input{intro_part2.tex}

	\subsection{Review of signal processing concepts}\label{sec:introduction:definitions}

	\input{intro_defs_part2.tex}

\section{Calibration of TPC cold electronics response function}\label{sec:calib_elec_resp}

\input{elec_resp_part2.tex}

	\subsection{MicroBooNE TPC cold electronics calibration system}\label{sec:calib_elec_resp:coldelec}

	\input{elec_resp_coldelec_part2.tex}

	\subsection{Electronics response parameterization}\label{sec:calib_elec_resp:parameterization}

	\input{elec_resp_parameterization_part2.tex}

	\subsection{Stability of electronics response}\label{sec:calib_elec_resp:stability}

	\input{elec_resp_stability_part2.tex}

	\subsection{Validation of electronics response correction}\label{sec:calib_elec_resp:validation}

	\input{elec_resp_validation_part2.tex}

\section{Validation of field response function simulation with data} \label{sec:fieldresp}

\input{field_resp_part2.tex}

	\subsection{Response extraction methodology} \label{sec:fieldresp:methodology}

	\input{field_resp_method_part2.tex}

	\subsection{Response in normal region} \label{sec:fieldresp:normalwires}

	\input{field_resp_normalregion_part2.tex}

	\subsection{Response in shorted-wire regions} \label{sec:fieldresp:shortedwires}

	\input{field_resp_shortedregions_part2.tex}

	\subsection{Signal processing for shorted-wire regions} \label{sec:fielderesp:L1shorted}

	\input{field_resp_L1shorted_part2.tex}

\section{Cleaning and filtering of data events}\label{sec:sigproc_data}

\input{sigproc_data_part2.tex}

	\subsection{Removal of PMT-induced TPC signals}\label{sec:sigproc_data:pmtinducedsig}

	\input{sigproc_data_pmtinducedsig_part2.tex}

	\subsection{Correction of ADC bit shifts}\label{sec:sigproc_data:adcbitshift} 

	\input{sigproc_data_adcbitshift_part2.tex}

	\subsection{Handling purity-monitor-induced burst noise and cathode burst events}\label{sec:sigproc_data:PMandCBnoise}

	\input{sigproc_data_PMandCBnoise_part2.tex}

\section{Evaluation of signal processing performance on MicroBooNE data}\label{sec:results}

\input{perf_eval_part2.tex}

	\subsection{Event displays}\label{sec:results:eventdisplays}

	\input{perf_eval_eventdisplays_part2.tex}

	\subsection{Demonstration of cross-plane charge matching}\label{sec:results:chargematching}

	\input{perf_eval_chargematching_part2.tex}

\section{Summary} \label{sec:summary}

\input{summary_part2.tex}


\acknowledgments

This document was prepared by the MicroBooNE collaboration using the
resources of the Fermi National Accelerator Laboratory (Fermilab), a
U.S. Department of Energy, Office of Science, HEP User Facility.
Fermilab is managed by Fermi Research Alliance, LLC (FRA), acting
under Contract No. DE-AC02-07CH11359.  MicroBooNE is supported by the
following: the U.S. Department of Energy, Office of Science, Offices
of High Energy Physics and Nuclear Physics; the U.S. National Science
Foundation; the Swiss National Science Foundation; the Science and
Technology Facilities Council of the United Kingdom; and The Royal
Society (United Kingdom).  Additional support for the laser
calibration system and cosmic ray tagger was provided by the Albert
Einstein Center for Fundamental Physics (Bern, Switzerland).

\appendix



\bibliographystyle{JHEP}

\bibliography{Signal_Processing}{}

\end{document}

%% file: uBauthorlistSP2.tex
\author[i]{C.~Adams}
\author[j]{R.~An}
\author[c]{J.~Anthony}
\author[aa]{J.~Asaadi}
\author[a]{M.~Auger}
\author[ee]{S.~Balasubramanian}
\author[h]{B.~Baller}
\author[p]{C.~Barnes}
\author[s]{G.~Barr}
\author[b]{M.~Bass}
% ...

\affiliation[a]{Universit{\"a}t Bern, Bern CH-3012, Switzerland}
\affiliation[b]{Brookhaven National Laboratory (BNL), Upton, NY, 11973, USA}
\affiliation[c]{University of Cambridge, Cambridge CB3 0HE, United Kingdom}
% ...
\affiliation[h]{Fermi National Accelerator Laboratory (FNAL), Batavia, IL 60510, USA}
\affiliation[i]{Harvard University, Cambridge, MA 02138, USA}
\affiliation[j]{Illinois Institute of Technology (IIT), Chicago, IL 60616, USA}
% ...
\affiliation[p]{University of Michigan, Ann Arbor, MI, 48109, USA}
% ...
\affiliation[s]{University of Oxford, Oxford OX1 3RH, United Kingdom}
% ...
\affiliation[aa]{University of Texas, Arlington, TX, 76019, USA}
% ...
\affiliation[ee]{Yale University, New Haven, CT, 06520, USA}



%% file: intro_part2.tex
MicroBooNE~\cite{uboone_1,uboone_2}, or the Micro Booster Neutrino Experiment, is
located at Fermilab in Batavia, Illinois, and has the goal of addressing the
low-energy electron-neutrino-like excess observed at MiniBooNE~\cite{miniboone_LEE}
on the same beam line (the Booster Neutrino Beam, or BNB).  In order to address
this anomaly, the experiment utilizes a different detector technology, a
liquid argon time projection chamber (LArTPC).  The LArTPC~\cite{rubbia77,Chen:1976pp,willis74,Nygren:1976fe}
provides fine-grained spatial resolution and an excellent calorimetric capability
that allows for discrimination between electrons and photons in the active
volume of the TPC~\cite{argoneut_nue}.  This discrimination power is key to the success of the
MicroBooNE physics program, as it allows one to identify the photon backgrounds which
complicate the MiniBooNE electron appearance search.

\begin{figure}[tb]
  \centering
  \includegraphics[width=\figwidth]{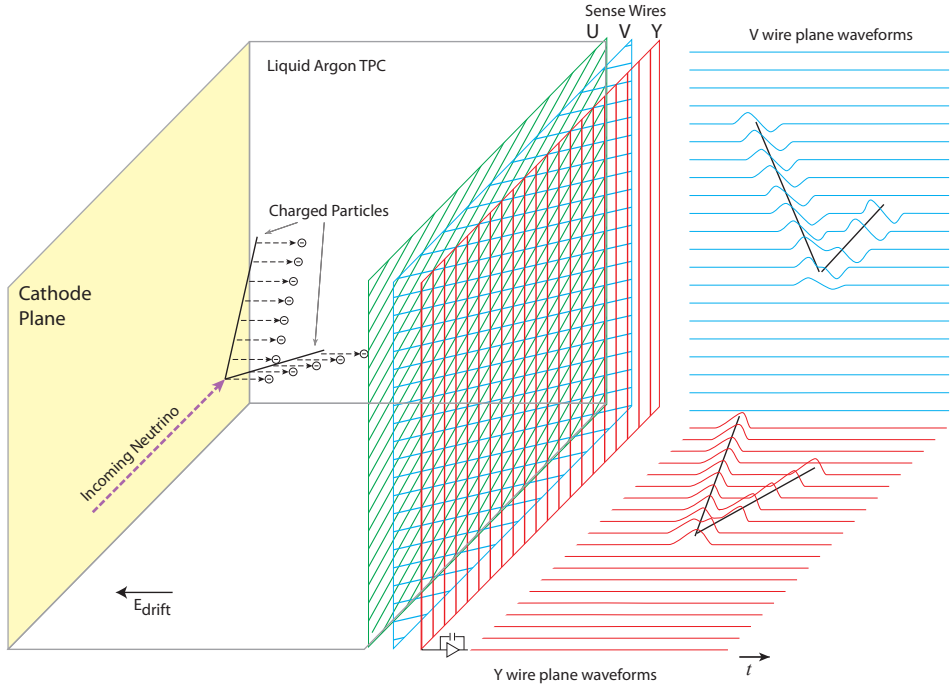}
  \caption{Illustration of signal formation in the MicroBooNE three-plane
  LArTPC, depicting the V plane (second induction plane) and Y plane
  (collection plane) TPC wire signals on the right of the image~\cite{Acciarri:2016smi}.}
  \label{fig:lartpc}
\end{figure}

The MicroBooNE LArTPC, illustrated in figure~\ref{fig:lartpc}, utilizes
three wire planes at the anode to obtain signals from drifting ionization
electrons produced from neutrino interactions within the active volume of the
detector.  Drifting ionization electrons first pass by the two induction planes
(the ``U plane'' and ``V plane''), inducing signals on the wires, and are
finally collected on the collection plane (the ``Y plane'').  The drift
electric field is approximately 273~V/cm and is maintained by a cathode
held at -70~kV as well as a field cage that surrounds the LArTPC active
volume and gradually steps down the voltage toward the anode plane via a
resistor chain.  The three wire planes are biased at specific voltages
(-110~V, 0~V, and +230~V for the U plane, V plane, and Y plane, respectively)
to ensure complete transparency for drifting ionization electrons, allowing for
all ionization electrons to reach the third (collection) wire plane without being
collected by one of the first two (induction) wire planes.  Bipolar
signals are produced on the induction planes and largely unipolar signals
are produced on the collection plane.

One primary asset of the LArTPC technology is its capability of discriminating between
electrons and photons in the detector.  This is strongly desired in accelerator neutrino
experiments as photons are often backgrounds to searches for electron neutrino
appearance.  Photons that convert into an electron-positron pair within the liquid argon
can be indentified as having roughly double the value of $dE/dx$ compared to a single
electron, measured using the calorimetric information (recovered from the charge
measured at the TPC wires) of the track-like beginning
of the particle signature in the detector.  In order to make use of the
electron versus photon discrimination capability of
the LArTPC technology, it is necessary to first demonstrate
the ability to extract charge precisely from these signals on the TPC wires; both imaging
and calorimetry in the TPC active volume depend on this precise charge extraction.

Three elements come into play for signal formation on TPC wires:
i) the distribution of the ionization charge cloud produced by charged particles
traveling through the TPC; ii) the field response describing charge
collection and induction on the TPC wires as ionization electrons drift through the
wire planes; and iii) the electronics response characterizing the
amplification and shaping of the induced current on the TPC wire in the
front-end electronics.  A detailed simulation of each of these effects is required to
realistically simulate particle interactions within the MicroBooNE detector for comparison
with data.  GEANT4~\cite{Geant4} is used to simulate the energy deposition of charged particles
traveling through the liquid argon and LArSoft~\cite{larsoft_paper} is utilized to simulate
ionization charge recombination, transport, and diffusion.  For the wire field response and
electronics amplification and shaping,
a dedicated simulation was developed for MicroBooNE~\cite{SP1_paper}.
The electronics response shape is taken from a parametrization of the
expected response of the front-end electronics at a given gain and peaking time,
while the wire field response is simulated using Garfield~\cite{garfield}.
This wire field response simulation includes a dependence on the impact parameter
of the ionization electron drift with respect to the wires, as well as the charge
induction on wires both close to and far from the drifting ionization electrons.

After the employment of software noise filtering techniques to eliminate
excess noise from external sources~\cite{noise_filter_paper}, a robust
signal processing chain~\cite{Baller:2017ugz, SP1_paper} is required
to extract the number of drifting ionization electrons passing through each wire plane.
It is necessary to validate this signal processing chain with
data from the MicroBooNE LArTPC in order to evaluate the performance of
drifted charge extraction from the TPC wires.
This work describes this validation, as well as the validation of the
TPC signal simulation described in Ref.~\cite{SP1_paper}, by making comparisons between
MicroBooNE data and simulation.  In addition, the performance of
charge-matching across the three wire planes of the MicroBooNE LArTPC is
illustrated in this work. This charge-matching can be utilized in novel
imaging techniques~\cite{wirecell} that take advantage of local cross-plane
charge correlations in order to construct three-dimensional images.
This approach simplifies charge clustering for electron and photon showers and is expected to help in the reconstruction of neutrino interactions.

This article is organized as follows.  First, in the remainder of the introduction,
a brief review of signal processing concepts referenced throughout this work is
presented.  In section~\ref{sec:calib_elec_resp} the calibration of the electronics
response function on a channel-by-channel basis, which
enables a uniform electronics response across all TPC channels, is described.
Next, in section~\ref{sec:fieldresp}, a comparison of data and
simulation at the waveform level is carried out, characterizing the
capability of the ionization signal simulation to reproduce the response of
the TPC wires and the electronics to drifted ionization charge.  This section also
includes discussion of a method used to account for the modified
response function in the shorted regions of the detector, where a group
of channels in the collection plane (Y plane) are shorted to a non-standard bias voltage. 
Special considerations necessary to employ the signal processing chain discussed
in Ref.~\cite{SP1_paper} on data are presented in section~\ref{sec:sigproc_data},
including the addressing of electronics/noise issues that complement the
studies highlighted in Ref.~\cite{noise_filter_paper}.  Results of
applying the signal processing chain on MicroBooNE data are shown in
section~\ref{sec:results}, including the characterization of its performance on
events in data and a demonstration of the precise extraction
of charge information from the induction planes of a LArTPC, which has previously
been shown to be feasible~\cite{corey_thesis}.  Emphasized in
this section is the matching of charge across the three planes of the MicroBooNE
LArTPC.  Finally, a summary of the investigations carried out in this work, along
with their significance, is presented in section~\ref{sec:summary}.

%% file: intro_defs_part2.tex
A number of fundamental signal processing concepts are referenced throughout this
article that are worth discussing briefly as background information.  For a more complete discussion
of LArTPC signal processing concepts and methodology, see Ref.~\cite{SP1_paper}.

The deconvolution technique is used to extract charge information from LArTPC waveform
data, and has been utilized previously in the data analysis of other LArTPC
experiments~\cite{icarus_sulej, Baller:2017ugz}.  More advanced deconvolution
techniques have been explored in MicroBooNE~\cite{SP1_paper} and are validated with data
events in this work.  In brief, the deconvolution is used to extract the original
ionization charge signal $S(t)$ from the measured signal $M(t')$.  The measured signal
is modeled as the convolution of the original signal $S(t)$ and a known detector
response function $R(t'-t)$:
\begin{equation}\label{eq:decon_1}
  M(t') = \int_{-\infty}^{\infty}  R(t'-t) \cdot S(t) dt.
\end{equation}
This ``full'' detector response $R(t'-t)$ includes both the wire field response
and the electronics response,
and can be either simulated or estimated from data using an independent
calibration sample (see section~\ref{sec:calib_elec_resp} and section~\ref{sec:fieldresp}).  
One can transform Eq.~\ref{eq:decon_1} into the frequency domain by applying a Fourier transformation, yielding
$M(\omega) = R(\omega) \cdot S(\omega)$, where $\omega$ is in units of
angular frequency.  
The original signal in the frequency domain $S(\omega)$ can be found by rearranging this equation:
\begin{equation}\label{eq:decon_2}
  S(\omega) = \frac{M(\omega)}{R(\omega)}.
\end{equation}

This solution is extended to account for additional detector and electronics
noise contributions through the introduction of a low-pass filter function $F(\omega)$,
yielding
\begin{equation}\label{eq:decon_filt}
  S(\omega) = \frac{M(\omega)}{R(\omega)} \cdot F(\omega).
\end{equation}
The purpose of the filter function is to attenuate problematic high-frequency noise that
is amplified by the deconvolution procedure (division of noise by the response function at frequencies
where the signal response is very small). 

The above deconvolution prescription is known as the ``one-dimensional (1D) deconvolution''
in the sense that a fast Fourier transform (FFT) is carried out in the time dimension only. 
This procedure assumes the induced current on each sense wire is independent of the topology of the charge distribution creating the signals on the wires.
However, as described in Ref.~\cite{SP1_paper}, 
the induced current on the sense wire receives additional topology-dependent contributions from ionization charge drifting past adjacent wires at roughly the same time.  
Accounting for these additional contributions and assuming the field response strength between wires is approximately constant allows Eq.~\eqref{eq:decon_1} to be generalized as
\begin{equation}\label{eq:decon_2d_1}
    M_i(t_0) = \int_{-\infty}^{\infty} \left( ... + R_1(t_0-t)\cdot S_{i-1}(t) + R_0(t_0-t) \cdot S_i(t) + 
    R_1(t_0-t) \cdot S_{i+1} (t) + ...\right) \cdot dt,
\end{equation}
where $M_i$ represents the measured signal from wire $i$.  $S_{i-1}$, $S_i$, and
$S_{i+1}$ represent the real signal inside the boundaries of wire $i$
and its adjacent intra-plane neighbors. $R_0$ and $R_1$ are the average full
response functions for ionization charge passing through the wire
region of interest and adjacent wire regions respectively. 

Converting this equation into the frequency domain by applying a Fourier transform yields
\begin{equation}\label{eq:decon_2d_2}
    M_i(\omega) = ... + R_1(\omega) \cdot S_{i-1}(\omega) + R_0(\omega) \cdot S_i(\omega) + R_1(\omega) \cdot S_{i+1} (\omega) + ... ,
\end{equation} 
which can be written in matrix notation as 
\begin{equation}
  \begin{pmatrix}
    M_1(\omega)\\
    M_2(\omega)\\
    \vdots\\
    M_{n-1}(\omega)\\
    M_{n}(\omega)
  \end{pmatrix}
  =
  \begin{pmatrix}
    R_0(\omega) & R_1(\omega) & \ldots & R_{n-2}(\omega) & R_{n-1}(\omega) \\
    R_1(\omega) & R_0(\omega) & \ldots & R_{n-3}(\omega) & R_{n-2}(\omega) \\
    \vdots  & \vdots      & \ddots & \vdots          & \vdots \\
    R_{n-2}(\omega) & R_{n-3}(\omega) & \ldots & R_0(\omega) & R_1(\omega) \\
    R_{n-1}(\omega) & R_{n-2}(\omega) & \ldots & R_1(\omega) & R_0(\omega) \\
  \end{pmatrix}
  \cdot
  \begin{pmatrix}
    S_1(\omega)\\
    S_2(\omega)\\
    \vdots\\
    S_{n-1}(\omega)\\
    S_{n}(\omega)
  \end{pmatrix}.
  \label{eq:matrix_expansion}
\end{equation}
Inverting the matrix $R$ via application of Fourier transforms allows for the determination
of the original signal vector $S$.  This ``two-dimensional (2D) deconvolution'' yields a more accurate recovery
of the ionization electron distribution using the known detector response in both the time and wire dimensions.

The 2D deconvolution procedure described above provides a robust and computationally efficient method to
extract the distribution of ionization electrons reaching each wire at the LArTPC anode.  While this
procedure works well for the collection plane, additional signal processing is necessary for the induction
planes due to the suppression of the induction plane wire response at low frequencies, a feature
associated with the bipolar signals observed on the waveforms.  This suppression leads to low-frequency
noise becoming substantially amplified in the 2D deconvolution.
Without mitigation, the amplification of low-frequency noise would lead to large uncertainties in
the estimation of charge induced on the waveforms.  In order to address this issue, signal regions of
interest (ROIs) are first selected on the induction plane waveforms using high-pass filters that
reduce the impact of the low-frequency noise when finding the signals.  Once the ROIs are found, the local
baseline of the signal contained in the ROI is defined by an interpolation using the signal sidebands,
and a Gaussian filter is utilized to extract the ionization charge.  This procedure is discussed in
more detail in Ref.~\cite{SP1_paper}.

%% file: elec_resp_part2.tex
Each wire is individually instrumented by an electronic circuit that amplifies and shapes its current signals~\cite{Radeka:2011zz}.
These pre-amplifiers are implemented as application-specific integrated circuits (ASICs) and are operated at cryogenic temperature inside the liquid argon cryostat in order to reduce 
electronic noise and simplify detector cabling.
The amplified analog voltage signals are routed out of the detector cryostat, amplified again by intermediate warm electronics and digitized at 2~\si{\MHz} by 12-bit analog to digital converters (ADCs)~\cite{Acciarri:2016smi}.

Cryogenic (cold) pre-amplifying electronics are a crucial element of the MicroBooNE LArTPC as the reduced electronic noise during cryogenic operation compensates for the absence of ionization charge amplification within the liquid argon. 
Charge measurements made with induction wire signals especially require low electronic noise due to the relatively small currents induced on the wires by the drifting ionization charge~\cite{WCT_SP}.
Excellent cold electronic noise performance was observed in the MicroBooNE detector following the start of regular operation, with an equivalent noise charge (ENC) measured to be less than or equal to 400~$e^{-}$ for the majority of channels compared to an expected signal size of $\sim$17000~$e^{-}$ from minimum ionizing particles traveling parallel to the wire plane and perpendicular to the wire direction~\cite{noise_filter_paper}.
%
%
%
%

Each individual cold electronics ASIC channel can be configured to use one of four different gain settings (4.7, 7.8, 14
and 25~\si{\milli\volt/\femto\coulomb}) and peaking time settings (0.5, 1, 2 and 3~\si{\micro\second}).
During normal operation, MicroBooNE uses the 14~\si{\milli\volt/\femto\coulomb} gain setting and 2~\si{\micro\second} peaking time, which satisfies the Nyquist criterion~\cite{Nyquist1928} for 2~\si{\MHz} digital sampling.
The output baseline voltage is set to $\sim$200~\si{\milli\volt} and $\sim$900~\si{\milli\volt} for the collection plane and induction planes, respectively, in order to maximize the dynamic range of signals on the waveform. 
The baseline voltages are well-matched to both the overall 1.6~\si{\volt} dynamic range of the cold electronics and the 2~\si{\volt} range of the ADCs.

This section describes the in-situ calibration of the MicroBooNE LArTPC cold electronics using an externally generated signal injected into the pre-amplifier channel inputs.
Calibrating the cold electronics response allows for a direct comparison between the ideal and real electronics responses.
This in-situ calibration can also be used to improve the accuracy of simulated data and signal processing algorithms.
It is also of interest to determine the uniformity of the electronics response.
Finally, repeating the calibration measurement periodically is required in order to evaluate the stability of the electronics response over time.

%% file: elec_resp_coldelec_part2.tex
\begin{figure}[htb]
  \centering
  \begin{subfigure}{.69\textwidth}
  \centering
  \includegraphics[width=1.\textwidth]{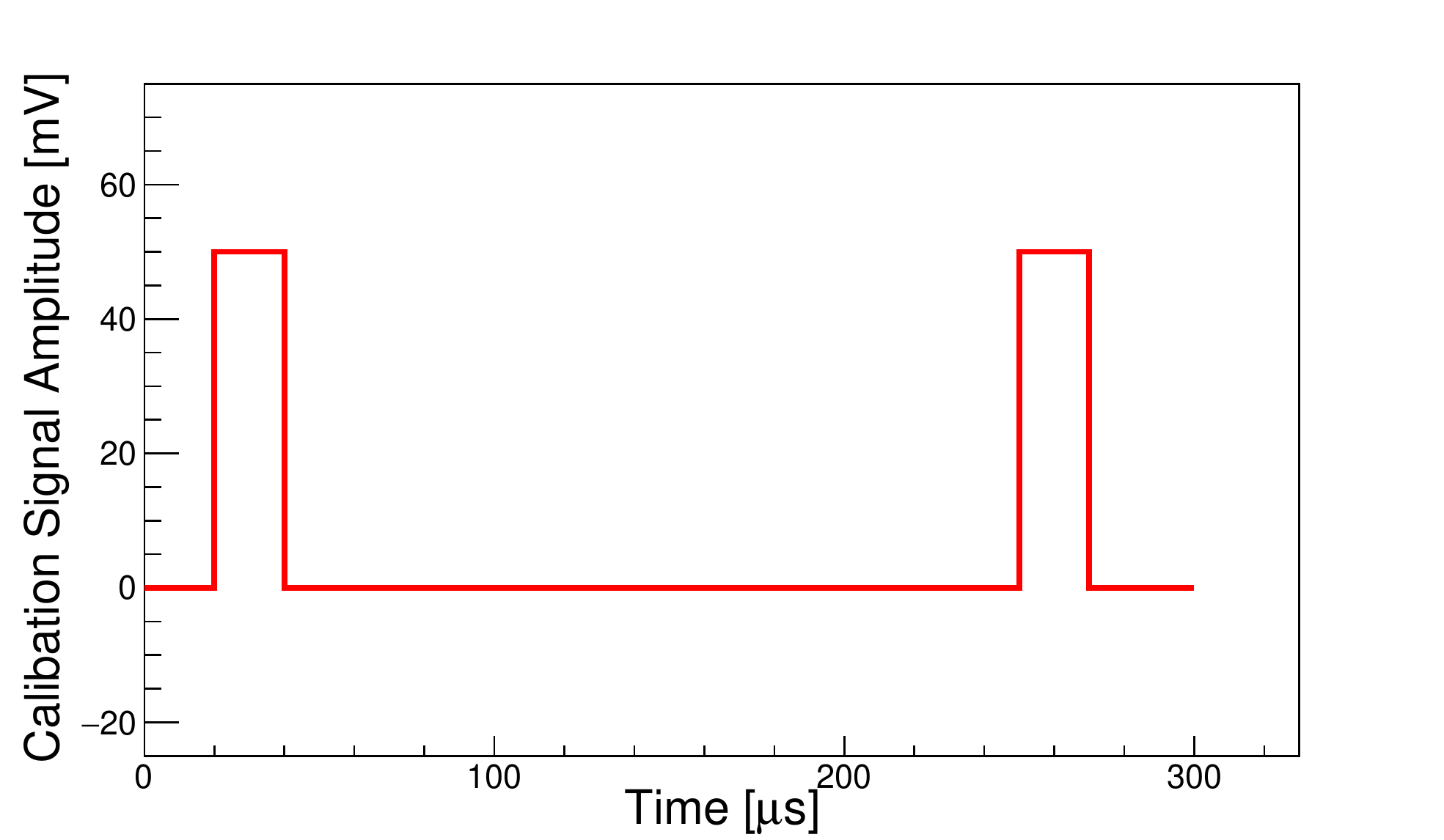}
  \end{subfigure}
  \hfill
  \begin{subfigure}{.29\textwidth}
  \centering
  \includegraphics[width=1.\textwidth]{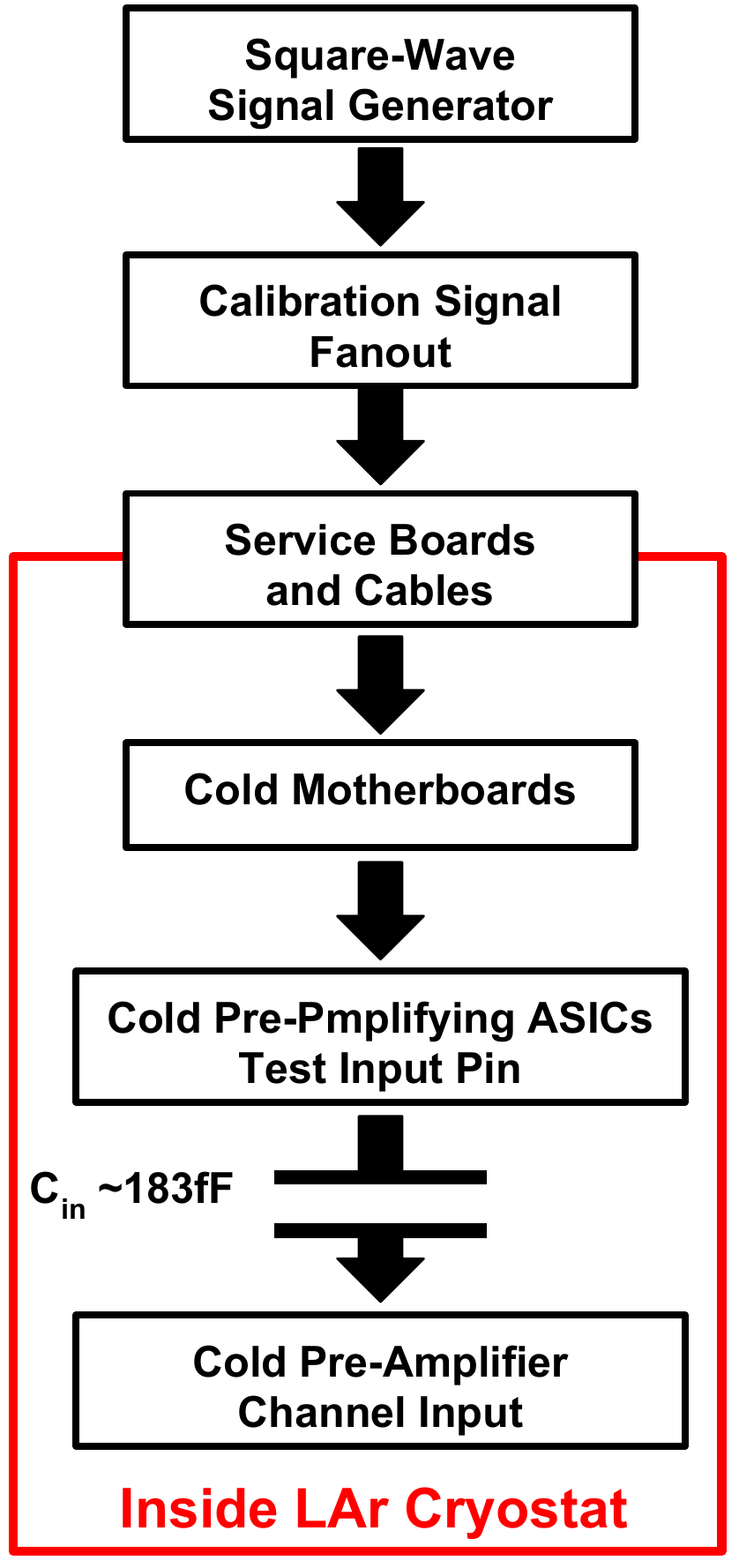}
  \end{subfigure}
  \caption{Schematic overview of the MicroBooNE cold electronics calibration signal system. The plot shows a $\sim$50 mV square wave voltage signal, generated outside the LAr cryostat and propagated to the cold electronics channel inputs via the calibration injection system, shown schematically in the graphic. Copies of the square wave voltage signal are generated with a fanout module and transmitted into the cryostat through dedicated service boards and cables. The signal is routed to every cold electronics ASIC test input pin and coupled to each pre-amplifier input through a $\sim$183 fF capacitor.}
  \label{fig:elecresp_calibSystem}
\end{figure}
\begin{figure}[htb]
  \centering
  \includegraphics[width=0.8\textwidth]{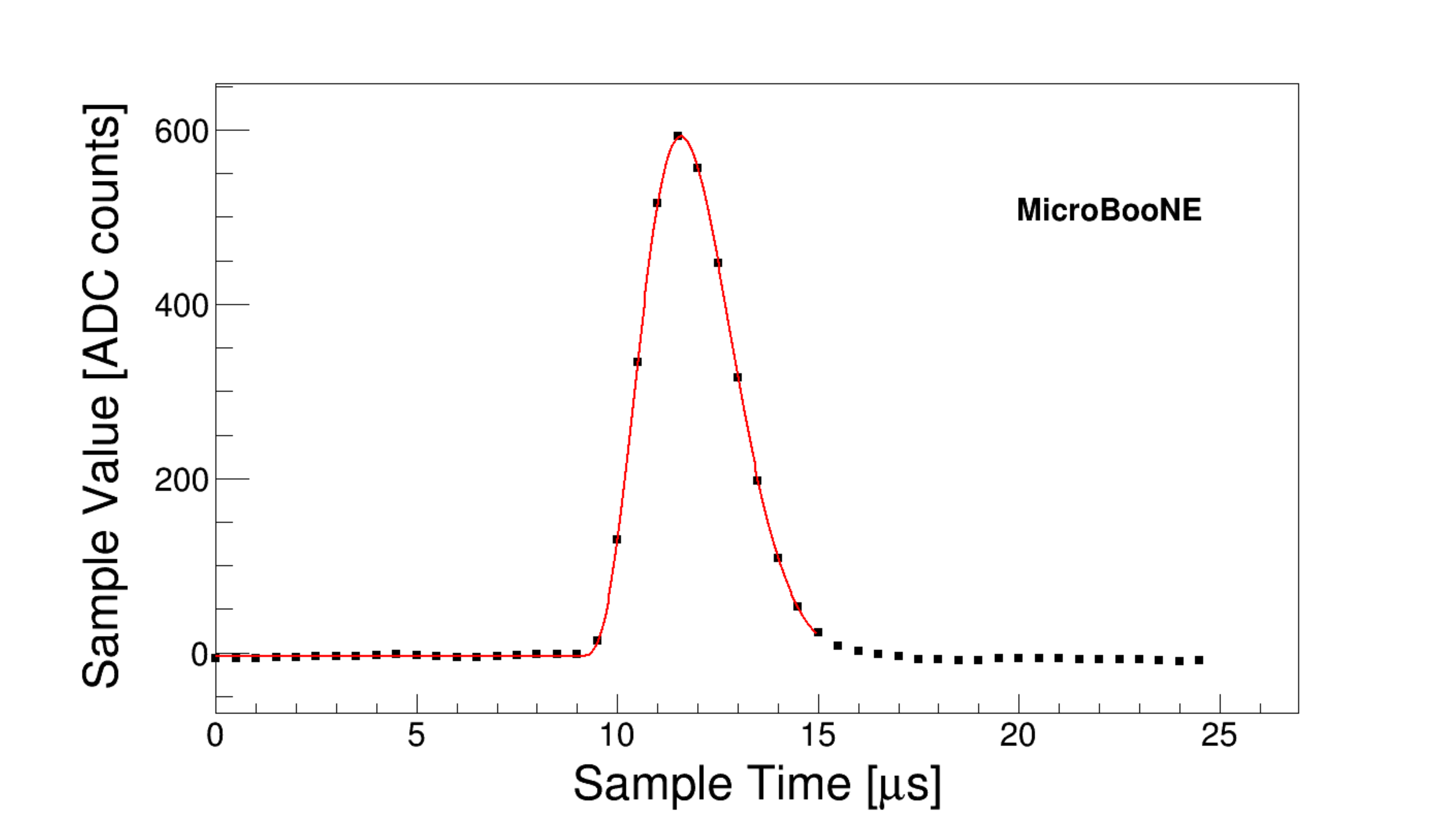}
  \caption{Example of a digitized waveform from an injected calibration signal on a single collection plane channel, including a fit to the electronics response function (overlaid in red). 
The fitted peaking time $t_p$ is $2.15 \pm 0.01$~\si{\micro\second} and the amplitude $A_0$ is $599 \pm 3$~ADCs.}
  \label{fig:elecresp_exampleCalibSignal}
\end{figure}

The MicroBooNE cold electronics calibration system uses a signal generator located outside the cryostat to provide a calibration signal to each
individual cold electronics channel input, as summarized in figure~\ref{fig:elecresp_calibSystem}.
The signal generator produces a square wave signal that is fanned out and propagated into the detector cryostat through dedicated service cables.
Service cables inside the cryostat connect to the ASIC carrier motherboards where the calibration signal is routed to each individual ASIC test input pin. 
Internally, the ASIC capacitively couples the test input signal to each pre-amplifier input individually through a $\sim$183~\si{\femto\farad} capacitor.
The ASICs must be configured to connect the test signal input pin to the calibration capacitors; during normal detector operation they are disconnected.
The capacitively coupled square wave calibration signal injects a known amount of charge into the pre-amplifier input on a timescale of $\sim$10~\si{\nano\second}, which is much shorter than the nominal pre-amplifier peaking time of 2~\si{\micro\second}.
Consequently the resulting signal waveform approximates the impulse response of the cold pre-amplifiers.
An example of a digitized calibration signal waveform is shown in figure~\ref{fig:elecresp_exampleCalibSignal}.


A number of channels of the cold electronics are considered non-functional and cannot be calibrated, with the specific causes discussed in Ref.~\cite{noise_filter_paper}.
Roughly 220~channels (out of 8256) of the cold electronics are located on ASICs that cannot be configured due to a hardware failure of the configuration chain from a service board through the cold cable to the cold electronics inside the cryostat.
These channels are still used to measure ionization charge but can only use the default 4.7 mV/fC gain and 1~\si{\micro\second} peaking time settings.
Calibration signals can not be injected into these channels as the ASICs must be specifically configured to enable the test input pin.
An additional $\sim$860~channels are not functioning correctly and similarly can not be calibrated. 
These channels are either unresponsive due to ASIC failure or to a short between different wire planes, and are approximately stable in number over time as discussed in Ref.~\cite{noise_filter_paper}.


%% file: elec_resp_parameterization_part2.tex
The impulse response of the TPC cold electronics is well described by a fifth-order semi-Gaussian anti-aliasing filter. The representation of this function in the time domain is shown in figure~\ref{fig:elecresp_exampleCalibSignal} and is parameterized by

\begin{equation}
\label{eq:ideal_time_resp}
\begin{aligned}
R(t,A_0,t_p) = &\,A_1 E_1 - A_2 E_2 ( \cos\lambda_1 + \cos\lambda_1 \cos\lambda_2 + \sin\lambda_1 \sin\lambda_2 ) \\
& + A_3 E_3 ( \cos\lambda_3 + \cos\lambda_3 \cos\lambda_4 + \sin\lambda_3 \sin\lambda_4 ) \\
& + A_4 E_2 ( \sin\lambda_1 - \cos\lambda_2 \sin\lambda_1 + \cos\lambda_1 \sin\lambda_2 ) \\
& - A_5 E_3 ( \sin\lambda_3 - \cos\lambda_4 \sin\lambda_3 + \cos\lambda_3 \sin\lambda_4 ).
\end{aligned}
\end{equation}

The parameters in Eq.~\ref{eq:ideal_time_resp} are obtained from a detailed simulation of the filter design and given by:
\begin{equation}
\label{eq:ideal_time_resp_par}
\begin{aligned}
A_1 &= 4.31054 A_0,
& A_2 &= 2.6202 A_0, \\
A_3 &= 0.464924 A_0,
& A_4 &= 0.762456 A_0,
& A_5 &= 0.327684 A_0, \\
E_1 &= e^{\frac{-2.94809 t}{t_p}},
& E_2 &= e^{\frac{-2.82833 t}{t_p}},
& E_3 &= e^{\frac{-2.40318 t}{t_p}}, \\
\lambda_1 &= 1.19361 \frac{t}{t_p},
& \lambda_2 &= 2.38722 \frac{t}{t_p}, \\
\lambda_3 &= 2.5928 \frac{t}{t_p},
& \lambda_4 &= 5.18561 \frac{t}{t_p}, \\
\end{aligned}
\end{equation}
where $t$ is time in \si{\micro\second}, $t_p$ is the peaking time constant in \si{\micro\second}, and $A_0$ is the amplitude parameter in units
of ADC counts for digitized waveforms.

Electronics response calibration data are recorded while the detector is in its normal operational state. 
The cold electronic pre-amplifier ASICs are configured to use their default settings with the exception
that the calibration signal input is enabled for every electronics channel in the detector. 
Calibration signal pulse shapes are identified in digitized waveform data and the leading edge fitted using the response function described in Eq.~\ref{eq:ideal_time_resp}. 
Calibration signal amplitude and peaking time parameters for each channel are obtained from these fits.
The results of these fits are used to parameterize the electronics response for each channel.

 

\begin{figure}[tb]
  \centering
  \begin{subfigure}[t]{0.495\textwidth}
    \centering
    \includegraphics[width=1.0\textwidth]{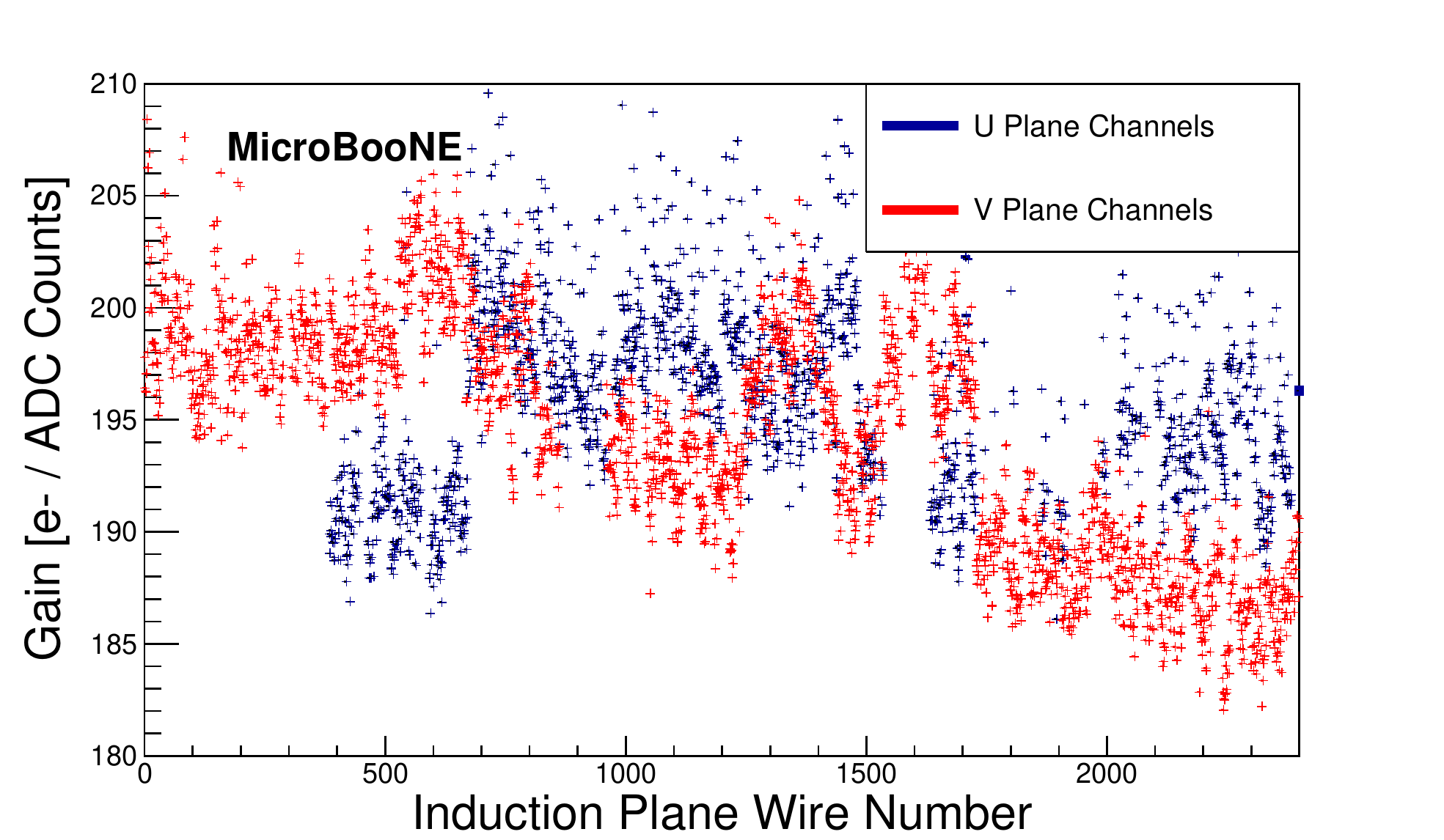}
    \caption{Uncorrected gain for induction plane channels.}
  \end{subfigure}
  \hfill
  \begin{subfigure}[t]{0.495\textwidth}
    \centering
    \includegraphics[width=1.0\textwidth]{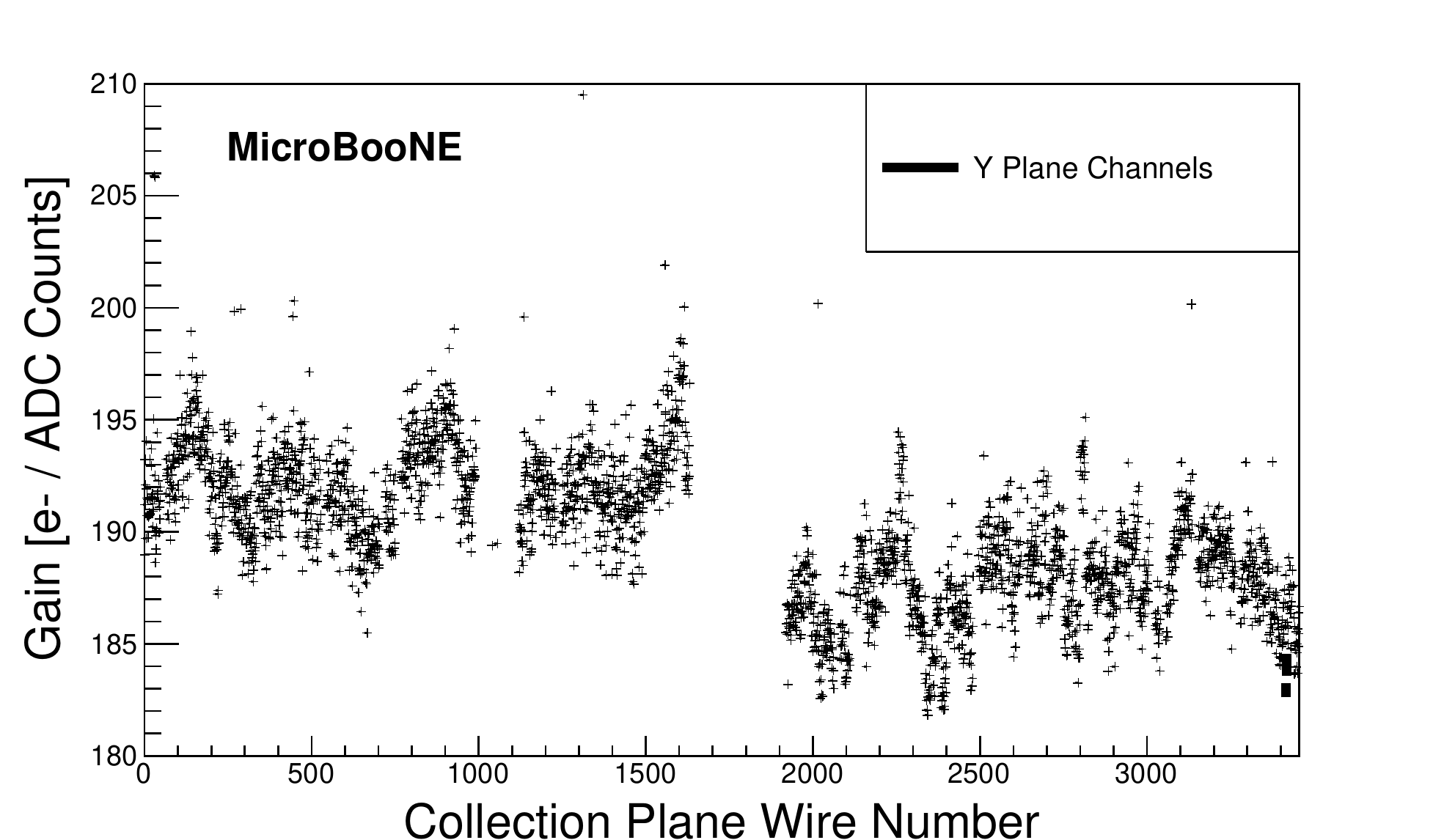}
    \caption{Uncorrected gain for collection plane channels.}
  \end{subfigure}
  \vspace{1cm}
  \begin{subfigure}[t]{0.495\textwidth}
    \centering
    \includegraphics[width=1.0\textwidth]{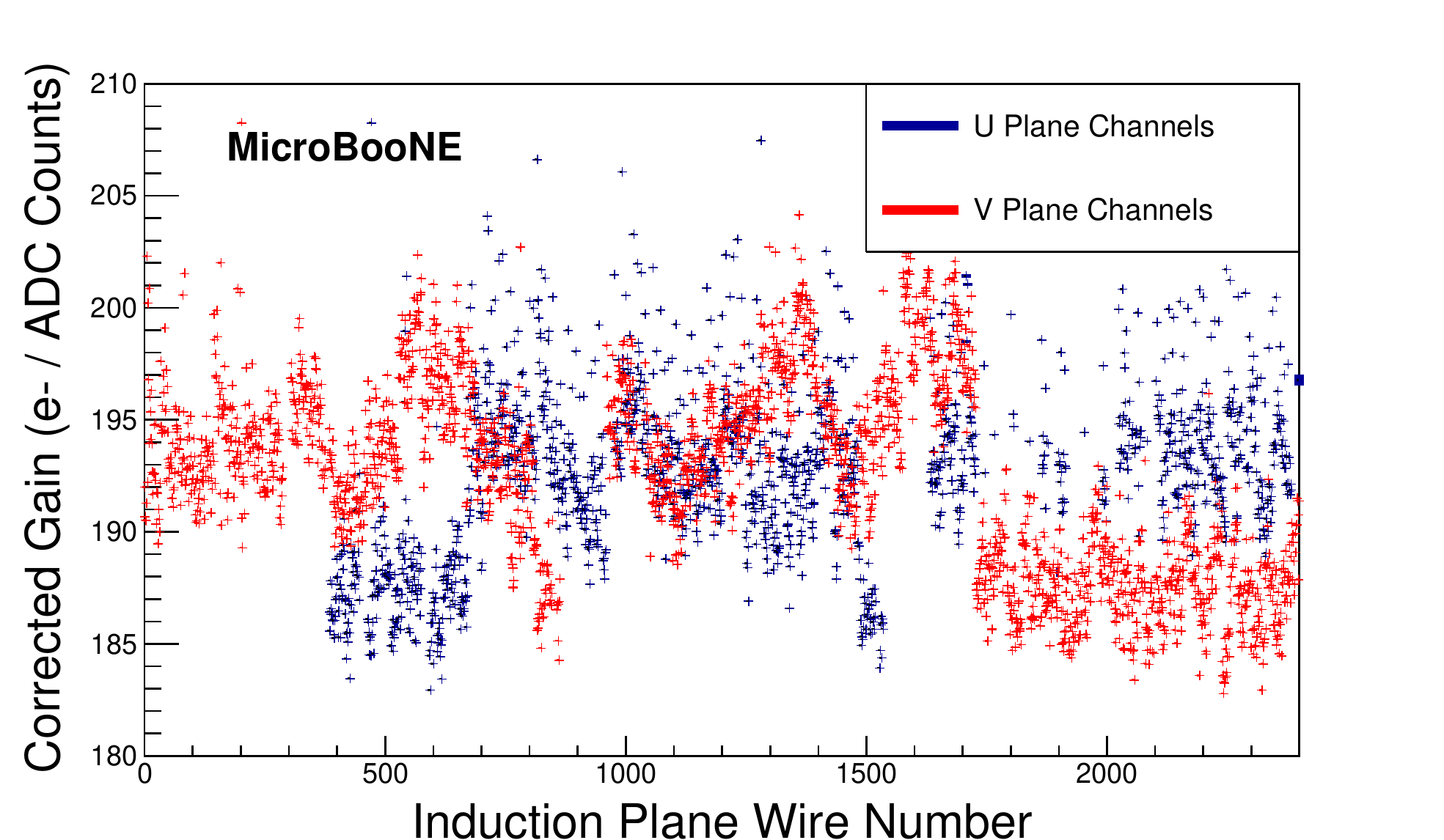}
    \caption{Corrected gain for induction plane channels.}
  \end{subfigure}
  \hfill
  \begin{subfigure}[t]{0.495\textwidth}
    \centering
    \includegraphics[width=1.0\textwidth]{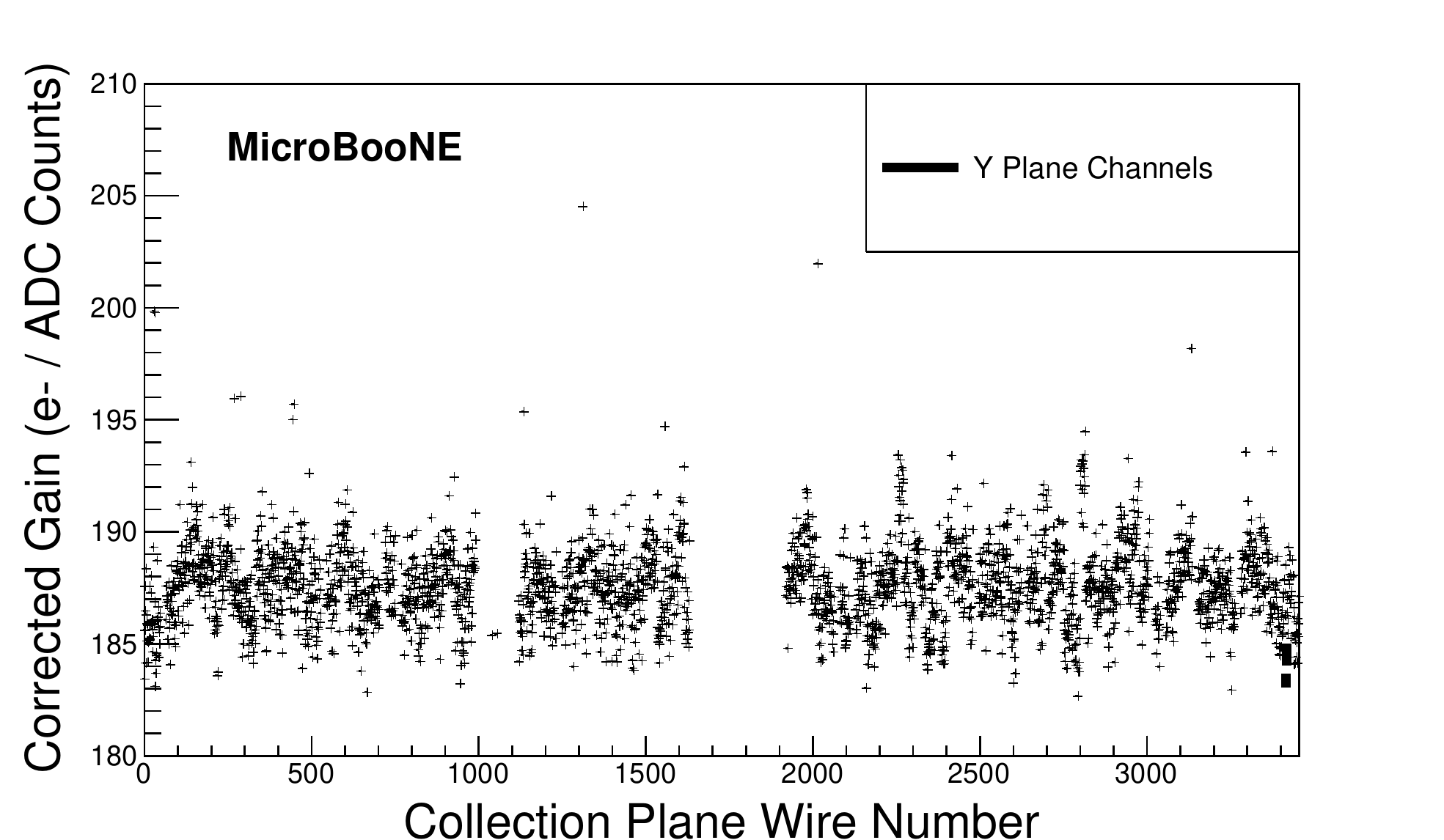}
    \caption{Corrected gain for collection plane channels.}
  \end{subfigure}

  \caption{Uncorrected gain measurement for every TPC induction (a) and collection (b) plane channel.  
Also shown is the induction (c) and collection (d) channel gain measurement after correcting for both calibration module attenuation and service cable dependence. 
The first few hundred U Plane channels cannot be calibrated due to the inability to configure the cold electronic ASICs, and are not plotted here.
A subset of induction plane channels are misconfigured to use the collection 200~\si{\milli\volt} baseline setting, which contributes to some of the residual variations in the induction wire corrected gain measurements.
}
  \label{fig:elecresp_gainCalib}
\end{figure}

Electronics channel gain can be derived from the fitted calibration pulse amplitude given the known amplitude of the injected signals.
These gain measurements need to be corrected to account for attenuation introduced by the calibration signal injection system.
The most important correction to electronics response gain measurements is due to the calibration signal fanout modules. 
A pair of modules receives the analog calibration signal from the function generator and propagates copies to each set of motherboards sharing common service cables.
The calibration signal input for one of these modules is output from the other, and this daisy-chained configuration attenuates the second calibration signal copy.
This can be observed directly by comparing calibration signal average pulse height measurements for sets of channels connected to each module, as in the top row of figure~\ref{fig:elecresp_gainCalib}.
A correction is applied to the gain measurements for channels receiving the calibration module attenuated signal to remove this effect.
A similar correction removes variations in gain measurements due to service cable attenuation.
These corrections are constant over time, although there are two sets of service cable attenuation corrections corresponding to the periods before and after a service board hardware upgrade was performed in summer 2016.
Calibration signal injection is observed to induce electronic crosstalk into adjacent channels with a bipolar signal size $\sim$1\% of the calibration pulse height, which is small enough to not require any correction.

\begin{figure}[tb]
  \centering
  \includegraphics[width=0.8\textwidth]{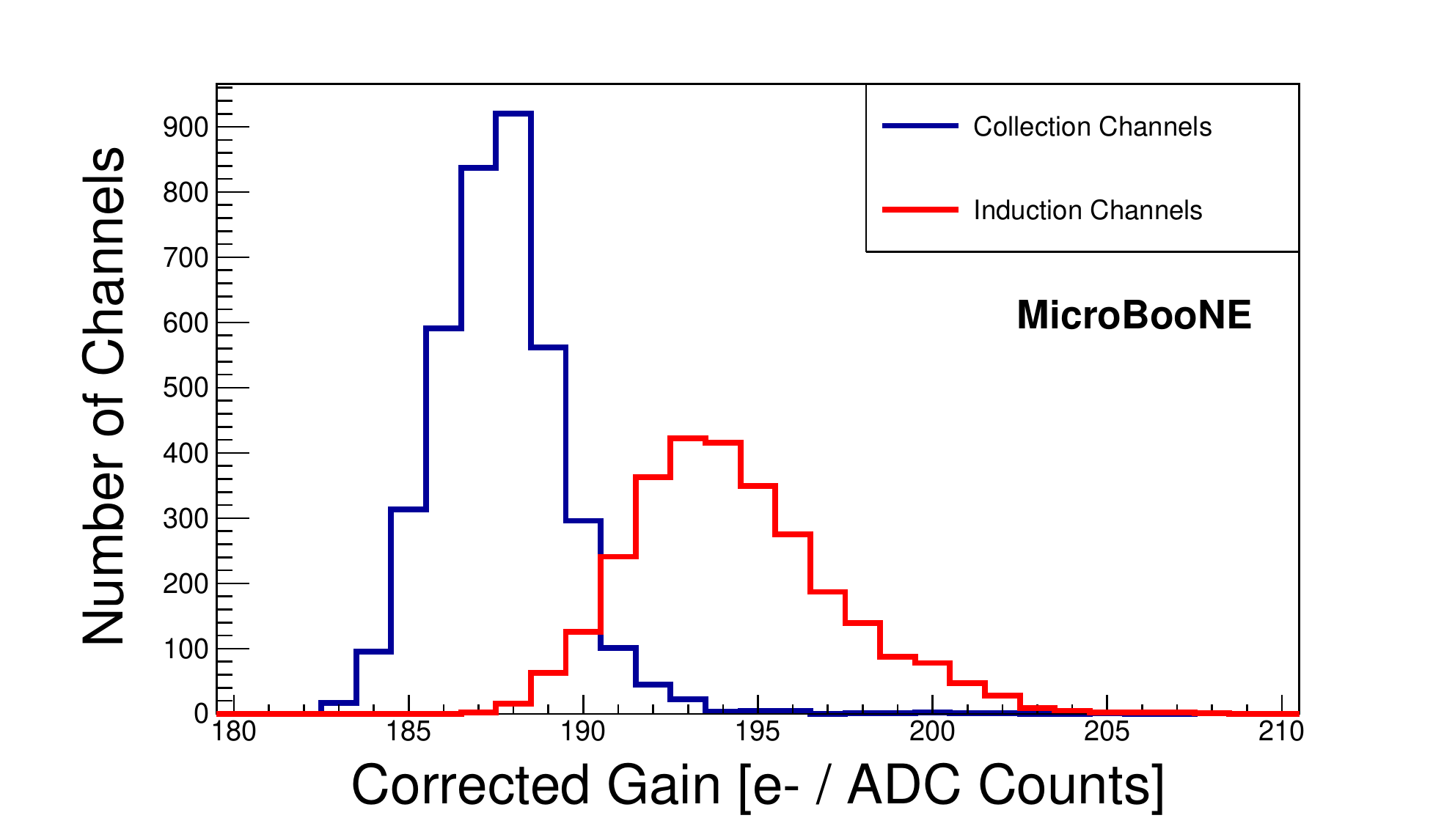}
  \caption{Measured TPC electronics pre-amplifier gain after applying calibration fanout module and service cable corrections.  A systematic discrepancy is observed between collection channels and induction channels, an understood feature of the pre-amplifier.  The average measured gain of the induction channels is $194.3 \pm 2.8$ [$e^{-}$/ADC] while the average measured gain of the collection channels is $187.6 \pm 1.7$ [$e^{-}$/ADC].}
  \label{fig:elecresp_corrGain}
\end{figure}

The gain measurements after applying the previously defined calibration module fanout and service cable corrections are shown in the bottom row of figure~\ref{fig:elecresp_gainCalib}.
There is a systematic discrepancy in measured gain between channels configured with the collection baseline setting and those configured with the induction baseline setting, as shown in figure~\ref{fig:elecresp_corrGain}, that is an intrinsic feature of the cold pre-amplifiers.
Channels configured to use the induction baseline setting have an average measured gain of $194.3 \pm 2.8$ [$e^{-}$/ADC] while the average collection-mode channel gain is $187.6 \pm 1.7$ [$e^{-}$/ADC].
This discrepancy contributes to some of the residual variations in the induction wire corrected gain measurements as a subset of induction plane channels are misconfigured to use the collection 200~\si{\milli\volt} baseline setting.
The variation in the $\sim$183~\si{\femto\farad} test signal injection capacitors is $\sim$0.5\%, which contributes to the observed variation in these gain measurements.
Additionally, these gain measurements are also expected to have a systematic bias due to the overall average attenuation introduced by the calibration injection system.
Calibration methods such as measuring energy deposition from minimimum ionizing particles are required to measure the absolute gain given the inability to measure this overall attenuation factor in-situ.

\begin{figure}[tb]
  \centering
  \includegraphics[width=0.92\textwidth]{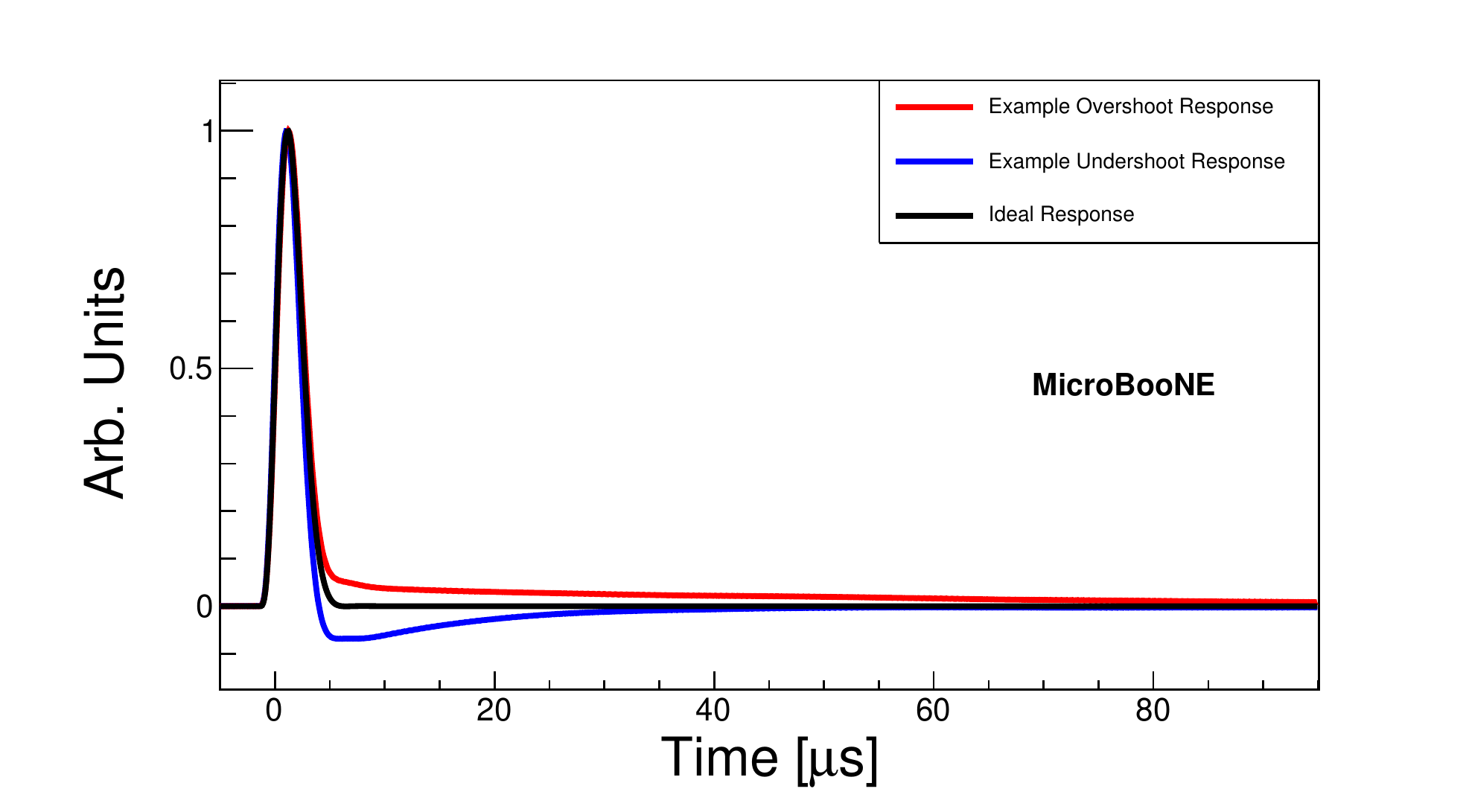}
  \caption{Example calibration signal waveforms, averaged over many signal pulses, with clearly visible extended tails. 
These extended tails are not well described by the ideal electronics response, which is also shown for comparison (black).  
This is a known feature in this version of the cold electronic ASICs and has been addressed in a subsequent version of the electronics~\cite{pole_zero_fix}.
These tails are corrected in the signal processing chain using the deconvolution-based method described in this section.
}
  \label{fig:elecresp_nonidealExamples}
\end{figure}

The electronics response peaking time parameter is obtained from fits to calibration signals for each cold pre-amplifier channel individually.
Peaking time parameters are not measured for misconfigured or non-functioning channels.
The mean peaking time is $2.18 \pm 0.08$~\si{\micro\second}, which is $\sim$10\% higher than the nominal value
of 2~\si{\micro\second}, due to the additional input capacitance from the TPC wires.
The overall variation in fitted pre-amplifier peaking time is $\sim$3\%, which is significantly less uniform than the electronics gain with $\sim$1\% variation.
This variation does not seem to be caused by the implementation of the calibration system but is a feature of the pre-amplfiier ASICs.
Figure ~\ref{fig:elecresp_nonidealExamples} suggests that the source of this variation is a disagreement in the shape of the ideal and real electronics response.
In particular, the real electronics response has much longer tails than does the ideal response, which is a known feature in this version of the cold electronic ASICs and has been addressed in a subsequent version of the electronics~\cite{pole_zero_fix}.

The observation of non-ideal components in the calibration signal waveforms suggests that parameterizing the response in terms of the
idealized response may introduce bias in charge measurements when comparing data to simulation.
As an alternative, it is possible to measure the average impulse electronics response shape from calibration signals and use this
in place of the ideal response for use in signal processing algorithms, as described in section~\ref{sec:calib_elec_resp:validation}.

%% file: elec_resp_stability_part2.tex
\begin{figure}[tb]
  \centering
  \includegraphics[width=0.825\textwidth]{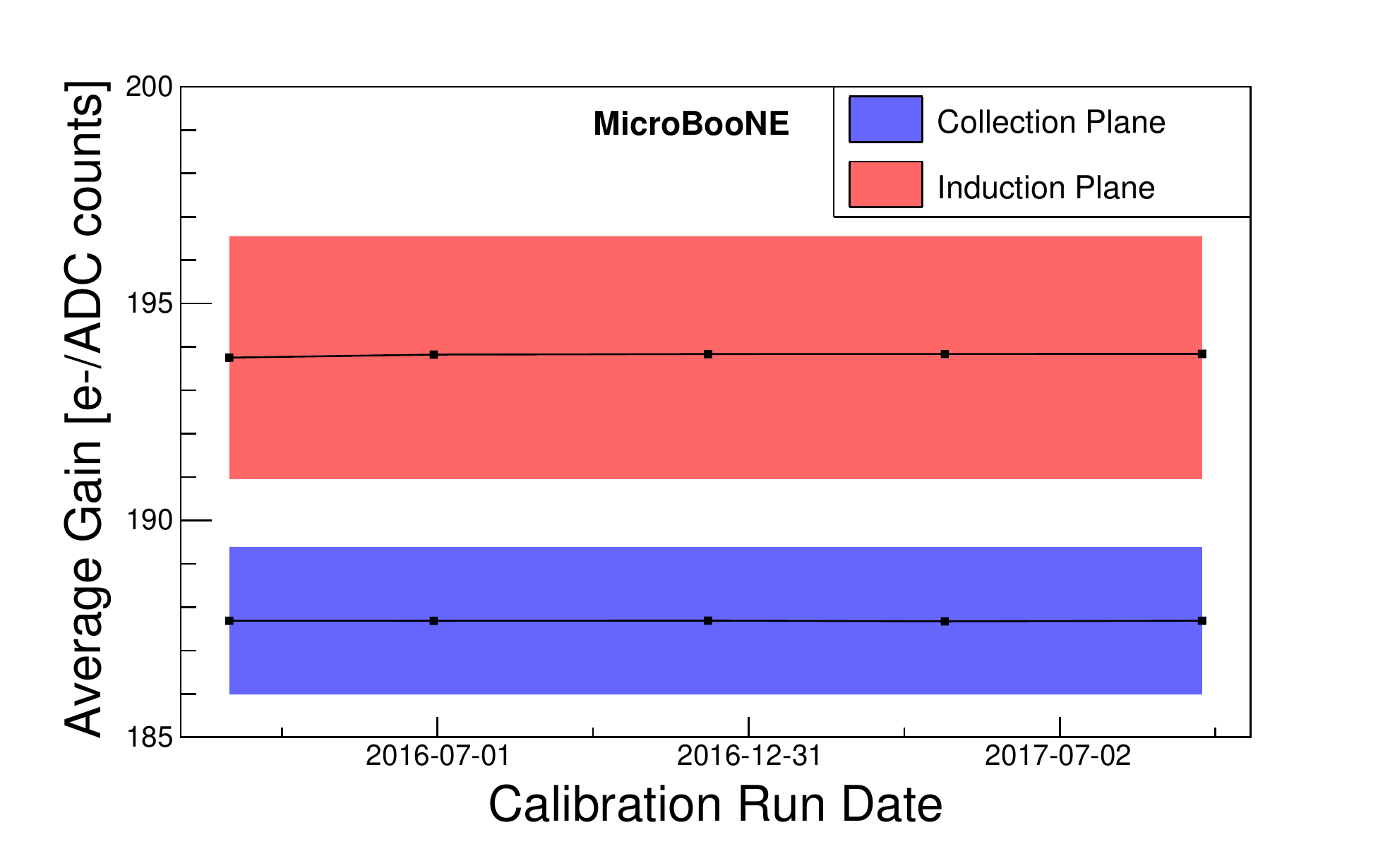}
  \caption{Gain measurements of the TPC cold electronics channels taken over time, with red and blue colored bands representing the RMS variation of the corrected channel gain in the induction and collection planes, respectively. The average discrepancy in channel gain measurements between calibration data runs is less than 0.2\% and is significantly smaller than the variation in channel gain within the induction and collection wire planes.
          }
  \label{fig:elecresp_gainStability}
\end{figure}

In order to apply electronics response measurements in signal processing algorithms, it is necessary to demonstrate that the response is stable over time.
If the response changes significantly over time, there is the potential to bias ionization charge measurements by using response parameters that are not appropriate for a given time period.
The stability of the electronics response is evaluated by comparing the measured electronics response parameters from calibration runs spaced several months apart.
It is observed that the measured cold pre-amplifier gain is very stable over this timescale, as shown in figure~\ref{fig:elecresp_gainStability}, with an average deviation less than 0.2\% across all measurements.

%% file: elec_resp_validation_part2.tex
In this section we demonstrate how the measured average response waveforms can correct the observed non-uniform and non-ideal electronics response.
This correction is achieved through a 1D deconvolution based on discrete-space Fourier transformation techniques described in section~\ref{sec:introduction:definitions}. 
The correction can be defined in the frequency domain as:
\begin{equation}
M_i^{corr}(\omega) = M_i(\omega) \cdot \frac{R_{ideal}(\omega)}{R_i(\omega)},
\label{eq:elec_resp_corr}
\end{equation}
where $\omega$ is in units of angular frequency, $R_{ideal}(\omega)$ is the Fourier transform of the ideal electronics response function and $R_i(\omega)$ is derived from the measured average electronics response for the $i$th channel.
Since the $R_{ideal}$ and $R_i$ are generally similar, the change in the electronics noise is minimal and can be safely ignored.
The above procedure only acts on a single channel at a time (i.e. independent of any other channels). 
Therefore, it can be applied before the general 2D deconvolution~\cite{SP1_paper} dealing with the long range effect of the field response across multiple wires. 

\begin{figure}[tb]
  \centering
  \includegraphics[width=0.8\textwidth]{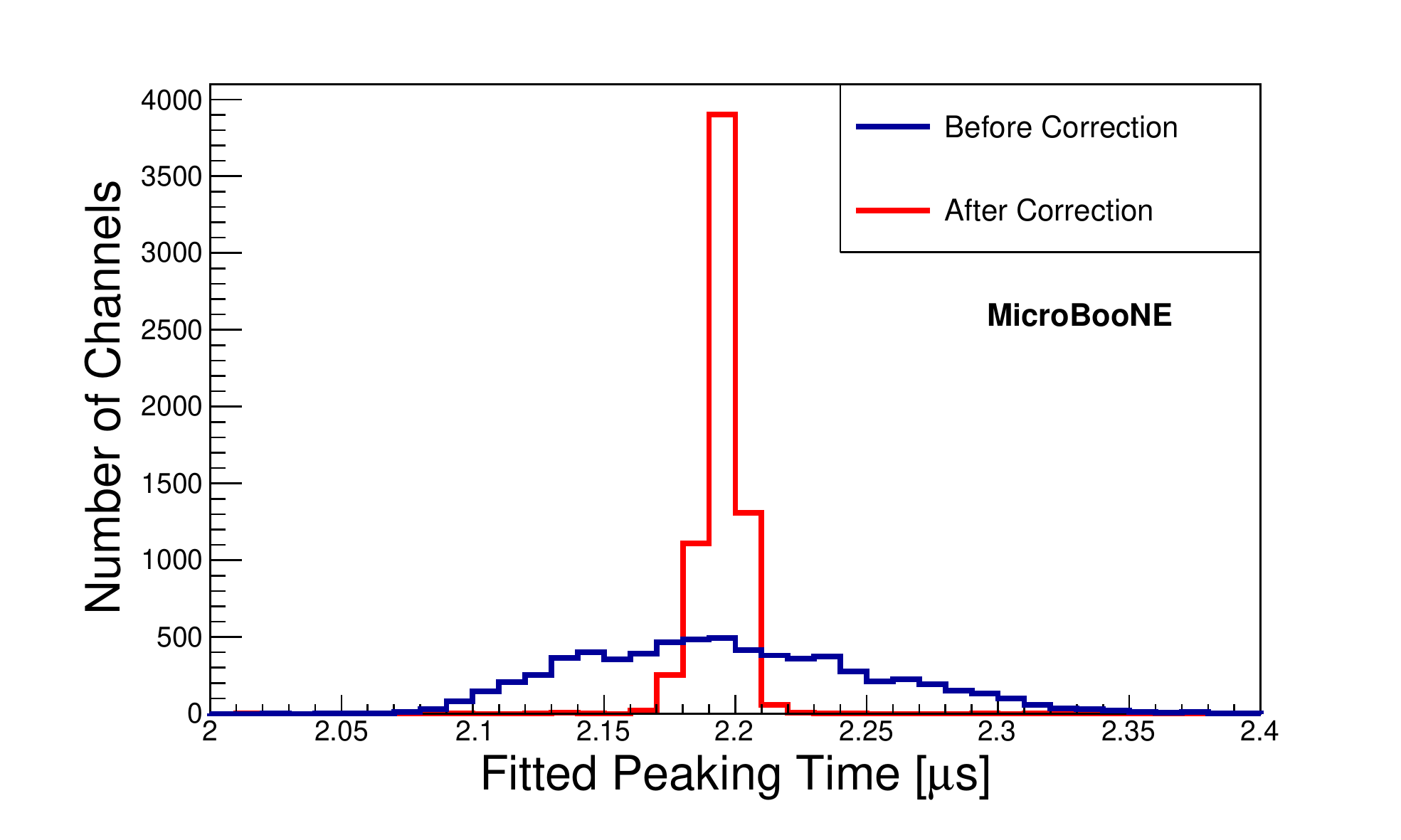}
  \caption{Peaking time measurement for the channels of the TPC cold electronics before (blue) and after (red) the electronics
response correction. 
The overall variation in the measured peaking time after applying the electronics response correction is $\sim$0.8\% compared to $\sim$3.5\% before the correction.
          }
  \label{fig:elecresp_prepostPulserShape}
\end{figure}

The electronics response correction defined in Eq.~\ref{eq:elec_resp_corr} is applied to injected calibration signals in order to evaluate the impact on the shape of the impulse response.
Injected calibration signals with and without the electronics response correction applied are fitted with the ideal electronics response function.
The distributions of the measured peaking times are summarized in figure~\ref{fig:elecresp_prepostPulserShape}. 
The calibration signal shapes are significantly more uniform after the electronics response correction is applied, 
with $\sim$0.8\% variation in the measured peaking time parameter after the correction compared to the original $\sim$3\% variation.

\begin{figure}[tb]
  \centering
  \begin{subfigure}{.99\textwidth}
  \centering
  \includegraphics[width=1.\textwidth]{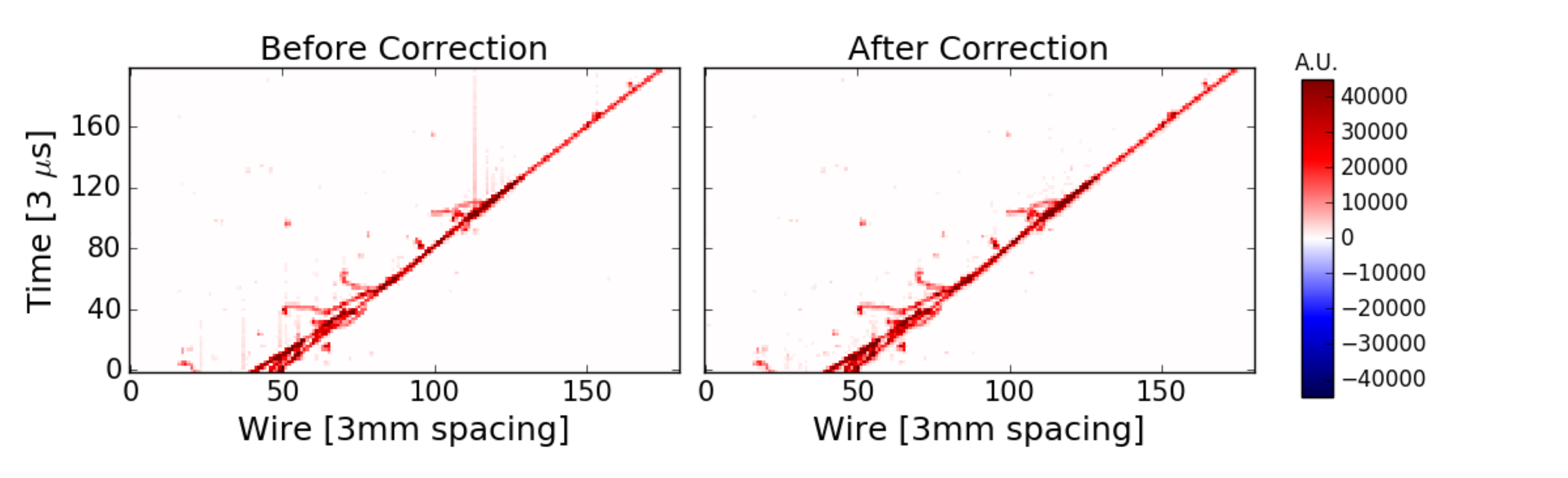}
    \Put(-170,230){\fontfamily{phv}\selectfont \textbf{MicroBooNE}}
  \caption{Event displays before and after electronic response correction.}
  \end{subfigure}
  \vspace{1cm}
  \begin{subfigure}{.59\textwidth}
  \centering
  \includegraphics[width=1.\textwidth]{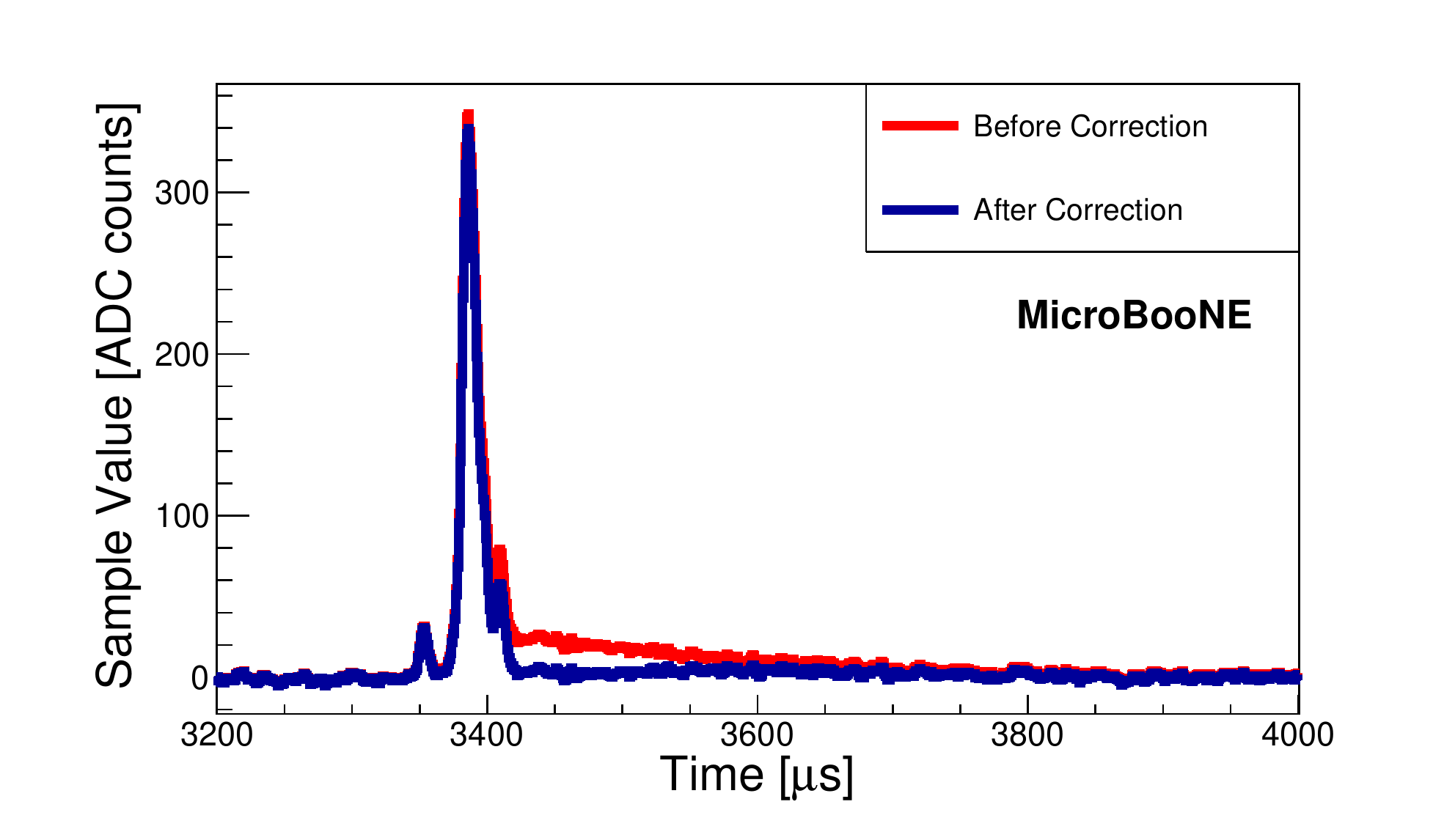}
  \caption{Waveform before and after electronic response correction.}
  \end{subfigure}
  \caption{Example event displays showing the Y plane view from Event 0, Run 3455 (a) before and after the application of the electronic response correction defined in Eq.~\ref{eq:elec_resp_corr} in units of average baseline subtracted ADC scaled by 250 per 3~\si{\micro\second}.
The long tails (faint vertical stripes going upward from the track) seen on certain collection plane channels following large charge signals are largely removed by the correction. The effect of the correction on an
individual digitized waveform (b) also clearly shows this tail removal.}
  \label{fig:elecresp_prepostCorr}
\end{figure}

This correction is also evaluated by applying it to signals produced by a cosmic muon track recorded during normal detector operations as shown in figure~\ref{fig:elecresp_prepostCorr}.
This and following event displays show digitized waveform samples from a subset of wires in a specified plane over a fraction of the total time period recorded for the event.
The channels and time window were chosen to highlight signals produced by the ionization charge of the cosmic muon, which produce a recognizable image of the track.
Signals before and after the response correction are both shown.
The horizontal axes correspond to the number of wires relative to the leftmost wire while the vertical axes correspond to units of time.
The vertical axes time bins are 3~\si{\micro\second} intervals and the contents are the sum of the six digitized samples within that interval with average baseline subtracted, which defines the vertical scale.
The relative spatial displacement away from the wire planes of ionization charge producing the observed signals can be inferred from the difference in time and assuming the nominal ionization electron drift velocity.
Of particular note in these event displays is that the long tails observed on several collection channels following large ionization charge deposition are removed.

%% file: field_resp_part2.tex
As discussed in section~\ref{sec:introduction}, the characteristics of TPC signals are dictated by the
initial ionization charge cloud formation within the detector bulk, the signals
created on the TPC wires due to the drift of the ionization electrons near the wires (field response),
and the shaping of the signals via the front-end electronics (electronics response).  In the previous
section, the calibration of the electronics response was discussed.  This section now
focuses on the nature of the wire field response, which must be probed with ionization signals as
opposed to signals from the electronics calibration system described in section~\ref{sec:calib_elec_resp}.

In order to validate the simulation of the wire field response and electronics
response~\cite{SP1_paper}, we compare the simulated
field response and electronics response to that seen in data.  This requires
utilizing a data-driven technique to produce comparisons of
simulation and data at the waveform level.  In section~\ref{sec:fieldresp:methodology},
the data-driven method used to make this comparison is described.  In particular,
comparisons of the full response, which is the convolution of the wire field
response and electronics response, are made.  Despite making comparisons of the full response,
this method primarily demonstrates how well we simulate the wire field response,
as the electronics response is first calibrated in data using the cold electronics
calibration system (see section~\ref{sec:calib_elec_resp}).
The consistency of the wire field response between data and simulation is also an
indirect validation of the signal processing chain that is used to ``remove'' the field
and electronics response from the waveform, which is done in order to extract the amount
of charge seen on the TPC wires.  Correctly modeling the wire field response and electronics
response allows for an unbiased estimation of ionization charge on the TPC wires in data.

\begin{figure}[tb]
  \centering
  \includegraphics[width=\figwidth]{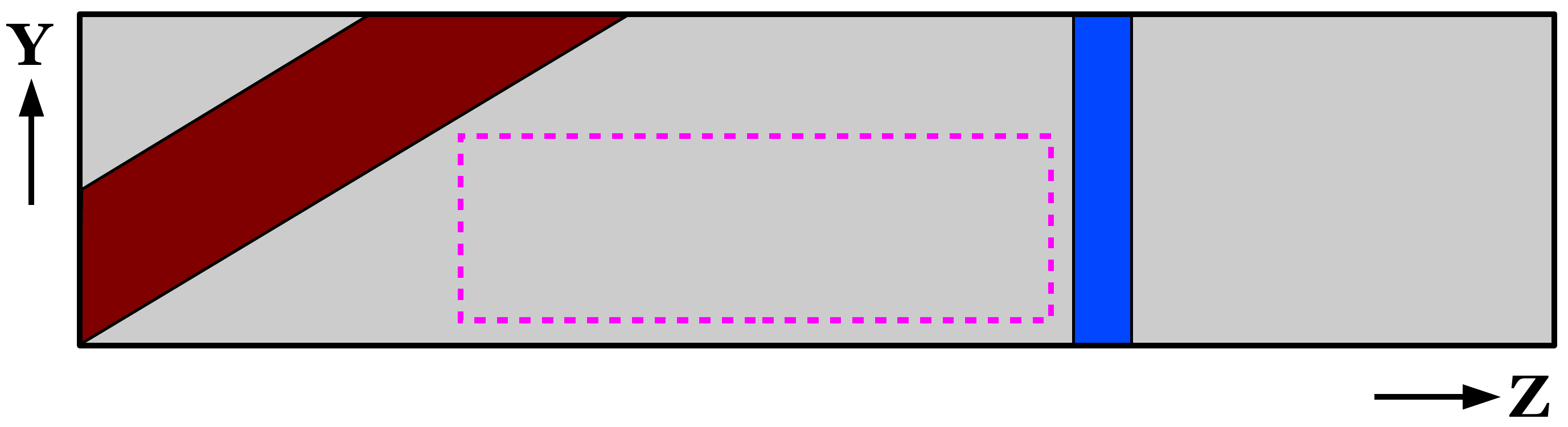}
  \caption{Illustration of the subregions of the TPC active volume that are used to
  study the different wire field responses from data, looking at a view of the anode plane
  ($y-z$ plane).  Depicted are the ``shorted-U'' and ``shorted-Y'' regions in red and blue,
  respectively.  Also shown is the subregion of the ``normal'' region used to study
  the nominal field response shapes (area inside dashed magenta box).  The TPC dimensions
  in the illustration are roughly 2.3~meters in the $y$~direction and 10.4~meters in the
  $z$ direction.}
  \label{fig:shortedregions}
\end{figure}

There is an additional complication due to there being multiple regions within
the TPC exhibiting non-standard wire field responses.  This is due to the shorting
of channels across wire planes in particular regions of the TPC, as described in Ref.~\cite{noise_filter_paper}.
The nature of the shorting of channels between the different
wire planes has not been observed to change since the beginning of data-taking at MicroBooNE.
The different bias voltages seen on channels in these regions lead to different
induced current distributions as ionization electrons move past the wires.
In addition to the ``normal'' region, where there are no shorted channels,
there is a ``shorted-U'' region where multiple U plane channels are shorted to one or more V plane channels,
and a ``shorted-Y'' region where multiple Y plane channels are shorted to one or more V plane channels.
The wire field responses are extracted separately in these different regions and
the response is assumed to be uniform within each of these regions.  These subregions
are illustrated in figure~\ref{fig:shortedregions}.  Similar to the normal region, the field response
simulation in the shorted-channel regions is done using Garfield with identical wire geometry as
described in Ref.~\cite{SP1_paper}, in which the signal formation mechanism has been
elaborated in detail. In contrast to the simulation in the normal region,
in the shorted-channel regions the bias voltages on the shorted channels are adjusted to
yield significant voltage reduction on the wires.  This tuning leads to better agreement
of the simulated wire field response to that observed in data.

After discussing the data-driven method used to extract the full response from data in
section~\ref{sec:fieldresp:methodology}, comparisons of the full responses in data and simulation
are made for the normal region in section~\ref{sec:fieldresp:normalwires} and for the shorted-U
and shorted-Y regions in section~\ref{sec:fieldresp:shortedwires}.  The detailed Garfield simulation
steps are also discussed in section~\ref{sec:fieldresp:shortedwires}, focusing on the differences between
the simulation in the normal and shorted-channel regions.  Finally, in section~\ref{sec:fielderesp:L1shorted}
the special treatment of the shorted-Y region in the signal processing chain is discussed.

%% file: field_resp_method_part2.tex
In order to compare the full response in data to that predicted by simulation,
a track-based data-driven method is used to extract the full response from off-beam
cosmic ray data (data recorded when it is known that
no beam-related neutrinos are passing through the MicroBooNE detector).  This is
done by i) determining the value of $t_{0}$, where $t_{0}$ refers to the time at
which the cosmic ray enters the TPC volume, ii) ensuring that a substantial
amount of light is seen by the MicroBooNE photomuliplier tubes (PMTs)~\cite{Conrad:2015xta} at
the time corresponding to the value of $t_{0}$ found, iii) correcting the drift coordinate,
$x$, of the cosmic ray track in the TPC to the location the cosmic ray actually
traversed through the TPC, and iv) coherently adding the waveforms associated
with the $t_{0}$-tagged track to recover a clean representation of the signal
shape from the waveform while suppressing the noise on the waveform.  The MicroBooNE PMT
system (32 PMTs located behind the anode plane) is used both for this $t_{0}$-tagging of
cosmic tracks as well as for triggering on
light associated with beam events within the detector, which is used to enhance the purity
of neutrino interaction signal events in the recorded data stream.  The motivation for
using $t_{0}$-tagged tracks for the data/simulation comparison comes from the desire
to minimize the effects of diffusion in
the full response estimation.  This can be done by utilizing only parts of tracks very
close to the anode plane where effects of diffusion are minimal.  One must know
the drift coordinate of the tracks in order to ensure this condition, necessitating the
use of $t_{0}$-tagged tracks.

\begin{figure}[tb]
  \centering
  \begin{subfigure}{0.49\textwidth}
    \centering
    \includegraphics[width=.99\textwidth]{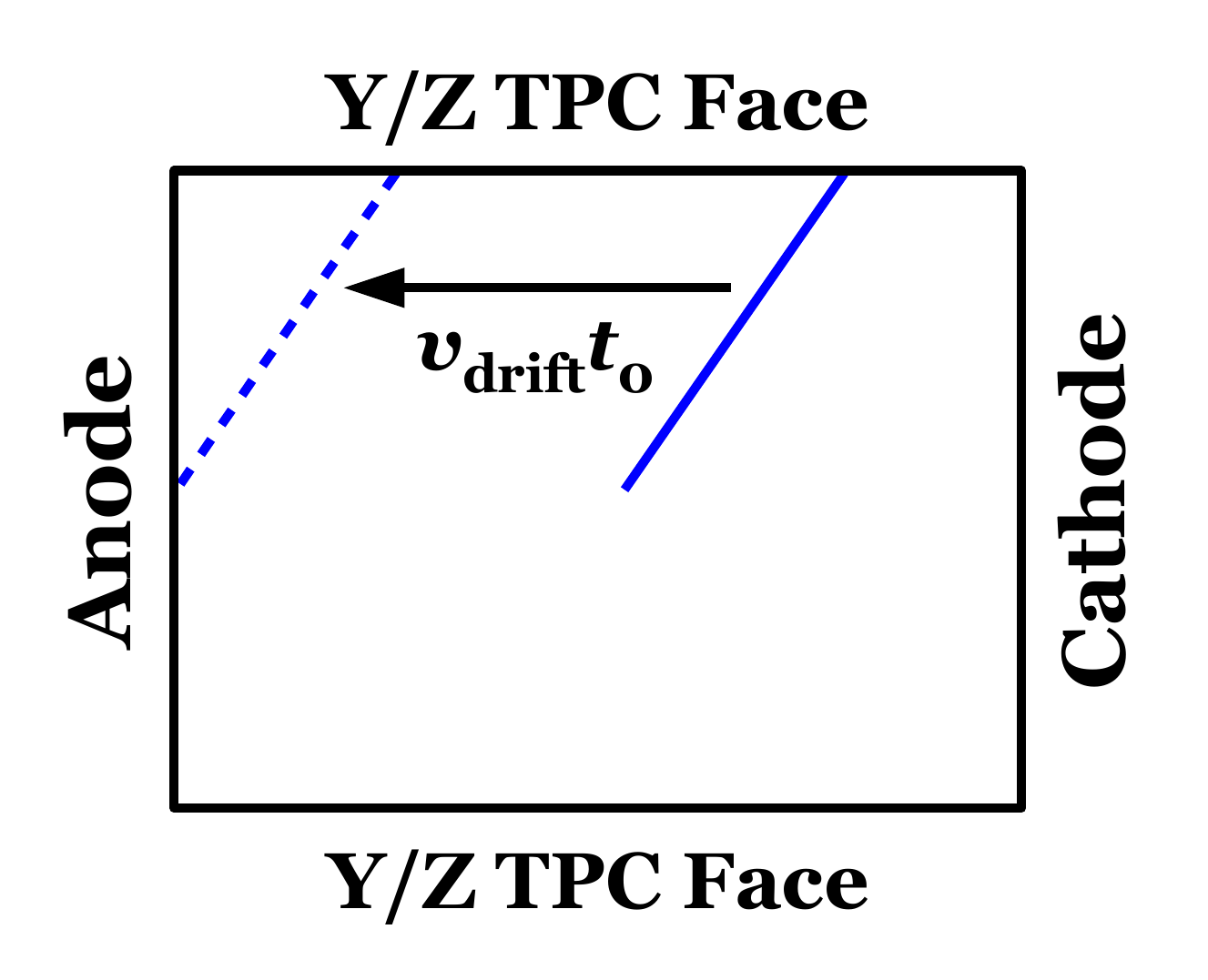}
    \caption{Example $t_{0}$-tagging of anode-piercing track.}
  \end{subfigure}
  \begin{subfigure}{0.49\textwidth}
    \centering
    \includegraphics[width=.99\textwidth]{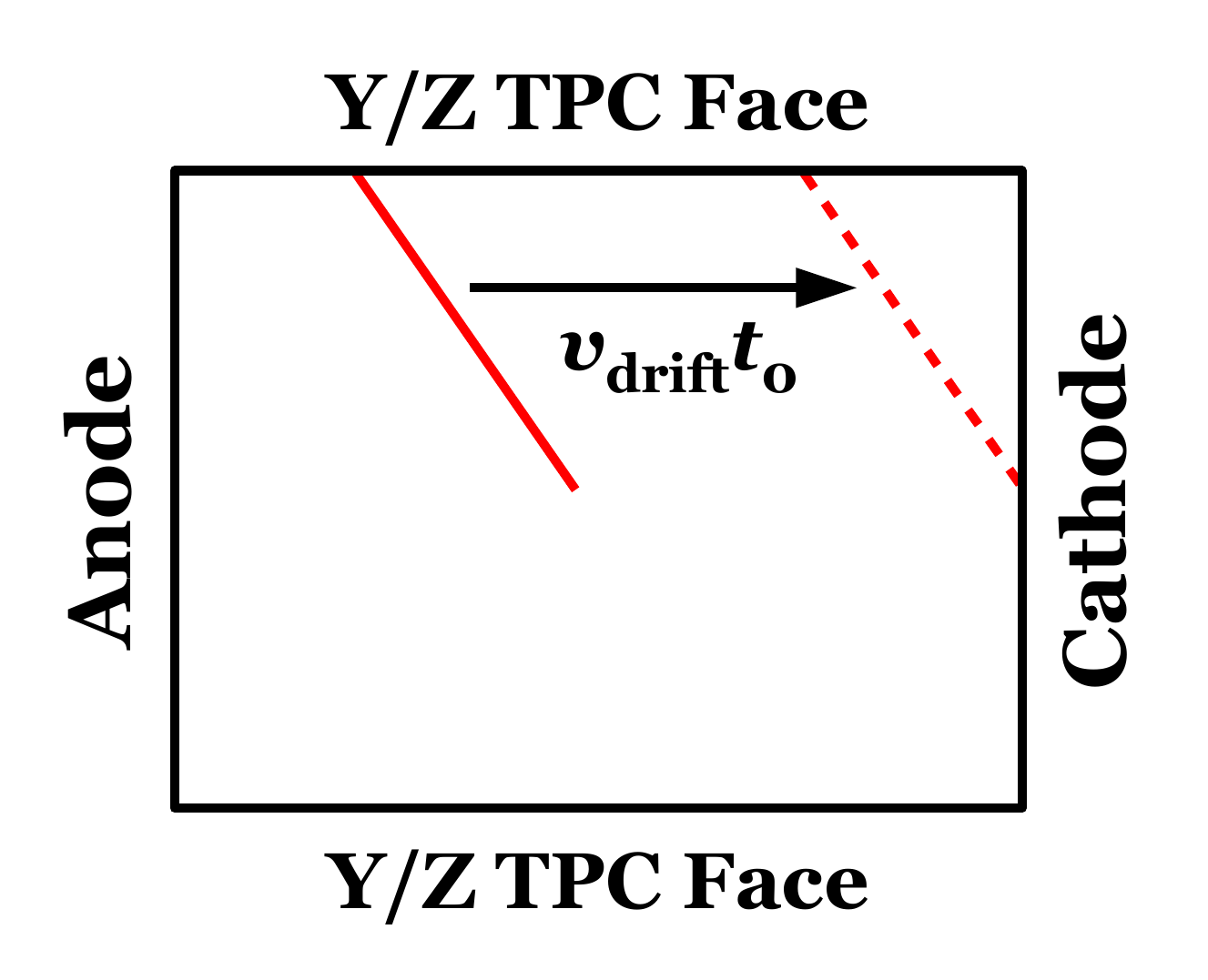}
    \caption{Example $t_{0}$-tagging of cathode-piercing track.}
  \end{subfigure}
  \caption{Illustration of the $t_{0}$-tagging method used in the studies presented in this work.  
The drift coordinate ($x$) of cosmic ray tracks are corrected using the assumption that the cosmic ray is through-going, requiring the track to pass through only one TPC face in $y$ or $z$, and using the angle of the track in the $y-x$ plane (or $z-x$ plane) to determine if the track is anode-piercing (a) or cathode-piercing (b).  The known ionization electron drift velocity, $v_{\mathrm{drift}}$, then allows for the correction of the drift coordinate of the cosmic ray track.  The track before and after the correction is shown by the solid and dashed lines, respectively.  Before the correction (when the $t_{0}$ of the track is assumed to be the trigger time), the track falsely appears to stop in the middle of the TPC.}
  \label{fig:t0tagging}
\end{figure}

The $t_{0}$-tagging method used in these studies is visualized in figure~\ref{fig:t0tagging}.
First, a sample of data events (on order of 10k) are collected from the MicroBooNE detector
using off-beam triggers (data collected at periodic intervals while the beam is not active),
providing a large set of cosmic rays (more than 100k) for study
without contamination from neutrino interaction events.  The track associated with
each cosmic ray is then reconstructed in three dimensions
(3D) using the Pandora multi-algorithm pattern recognition~\cite{Acciarri:2017hat} employed on
TPC data.  The cosmic ray is assumed to be through-going (passing through two faces of the TPC).
This is a good assumption as most cosmic rays have high enough momentum to
pass through the TPC completely without stopping.  Next, the track is required to either enter
or exit a TPC face in $y$ (top or bottom of TPC) or $z$ (front or back of TPC, with respect
to direction of the beam), but not both.  If this condition is met,
and the cosmic ray is through-going (as assumed), the cosmic ray must either enter or exit 
the anode or cathode.  Depending on the angle that the reconstructed track makes in the
$y-x$ plane, the cosmic ray is then determined to be either ``anode-piercing'' or ``cathode-piercing''
(see figure~\ref{fig:t0tagging}), and the value of $t_{0}$ is assigned 
based upon the time tick associated with the signal on the
waveform at the part of the track closest to either the anode or cathode, respectively.  A ``flash''
of light seen in several of MicroBooNE's 32 eight-inch PMTs is required to be found at the same
time as the determined $t_{0}$ value in order to increase the purity of the $t_{0}$-tagging technique.  
Finally, the drift coordinate of the reconstructed track is adjusted accordingly using the known drift
velocity ($v_{\mathrm{drift}}$) of ionization electrons in liquid argon at an electric field of 273~V/cm,
roughly 1.1~\si{\milli\meter/\micro\second}~\cite{yichen}.  Validations using Monte Carlo
simulation and data-driven validations utilizing an external small cosmic
ray tagger~\cite{Acciarri:2017rnj} have shown this $t_{0}$-tagging method to reconstruct the track
$t_{0}$ correctly at least 98\% of the time for anode-piercing tracks and at least 97\% of the time
for cathode-piercing tracks.

With the drift coordinate of the cosmic ray track corrected, one can proceed to use
the subset of anode-piercing tracks (roughly one per recorded off-beam event) to
estimate the full response via the method described above.  Only waveforms associated with the
portion of the track between 2~\si{\centi\meter} and 10~\si{\centi\meter} away from
the anode plane are used in order to minimize the influence of diffusion, which is 
why anode-piercing tracks are used instead of cathode-piercing tracks.  Waveform signals are lined up
in time across many different wires (and tracks from different events) and are added together.  This
can be done as cosmic muons are well-approximated by straight lines, so the signal can be assumed to
be coherent across a short distance of the cosmic muon track.  For 
unipolar signals, such as those observed by the collection plane wires, the positive signal peak 
is the feature lined up in time; in contrast, for the bipolar signals of the induction plane wires, 
the negative dip of the signal is instead lined up in time.  The maximum (minimum) ADC value in the
region of the waveform containing the track is used to line up the collection (induction) signals
in time.  The coherent nature of the
signal across waveforms associated with different wires leads to the signal response being
preserved when the waveforms are added together.  Conversely, noise on the waveform averages out
as it is incoherent across different wires.  The result is a data-driven estimate of the full
response for each plane in each of the regions.  

\begin{figure}[tbp]
    \centering
    \begin{subfigure}[t]{0.8\textwidth}
    \includegraphics[width=1.0\textwidth]{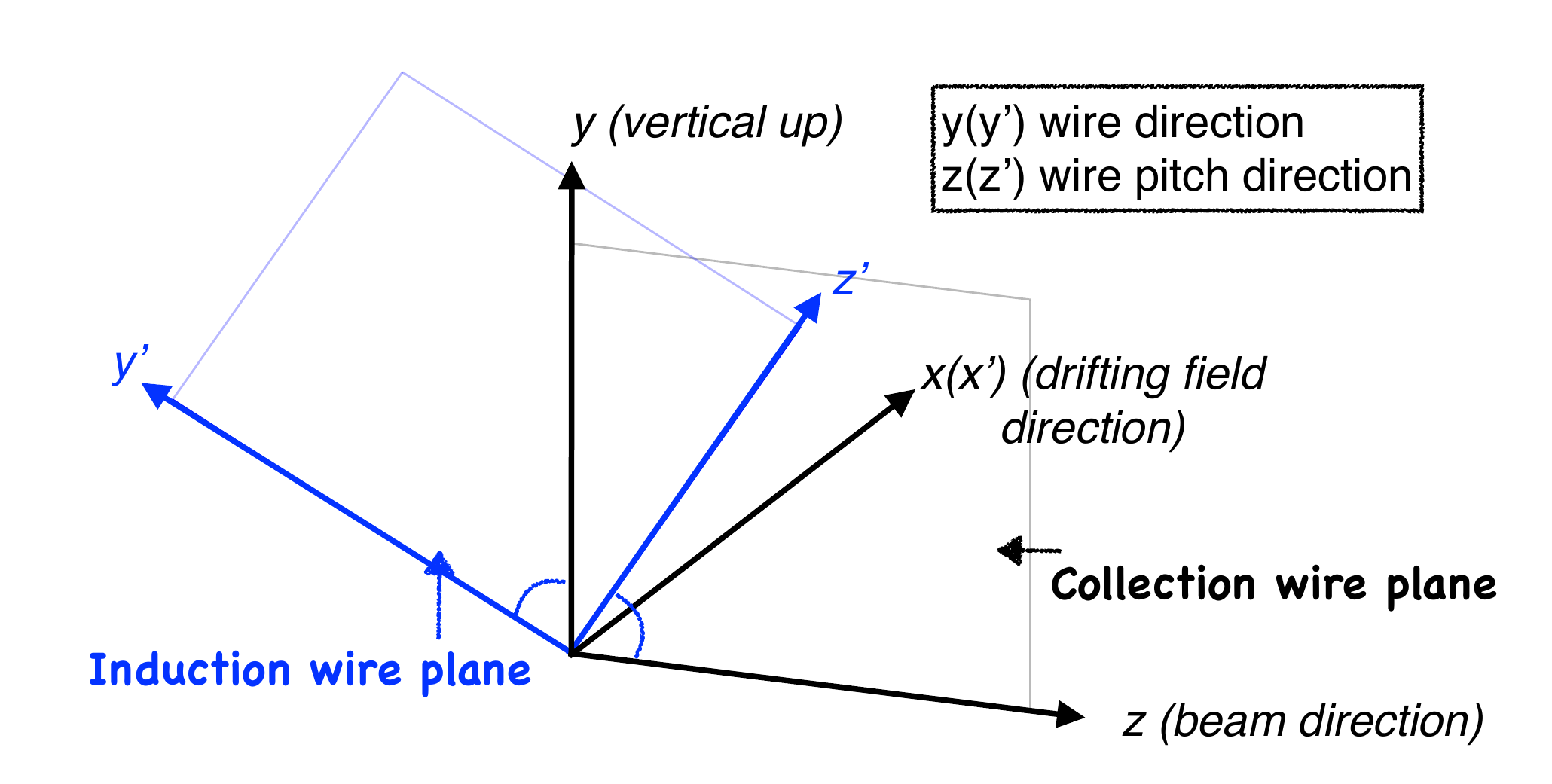}
    \caption{Coordinates for collection and induction planes. The $y'$ ($z'$)
    axis is rotated by 60$^{\circ}$} around $x$ axis from $y$ ($z$) axis. 
    \label{fig:coordinate}
    \end{subfigure}
    \begin{subfigure}[t]{0.8\textwidth}
    \includegraphics[width=1.0\textwidth]{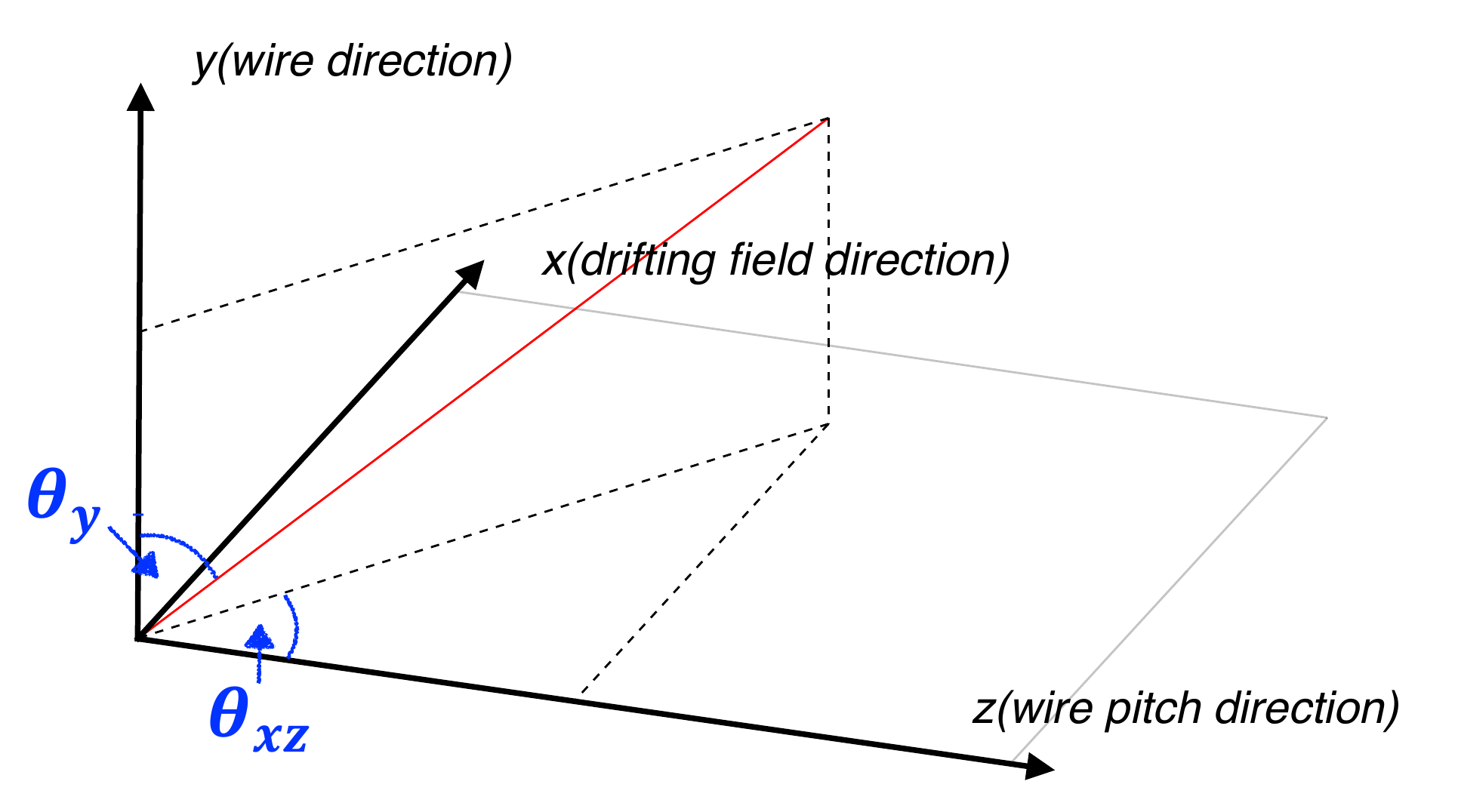}
    \caption{Definition of two angles, $\theta_{xz}$ and $\theta_y$, which define the direction of a track
    in the TPC.}
    \label{fig:angle}
    \end{subfigure}
    \caption{Geometric coordinates (a) and angles for description of track topology (b).}
    \label{fig:geometry}
\end{figure}

The coordinate system used for these studies is described in figure~\ref{fig:coordinate}.
For each wire plane, the $x$-axis is along the drifting 
field direction (opposite to the electron drift direction), the $y$-axis is along the wire orientation,
and the $z$-axis is along the wire pitch
direction.  Different coordinates are used for each wire plane, as the wire directions of the different
planes are at an angle to one another; the induction planes use the ``primed'' coordinate system
as shown in figure~\ref{fig:coordinate}.  The nominal (default) detector coordinate system is
identical to the Y plane's coordinate system for which the $y$-axis is vertical in the upwards
direction (toward the sky) and the $z$-axis is along the direction of the neutrino beam.
The origin of the coordinate system is located at the center of the U plane's upstream edge (the edge
closest to the source of the neutrino beam).
Based on the predefined coordinate for each wire plane, two angles define the direction of the track.
As shown in figure~\ref{fig:angle}, $\theta_y$ is the angle between the track and 
the $y$-axis, and $\theta_{xz}$ is the angle between the projection onto the $x-z$ plane and the 
$z$-axis.

In estimating the full response, only tracks with $5^{\circ} < \theta_{xz} < 15^{\circ}$ are used in order
to ensure that the impulse response of the drifting ionization electrons is well approximated, though
cross-checks of the signal simulation against data are also performed at higher track angles.
By repeating the measurement using a Monte Carlo simulation of cosmic ray tracks, it is
determined that there is a Gaussian smearing of approximately 1~\si{\micro\second} introduced
by the method itself.  The origin of this smearing is from the alignment of signals in time across many
different waveforms, which introduces a slight broadening to the resulting average response.  While
delta rays from the cosmic tracks are not excluded in the analysis, this Monte Carlo study finds that
the impact from delta rays to the broadening of the signal is negligible.  When making comparisons of the
full response between data and simulation, the 1~\si{\micro\second} Gaussian smearing is applied to the
simulation.


%% file: field_resp_normalregion_part2.tex
\begin{figure}
\centering
\includegraphics[width=.7\textwidth]{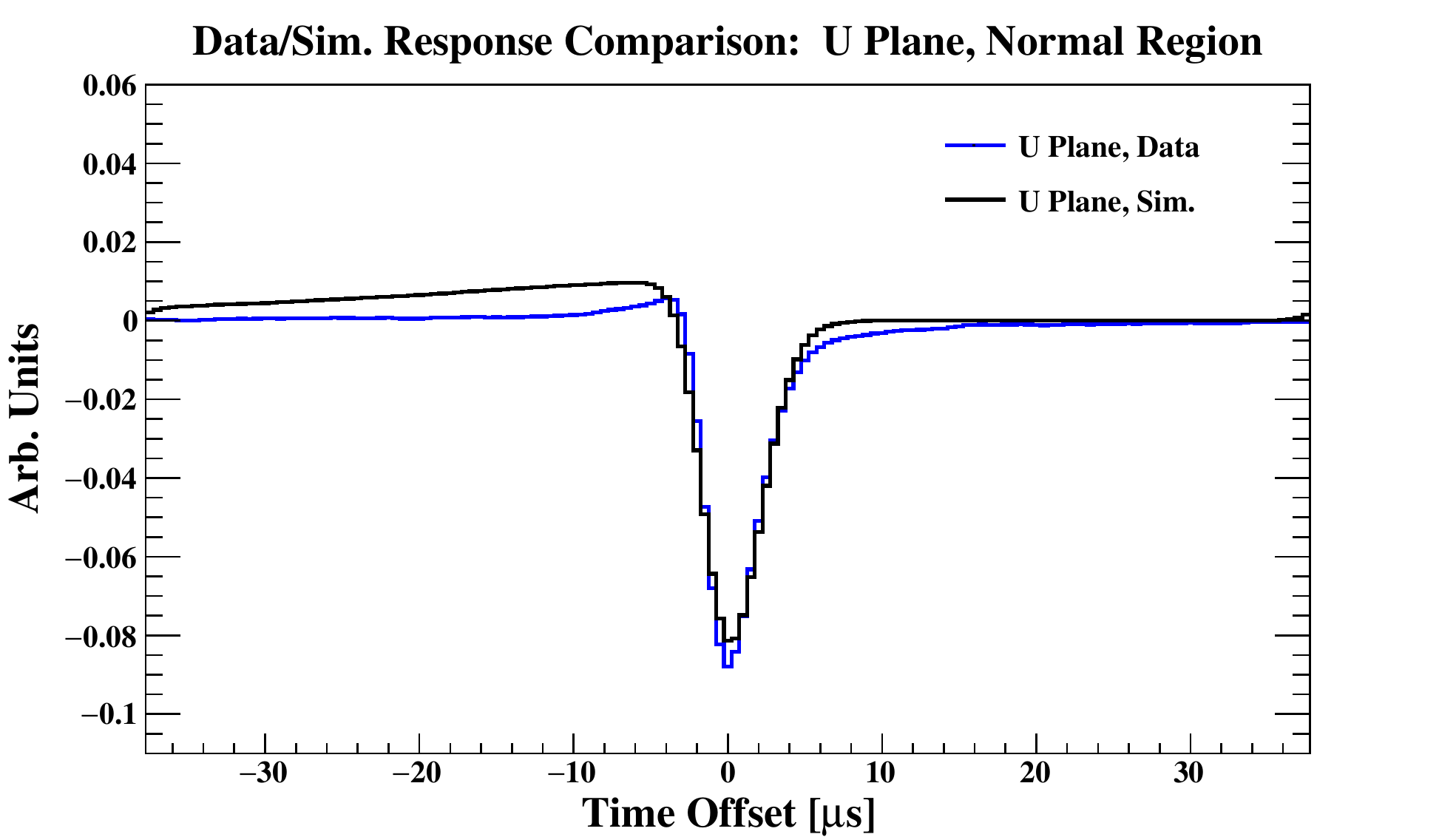}
\\
\vspace{3mm}
\includegraphics[width=.7\textwidth]{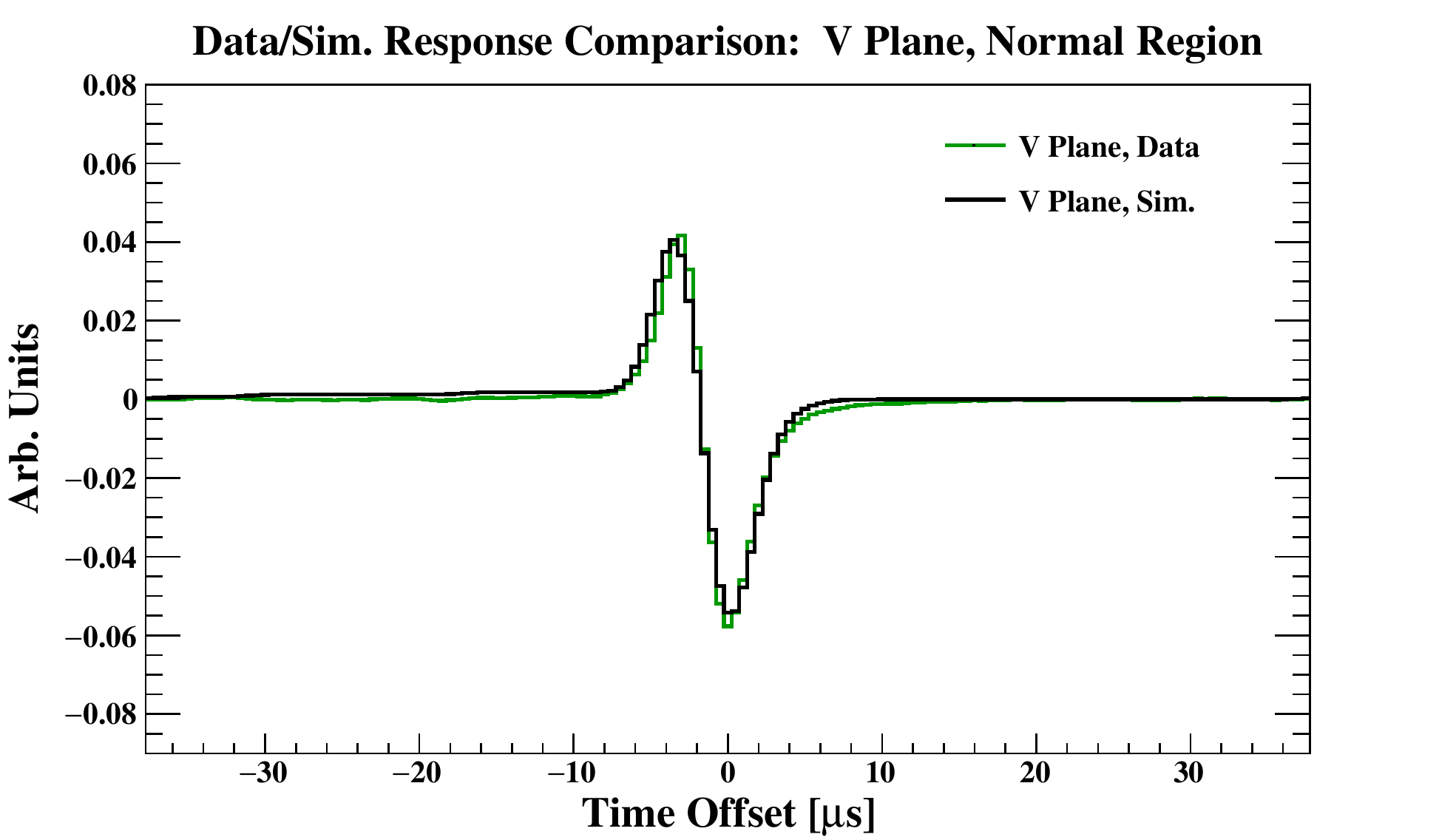}
\\
\vspace{3mm}
\includegraphics[width=.7\textwidth]{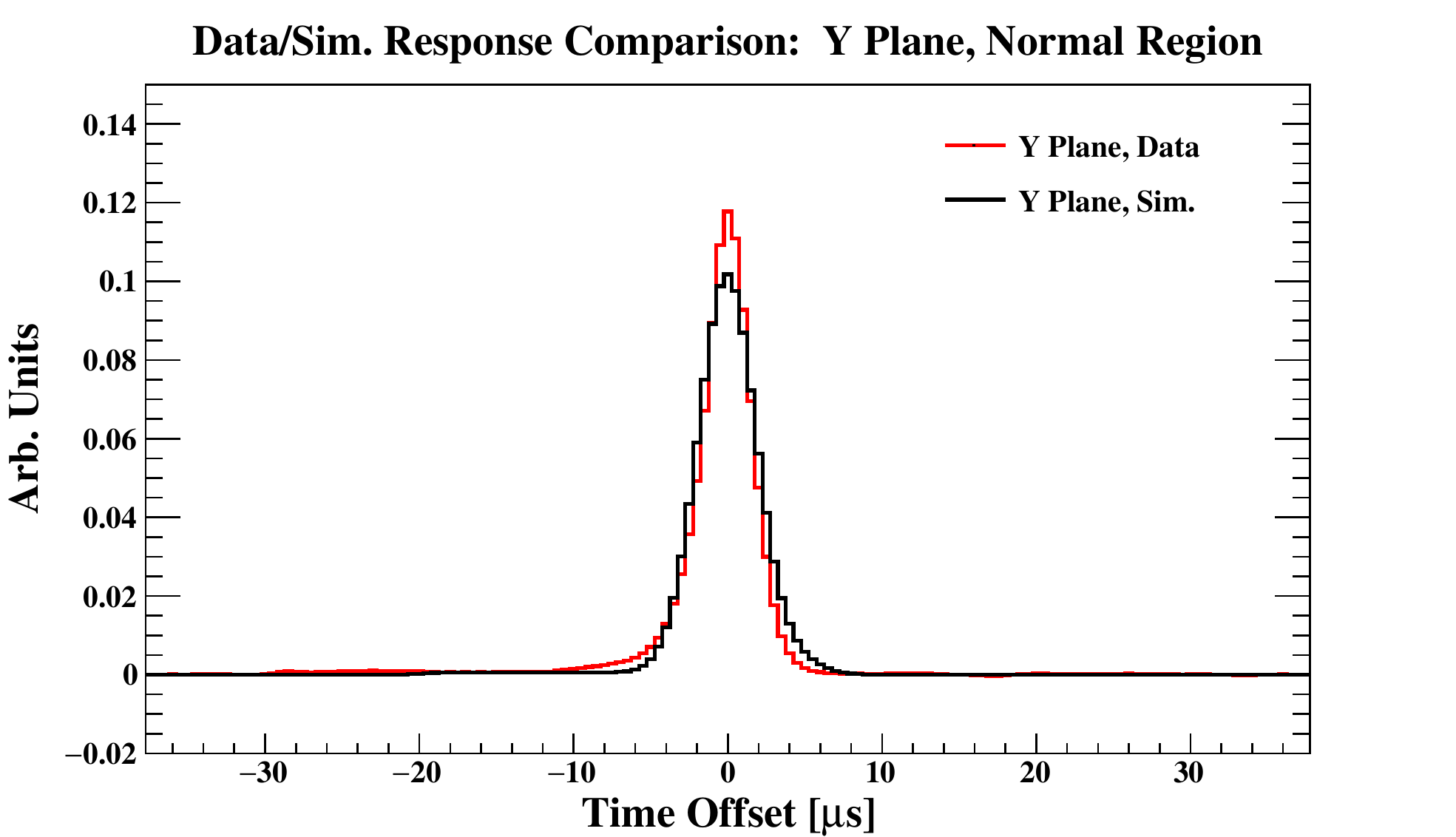}
\vspace{2mm}
\Put(-255,1000){\fontfamily{phv}\selectfont \textbf{MicroBooNE}}
\caption{Data/simulation comparison of the full response for the different wire planes in the normal region of the MicroBooNE TPC.  
Simulation is shown in black, while the colored curves represent the full response extracted from data.  
In general there is good agreement between data and simulation, but differences are observed between data and simulation in the ``front porch'' of the U plane response.  
This discrepancy is reduced by adding more wires in the wire field response simulation (see text for more details).} \label{fig:DataMCRespComp_NormalRegion}
\end{figure}

By applying the methodology discussed earlier in this section, the full response in the normal region is extracted from data and compared to the convolution of the wire field response simulated by Garfield with a parametrization of the electronics response function for a given peaking time (2.2~\si{\micro\second} in this case to match the effective peaking time seen in data).  The comparison of the full response for data to that for the simulation in the normal region is illustrated in figure~\ref{fig:DataMCRespComp_NormalRegion}.  The absolute normalization of the full responses is arbitrary and only fixed relative to the integral of the Y plane response, which is set to unity.

\begin{figure}
\centering
\includegraphics[width=.8\textwidth]{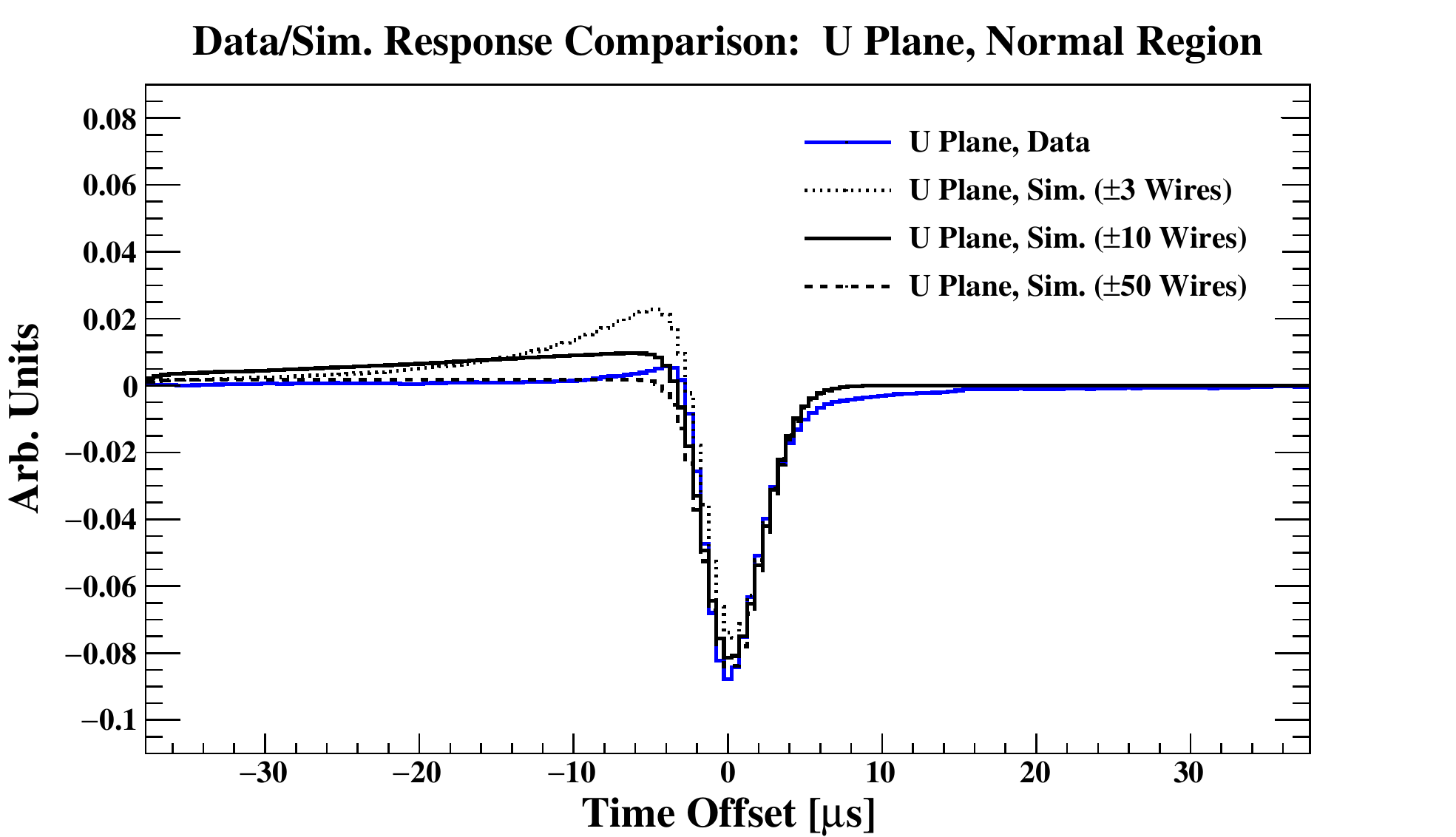}
\vspace{2mm}
\Put(-285,300){\fontfamily{phv}\selectfont \textbf{MicroBooNE}}
\caption{Comparison of U plane full response (electronics response convolved with wire field response) simulation to data, using a variable number of wires adjacent to the wire closest to the ionization electrons in the Garfield simulation:  $\pm 3$, $\pm 10$, and $\pm 50$.  This comparison is made in the region of the TPC unimpaired by shorted channels (normal region).  The ``front porch" of the U plane wire field response becomes less pronounced when the simulation utilizes a greater number of wires, converging to what is seen in data.  A small positive peak before the ``dip'' of the U plane response is observed in data and is not reproduced in the simulation when utilizing any number of wires.  This discrepancy is likely due to signal fluctuations present in data that are not included in the simulation.} \label{fig:UplaneFrontPorchComp}
\end{figure}

In general there is quite good agreement between data and simulation, though there are a few differences worth pointing out.  First of all, the U plane ``front porch'' or the slow-rising feature of the waveform in front of (at lower values of time) the primary ``dip'' exhibits some disagreement between data and simulation.  This feature is due to the absence of a shield plane in front of the U plane in MicroBooNE, allowing charge further away from the anode plane to produce a signal on the unshielded U plane wires during drift (if there were a shield plane in front of the U plane, the U plane and V plane wire field responses would look very similar).  The disagreement between data and simulation can be attributed to the finite number of wires that are adjacent to the wire closest to the drifting ionization electrons included in the Garfield simulation of the wire field response (10 wires on either side of the primary wire in the results shown).  The impact of changing the number of adjacent wires in the Garfield simulation on the U plane full response is shown in figure~\ref{fig:UplaneFrontPorchComp}.  Increasing the number of adjacent wires would likely reduce the front porch feature even more in the simulation, at the cost of increased computation time.  However, the impact on charge reconstruction from this effect is expected to be small.

In addition to differences in the ``front porch'' feature of the U plane, in figure~\ref{fig:DataMCRespComp_NormalRegion} we can see a small discrepancy in the width of the signals between data and simulation.  This difference in signal width is most noticeable on the Y plane but can also be seen on the U and V planes.  The difference can potentially be explained by 3D effects that are not included in the Garfield simulation, which assumes that the wires of the different wire planes are parallel to one another and neglects the complicated drift of the ionization electrons through the wire planes in three dimensions.  
A potential improvement in the future is to use a full 3D simulation of the wire field response that is currently computationally expensive and too imprecise to utilize at present.  The development of a sufficiently precise 3D simulation of the wire field response is in progress.

\begin{figure}
\centering
\includegraphics[width=.7\textwidth]{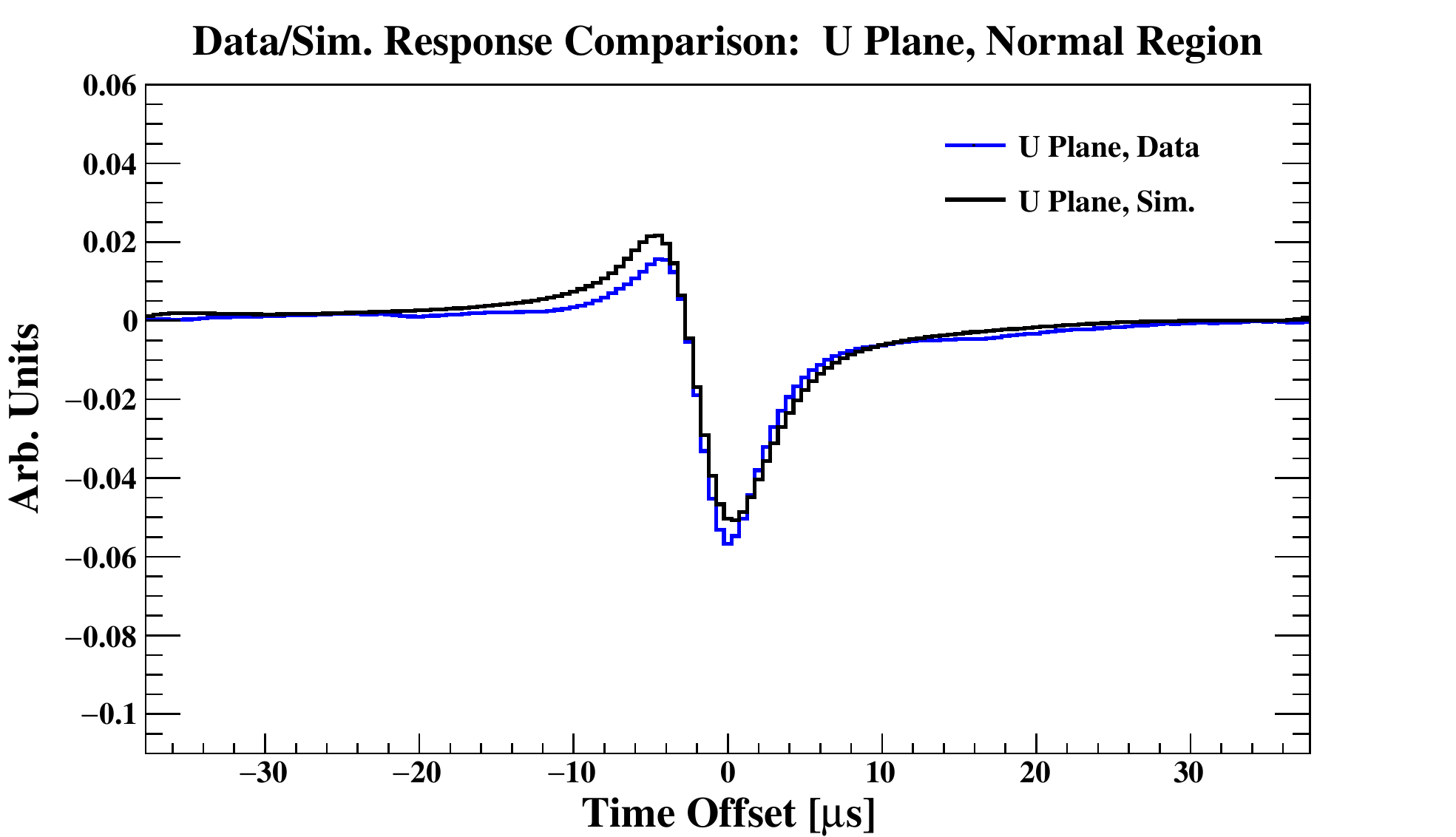}
\\
\vspace{3mm}
\includegraphics[width=.7\textwidth]{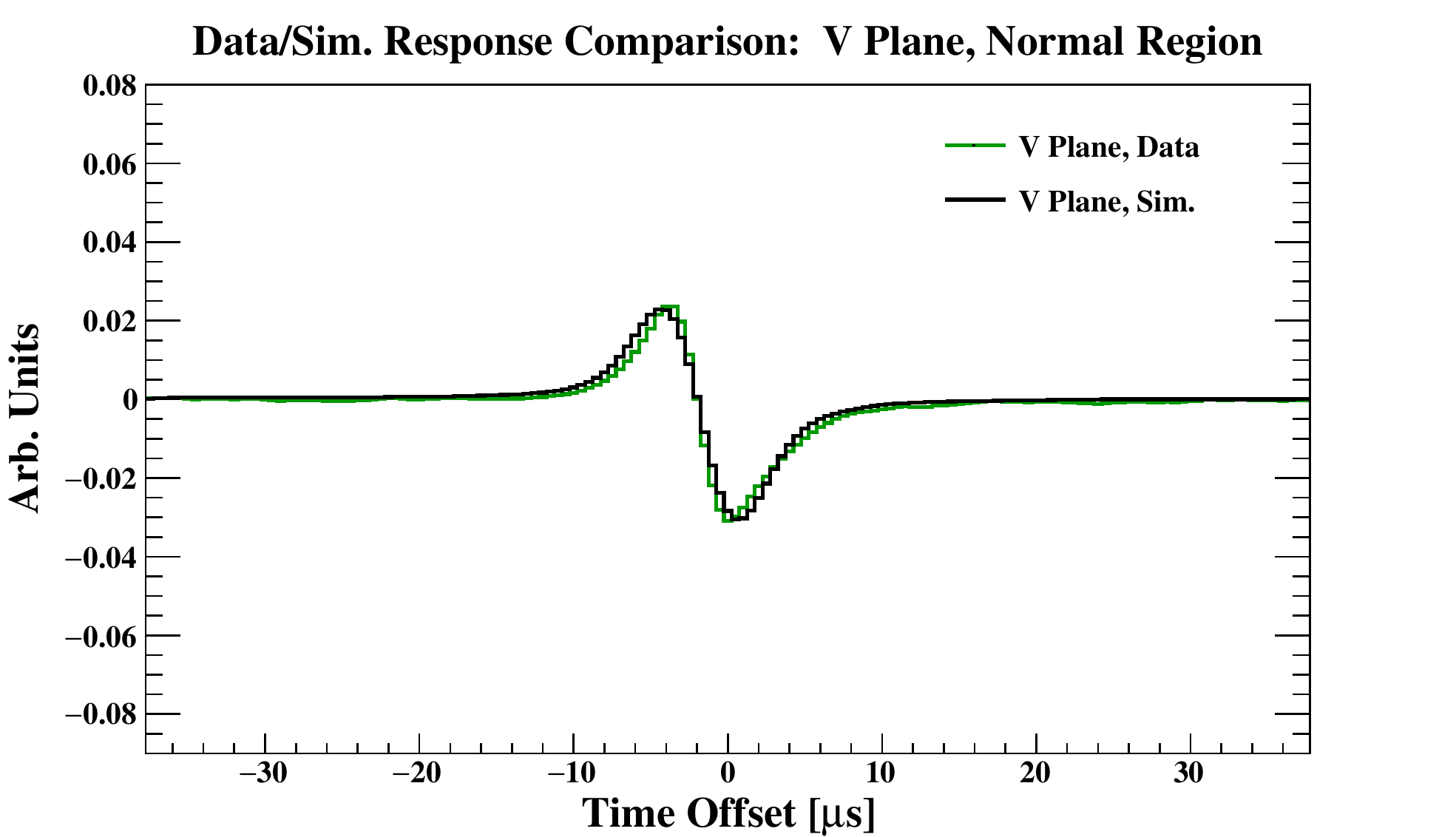}
\\
\vspace{3mm}
\includegraphics[width=.7\textwidth]{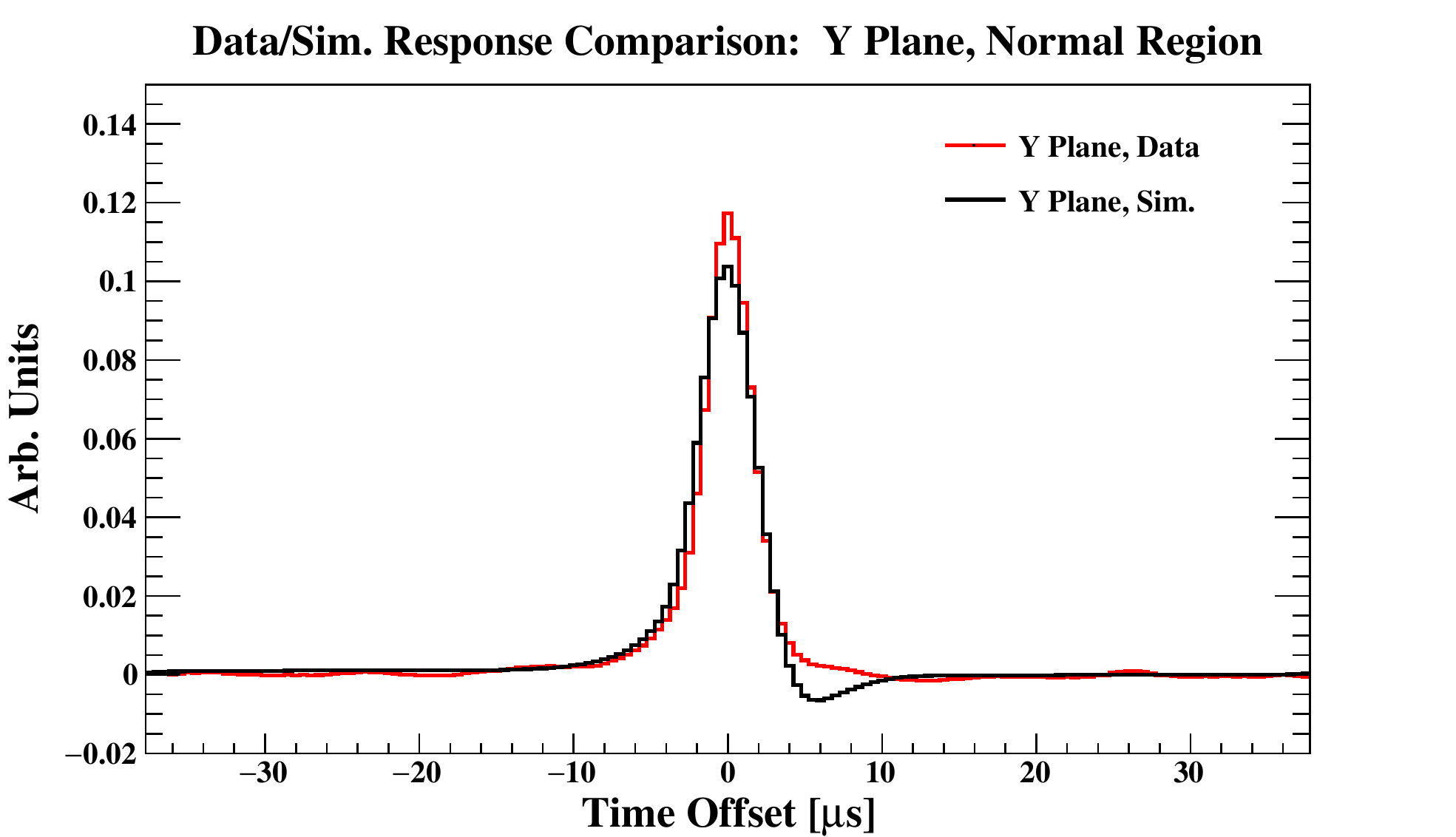}
\vspace{2mm}
\Put(-255,1000){\fontfamily{phv}\selectfont \textbf{MicroBooNE}}
\caption{Comparison of the signal simulation for the different wire planes to data using tracks with~$40^{\circ} < \theta_{xz} < 50^{\circ}$, looking at the region of the TPC unimpaired by shorted channels (normal region).  Simulation is shown in black, while the colored curves represent the full response extracted from data.} \label{fig:DataMCRespComp_NormalRegion_45deg}
\end{figure}

As described in Ref.~\cite{SP1_paper}, charge induction on wires adjacent to the wire that is closest to drifting ionization charge can lead to large contributions to the signals observed on TPC wires.    This effect is more noticeable when an ionization track is at higher angle with respect to the anode plane (i.e.~higher $\theta_{xz}$).  In order to probe how well the simulation performs in reproducing this effect as seen in data, a similar comparison to that illustrated in figure~\ref{fig:DataMCRespComp_NormalRegion} is shown in figure~\ref{fig:DataMCRespComp_NormalRegion_45deg} using tracks with~$40^{\circ} < \theta_{xz} < 50^{\circ}$.  The comparison shown in figure~\ref{fig:DataMCRespComp_NormalRegion_45deg} indicates that the simulation performs quite well in reproducing the effect of charge induced on wires neighboring the wire closest to drifting ionization.  There are minor discrepancies that would likely be reduced with an improved simulation of wire field response, as discussed above.  This includes a disagreement in the ``front porch'' region of the U plane wire field response and an additional discrepancy between data and simulation for the Y plane wire field response, for which the simulation predicts a small dip in the later part (at $t \approx 5$~\si{\micro\second}) of the waveform that is not observed in data.

%% file: field_resp_shortedregions_part2.tex
\begin{figure}
\centering
\begin{subfigure}{0.49\textwidth}
  \centering
  \includegraphics[width=.85\textwidth]{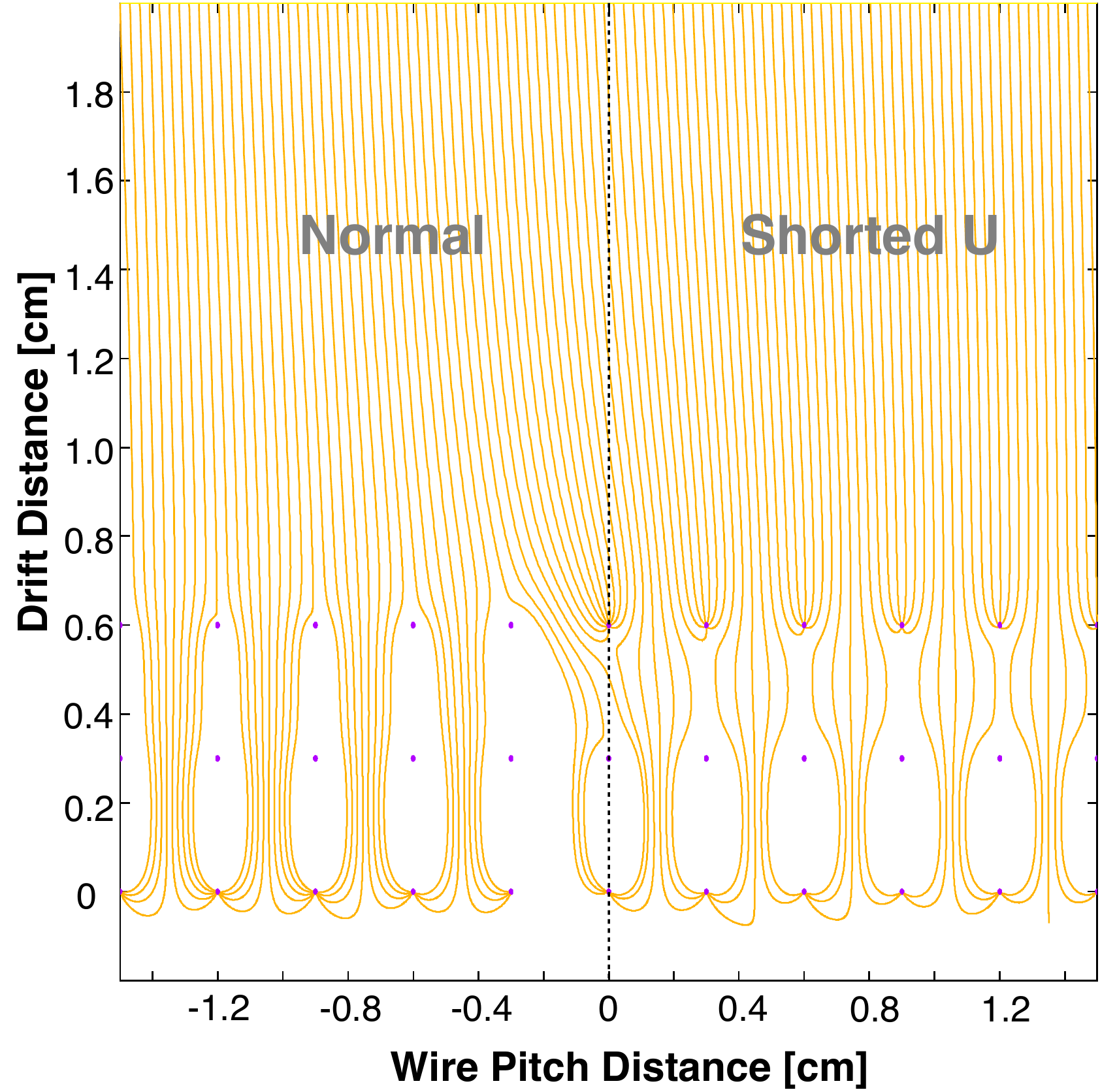}
  \caption{Shorted-U region.}
\end{subfigure}
\begin{subfigure}{0.49\textwidth}
  \centering
  \includegraphics[width=.85\textwidth]{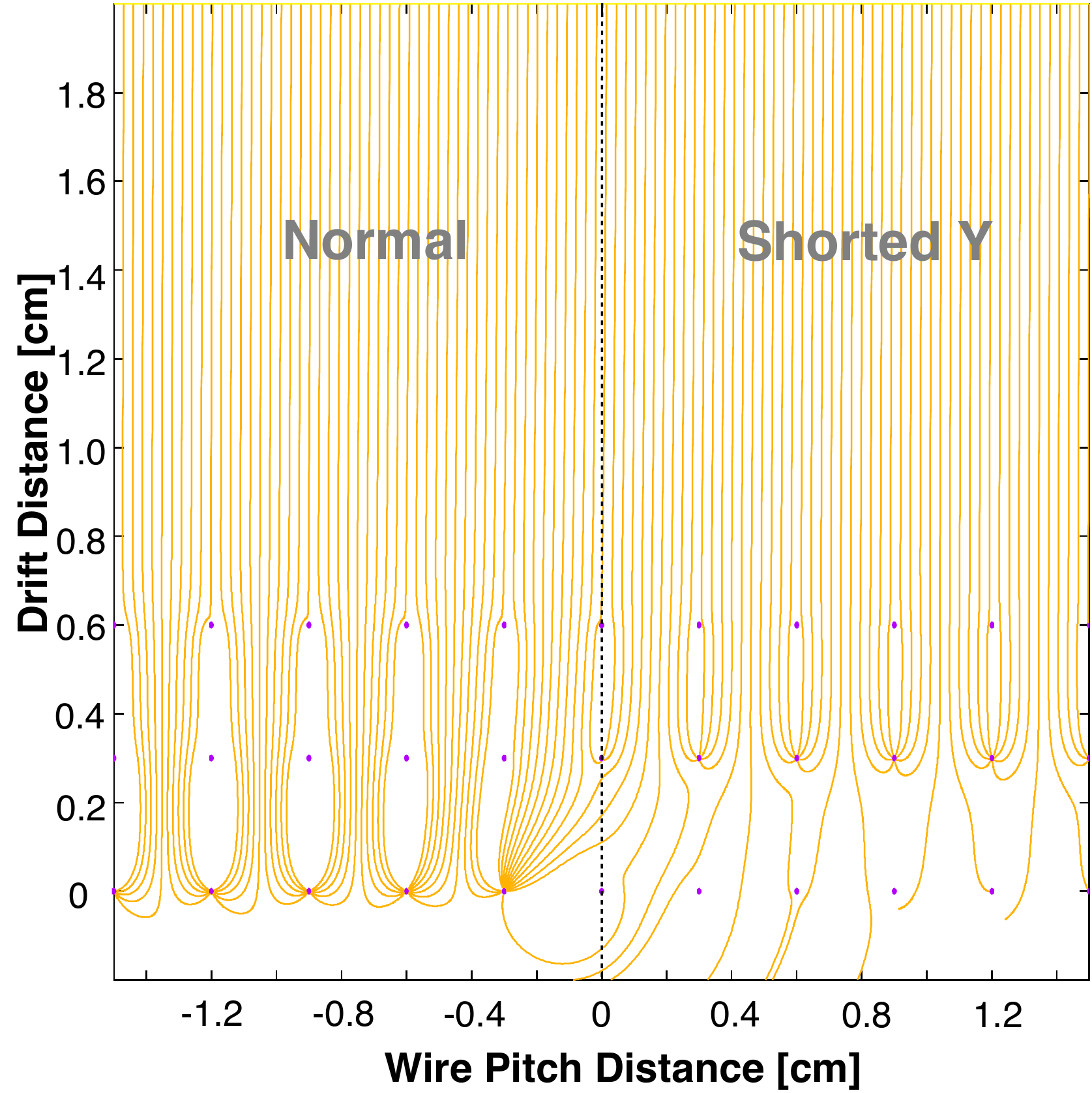}
  \caption{Shorted-Y region.}
\end{subfigure}
\caption{The trajectories of ionization electron drift as a demonstration of the simulation of the electric field near the wire planes using Garfield.  The ionization drift lines associated with the shorted-U region (a) and shorted-Y region (b) are shown, with a comparison made to the normal region in both cases.  A couple of features are worth noting.  In the case of the shorted-U region, a fraction of the charge that would be collected by the Y plane is instead collected by the unresponsive U plane wires, leading to a decreased transparency of the ionization electrons as seen by the V plane and Y plane wires.  In the case of the shorted-Y region, ionization electrons that are normally collected by the Y plane are instead collected on the V plane wires, leading to unipolar (as opposed to bipolar) signals on the V plane wires in this region of the detector.}\label{fig:DataMCRespComp_ShortedURegion_Field}
\end{figure}

The comparison of the full response in data and simulation made
in section~\ref{sec:fieldresp:normalwires} is repeated for the the shorted-U
and shorted-Y regions.  In these regions,
while the electronics response remains the same as in the normal region,
the wire field response differs dramatically, as shown in
figure~\ref{fig:DataMCRespComp_ShortedURegion_Field}.  This is due to the different
drift paths and velocities that ionization electrons experience as they drift through the
anode plane wires, as a result of the different bias voltages on the shorted channels.
Given the lack of information available regarding
the nature of the short, data must be utilized in order to best model the true
detector condition.

The short mechanism in the detector is still not completely clear.  The simplest
explanation of the short is that one or more wires
in the V plane physically touch a group of wires in the U plane or Y plane; however,
a visual inspection of the wire planes (prior to filling the cryostat with argon)
via the use of a camera inserted into the cryostat yielded no evidence
of wires physically touching each other~\cite{Carls:2015spa}.  In principle, the bias
voltage on the wires shorted by this cross-plane contact should be
reduced to the ground level because of the contact with the V plane, which is held at
detector ground.  However, the fact that the sensitive wire is connected
to a pair of diodes with 1.8~V bias for the pre-amplifier and
ground, respectively, provides a hint that the actual bias on the wires may be different
from ground.  An imperfect short may cause
residual voltage on the shorted channel significantly different from ground, with the
supply bias set at -110~V and +230~V for the U plane and Y plane, respectively.  Thus,
the bias voltages on the shorted channels are set to values different from ground in the Garfield simulation
in such a way as to reproduce the field response function shape as seen in data.  
For the shorted-U region, it was found that setting the bias voltage to -45~V leads
to good agreement between data and simulation.  A bias voltage of +20~V on the Y plane
wires in the shorted-Y region is necessary to best match the simulation to data.

\begin{figure}
\centering
\includegraphics[width=.7\textwidth]{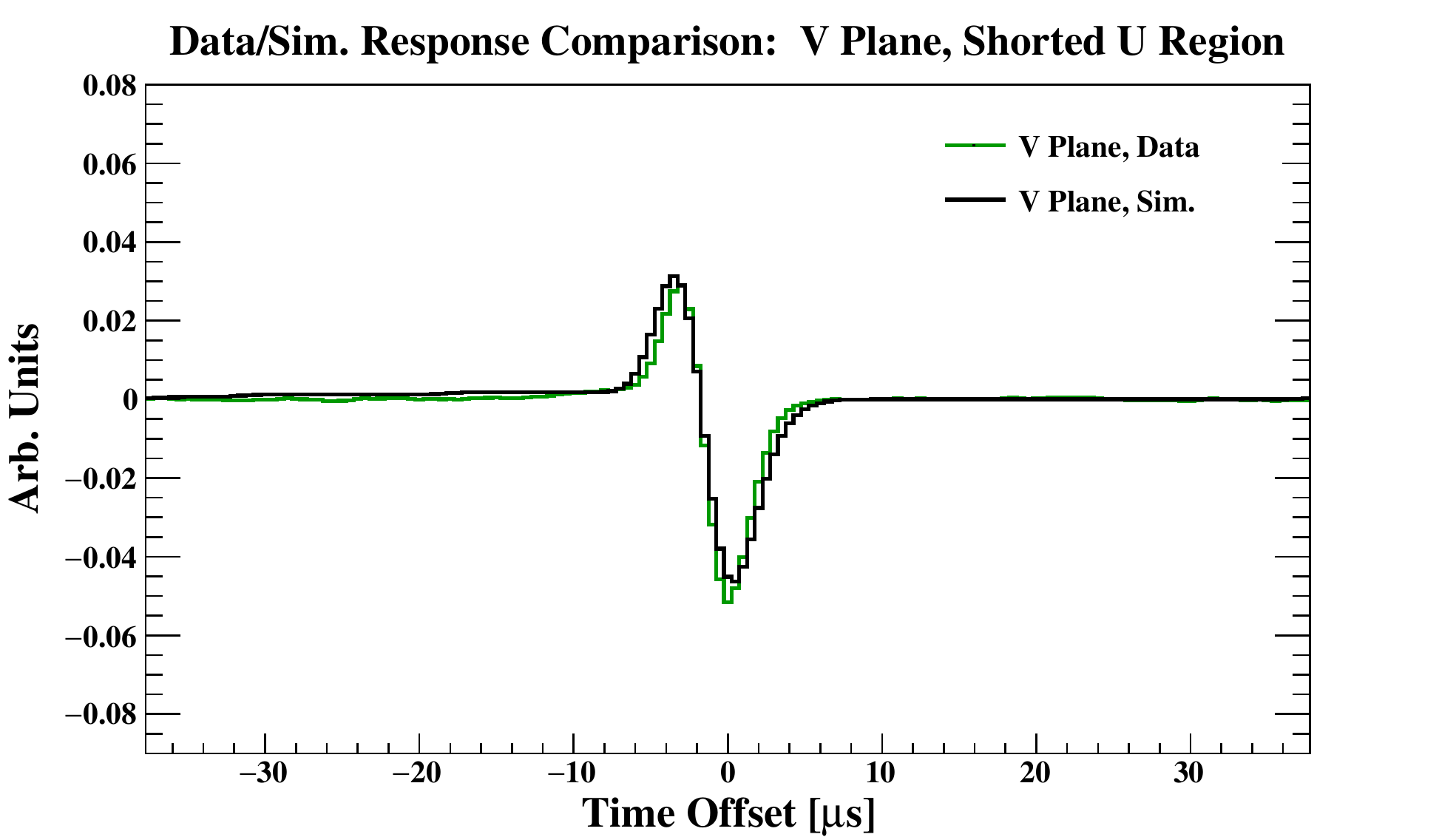}
\\
\vspace{3mm}
\includegraphics[width=.7\textwidth]{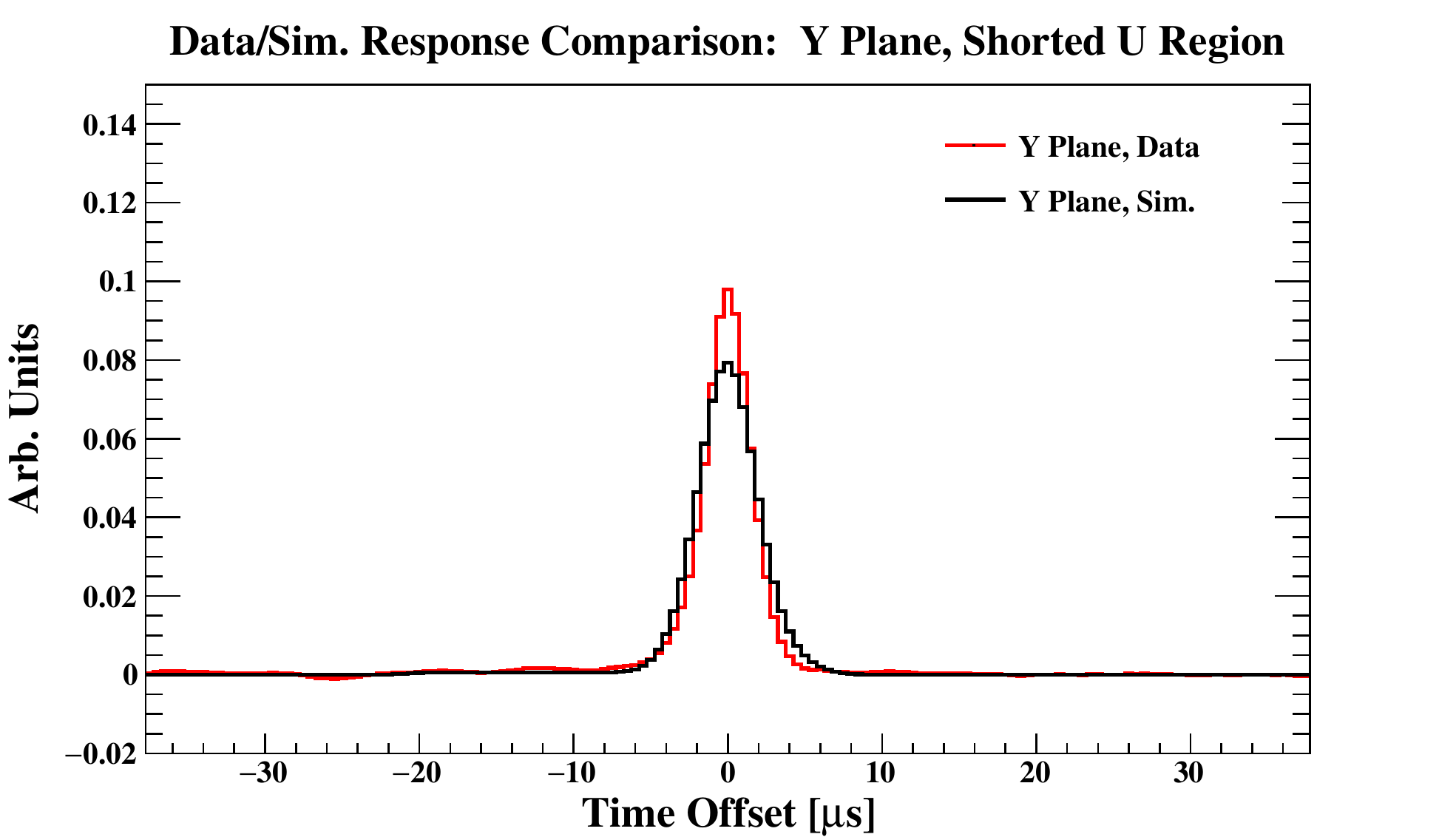}
\vspace{2mm}
\Put(-255,630){\fontfamily{phv}\selectfont \textbf{MicroBooNE}}
\caption{Data/simulation comparison of the full response for the different wire planes in the shorted-U region.  
Simulation is shown in black, while the colored curves represent the full response extracted from data using tracks with $5^{\circ} < \theta_{xz} < 15^{\circ}$.  
The simulation predicts a slightly broader response than that observed in data, potentially due to the simulation neglecting ionization electron drift in three dimensions (see section~\ref{sec:fieldresp:normalwires}).} \label{fig:DataMCRespComp_ShortedURegion}
\end{figure}

The data/simulation comparison for the
shorted-U region is made in figure~\ref{fig:DataMCRespComp_ShortedURegion}, utilizing tracks
with $5^{\circ} < \theta_{xz} < 15^{\circ}$.
As for the case of the normal region, the absolute normalization is arbitrary
and only fixed relative to the integral of the Y plane response in the
normal region (both for data and simulation, independently),
which is set to unity as mentioned previously.  Most of the shape features
observed in data are reproduced with the modified Garfield simulation.  As mentioned
above, the transparency to ionization drift in this region of the TPC is reduced
due to some of the ionization electrons being collected by the U plane wires, leaving
less charge to induce a signal on the V plane and to be collected by the Y plane.  The
U plane response is not shown in this region as the U plane wires are unresponsive
due to the short between the U and V planes.

\begin{figure}
\centering
\includegraphics[width=.7\textwidth]{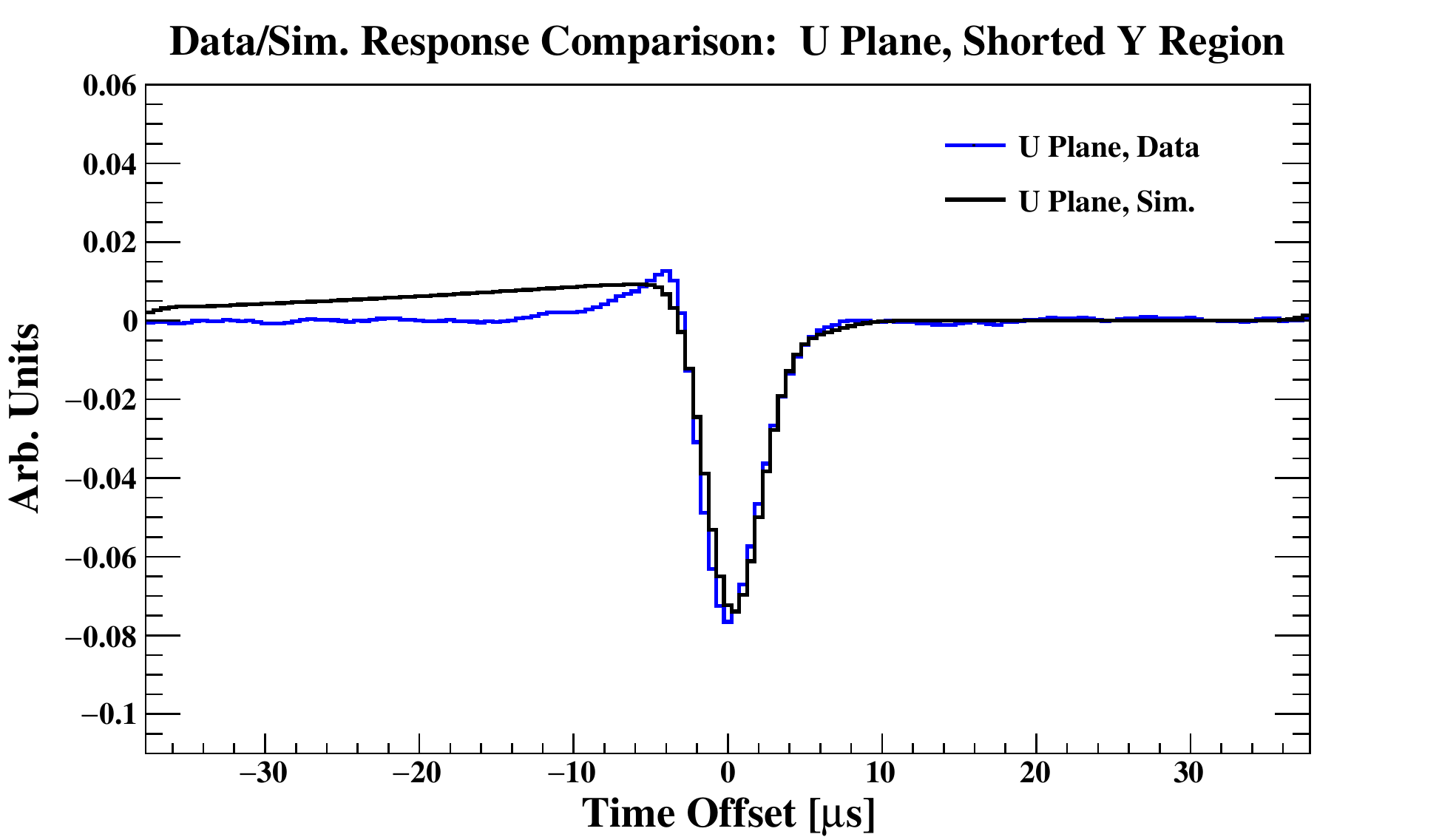}
\\
\vspace{3mm}
\includegraphics[width=.7\textwidth]{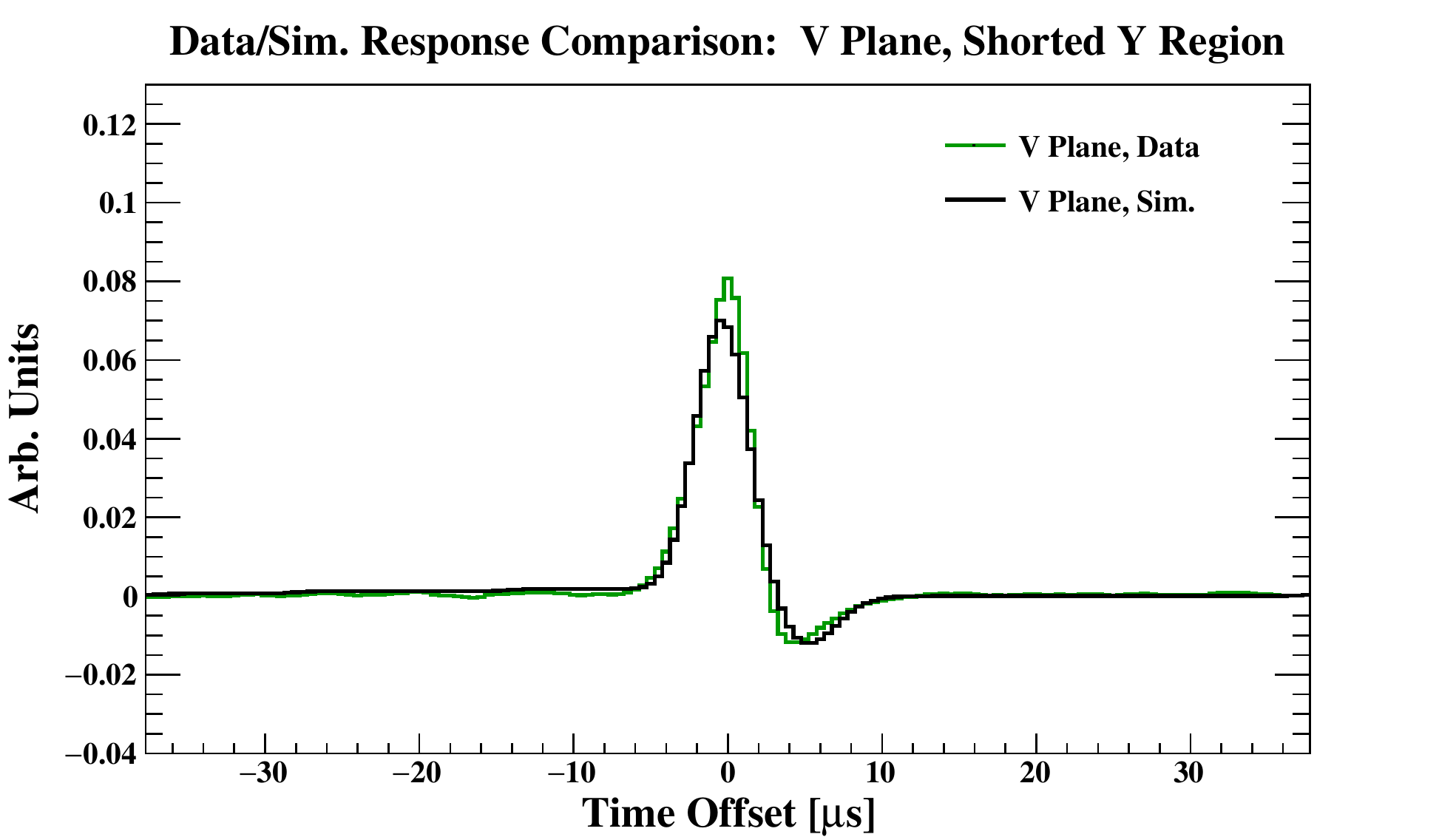}
\vspace{2mm}
\Put(-255,630){\fontfamily{phv}\selectfont \textbf{MicroBooNE}}
\caption{Data/simulation comparison of the full response for the different wire planes in the shorted-Y region.  
Simulation is shown in black, while the colored curves represent the full response extracted from data using tracks with $5^{\circ} < \theta_{xz} < 15^{\circ}$.  
As in the normal region, a small discrepancy is observed between data and simulation in the ``front porch'' of the U plane response that can be reduced by adding more wires into the wire field response simulation.} \label{fig:DataMCRespComp_ShortedYRegion}
\end{figure}

The same comparison is made for the shorted-Y region in figure~\ref{fig:DataMCRespComp_ShortedYRegion}, again
using tracks with $5^{\circ} < \theta_{xz} < 15^{\circ}$.
As expected, the U plane response is largely the same as in the normal region due to the short occurring
between the V and Y planes in this region.  However, the V plane full response is dramatically different,
unipolar instead of bipolar in shape.  This is because ionization electrons are being collected on the
V plane wires due to the Y plane wires in this region seeing a dramatically lower positive
voltage (zero in the case of a complete short to ground), leading to a loss of transparency for
ionization electrons drifting through the anode wire planes.  The modified Garfield simulation can
reproduce the features seen in data in this region, in particular the shape of the unipolar V plane
response.  The disagreement between data and simulation in the ``front porch'' of the U plane response is
similar to that observed in the normal region, discussed in section~\ref{sec:fieldresp:normalwires}.
The Y plane response is not shown in figure~\ref{fig:DataMCRespComp_ShortedYRegion} because the Y plane
wires are unresponsive in this region.

%% file: field_resp_L1shorted_part2.tex
As described in the previous section, there are two regions in the MicroBooNE
TPC where the bias voltages on the wire planes are distorted such that the
ionization electron drift paths are modified.  Such alterations change
the shapes of induced current produced on wires, requiring modifications to the signal
processing chain.  This is necessary in order to accurately determine the ionization electron
distributions arriving at the TPC wires in these parts of the detector.

For the shorted-U region, the change in bias voltage in the U plane wires leads
to a reduced number of ionization electrons passing through the V plane and being
collected by the Y plane.  Although there is clearly a reduction in the strength of the
induced current as shown in the full responses extracted from data, the change in the shape of
the field response function is limited.  Therefore, no modification is made to the standard signal
processing chain for the V plane and Y plane wires going through this region of the detector.  Once the 3D
positions of the ionization charge are deduced, the charge reconstructed in the shorted-U
region is scaled up in order to compensate for the lost ionization electrons that
are instead collected by the unresponsive U plane wires.

For the shorted-Y region, the situation is much more complicated. As the collection plane wires lose 
their bias voltage, some of the drifting charge is collected by the V plane wires positioned in front of them. 
The collection of ionization charge on these V plane wires leads to the formation of unipolar signals as 
opposed to the bipolar signals that are nominally seen.  Meanwhile, these V plane wires also span the 
normal region where the passage of the ionization electrons through the V plane would lead to bipolar
signals being produced.  Therefore, both unipolar and bipolar signals can simultaneously
exist on the set of V planes wires that physically overlap the shorted Y plane wires in this region. 
The standard signal processing chain attempts to interpret the measured signals with bipolar 
response functions.  In this case, the existence of unipolar signals leads to large distortions in the 
deconvolved signal.  These distortions propagate into the downstream event reconstruction chain,
and are not desired.  To deal with these situations, a special signal processing method based on 
compressed sensing techniques~\cite{lasso,bp,cs} was developed.  These techniques were
originally introduced into LArTPC event reconstruction~\cite{chao_l1} to improve the
speed of the Wire-Cell 3D imaging reconstruction~\cite{wirecell}.
The central idea of compressed sensing is that an underdetermined linear system can be correctly solved by making assumptions about the intrinsic properties of potential solutions, such as sparsity and nonnegativity.

In the normal region, the measured signal is a vector with $n$ time bins. The
ionization charge signal $S_1$ to be deduced can also be expressed as a vector with $n$
time bins.  As discussed in section~\ref{sec:introduction:definitions}, they are connected by a
presumably known bipolar response $R_1$, which is an $n\times n$ matrix:
\begin{equation}\label{eq:normal}
M = R_1\cdot S_1.
\end{equation}
Given $n$ measurements and $n$ number of unknowns, the extraction of $S_1$ essentially
requires an inversion of the $R_1$ matrix, which in practice is performed through a
discrete Fourier transformation. However, in the case of V plane wires going through
the shorted-Y region, there are two types of signals and responses: 
\begin{equation}\label{eq:shorted}
M = R_1 \cdot S_1 + R_2 \cdot S_2.
\end{equation}
Here, $S_2$ represents the ionization electrons collected by the V plane wires with the
unipolar field response, $R_2$.  In this case, there are $n$ measurements, but the
number of unknowns is increased to $2 n$, leading to an underdetermined system.  As a result,
there are in principle an infinite number of solutions $(S_1,S_2)$ that can explain the same 
measurement $M$.  Fortunately, Eq.~\ref{eq:shorted} can be solved with compressed sensing techniques~\cite{lasso,bp,cs} through the application of L1 regularization.  The solution to 
Eq.~\ref{eq:shorted} is obtained through a minimization of
\begin{equation}\label{eq:l1chi2}
\chi^2 = \left( M - R_1 \cdot S_1 - R_2 \cdot S_2 \right)^2 + \lambda \cdot \sum_i \left( |S_{1i}| + |S_{2i}|\right).
\end{equation}
Here, $|S_{1i}|$ represents the absolute value of the $i$th element of the unknown vector $S_1$
and $\lambda$ is the regularization strength, a parameter to be chosen before minimization.
The minimization of Eq.~\ref{eq:l1chi2} can be achieved through a coordinate descent
algorithm~\cite{coordesc}, which is an iterative procedure. The final charge observed on the V 
plane wires would be the summation of $S_1$ and $S_2$.  Due to the computational limitation of
the coordinate descent algorithm, the signal processing used for the V plane wires going through the
shorted-Y region (also referred to as ``L1SP'') is limited to a 1D deconvolution instead of the 2D deconvolution that is used elsewhere.  In addition, L1SP is only applied to the signal ROI (region of interest) instead 
of the entire waveform to limit the number of unknowns.  $R_1$ contains the average 
bipolar field response and $R_2$ contains the average unipolar field response, with the 
regularization strength chosen to achieve an optimal result.  The deconvolved signal from the
standard signal processing chain is only replaced by that from L1SP if the strength
of $S_2$ (unipolar) is significantly large.  

\begin{figure}[tb]
\centering
  \includegraphics[width=0.7\textwidth]{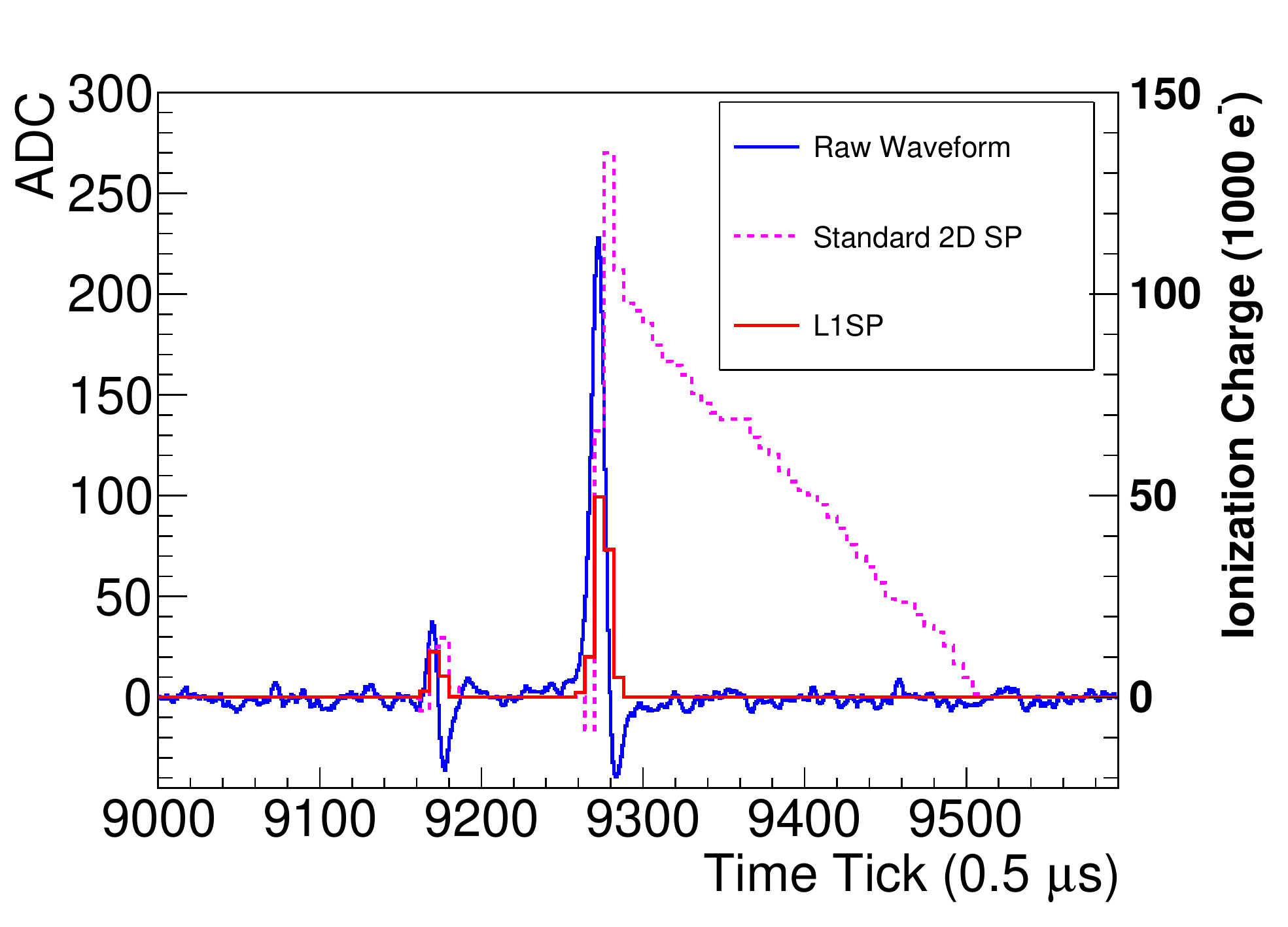}
    \Put(-250,350){\fontfamily{phv}\selectfont \textbf{MicroBooNE}}
  \caption{Illustration of the utilization of L1 signal processing (L1SP). A raw waveform (blue) is shown along with a comparison of two different signal processing algorithms performed on the waveform: the standard 2D deconvolution (pink, dashed) and the L1SP deconvolution (red). The first signal in the raw
  waveform is clearly bipolar and corresponds to a cloud of ionization electrons produced in
  the normal region. The second signal in the raw waveform is close to a unipolar
  shape and corresponds to another cloud of ionization electrons produced in the shorted-Y
  region, which is largely collected by V plane wires in this region of the detector. The deconvolved 
  results of L1SP, measured in units of 1000 electrons after combining every six time ticks of the waveform, 
  is compared with that of the standard 2D signal processing (2DSP), which contains a large distortion
  due to the incompatible field response used in the deconvolution.  Improvement in the L1SP case is observed.}
  \label{fig:l1sp_1D}
\end{figure}

\begin{figure}[tb]
\centering
  \includegraphics[width=0.9\textwidth]{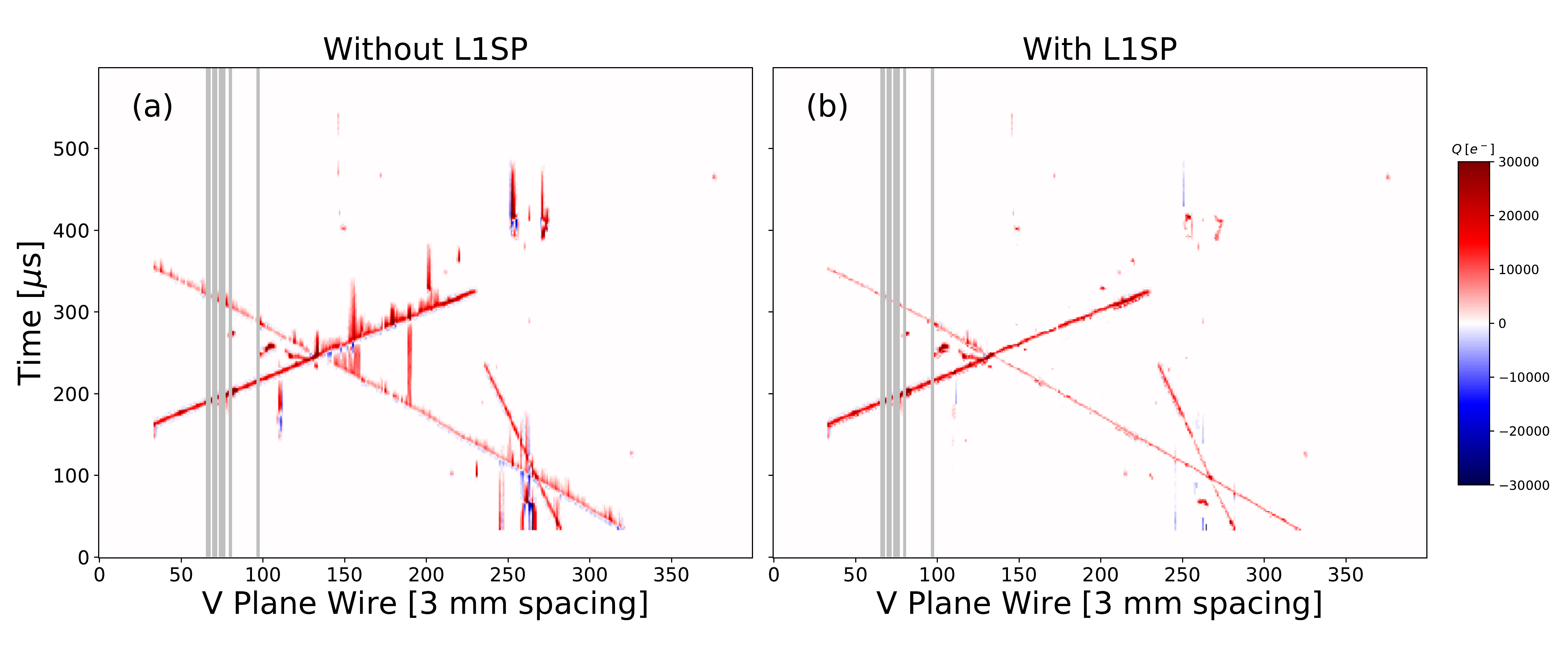}
    \Put(-110,260){\fontfamily{phv}\selectfont \textbf{MicroBooNE}}
    \Put(-280,260){\fontfamily{phv}\selectfont \textbf{MicroBooNE}}
  \caption{Event display highlighting the shorted-Y region of the detector, looking at the V plane wires 
  post-deconvolution.  Noisy and unresponsive channels are shaded grey in this illustration.  Shown is the result for the standard 2D signal processing treatment (a) as well 
  as results for the L1SP technique (b) in units of electrons per 3~\si{\micro\second}.  The long tails introduced by the standard treatment are clearly removed when the 
  L1SP technique is utilized for signal processing of the V plane waveform data in this region of the detector.}
  \label{fig:L1SP_EVD}
\end{figure}

Figure~\ref{fig:l1sp_1D} illustrates the improved performance of the L1SP technique using waveform data 
associated with a V plane wire that overlaps the shorted-Y region.  Compared to the results of standard 2D 
signal processing chain, the L1SP technique shows clear advantages in dealing with signals from the 
shorted-Y region, allowing for unbiased charge estimation in this region of the detector.  An event 
display is shown in figure~\ref{fig:L1SP_EVD}, highlighting a region of the detector where there are ionization 
tracks produced in the shorted-Y region and looking at the deconvolved V plane wire signals with and without 
the utilization of L1SP for the signal processing.  
The long tails introduced by the standard 2D signal processing chain are clearly removed using the L1SP technique.

%% file: sigproc_data_part2.tex
Additional steps are required in order to apply the signal processing chain described in
section~\ref{sec:introduction} to actual data collected by the MicroBooNE LArTPC.
Various features unique to events in data are accounted for in order to put
data and simulation on equal footing.  Ultimately, this care is taken to improve the
accuracy of physics measurements performed using MicroBooNE data.

A number of features have been observed in data events since the beginning of data-taking
with the detector.  These features are either difficult to introduce into simulation,
or are best handled by correcting the data events directly (as negligible residual bias
is expected after the correction).  For instance, removal of signals on the TPC wires
introduced by capacitive coupling to the PMT system is described in
section~\ref{sec:sigproc_data:pmtinducedsig}.  Another correction that addresses the shifting
of bits in the ADCs is next discussed in section~\ref{sec:sigproc_data:adcbitshift}.  Finally,
covered in section~\ref{sec:sigproc_data:PMandCBnoise} is the identification of ``burst-like''
signals on the TPC wires introduced by either the firing of the purity monitors in the
detector or capacitive coupling to activity on the cathode or field cage.

%% file: sigproc_data_pmtinducedsig_part2.tex
The PMTs serve as the light collection system for MicroBooNE, enabling the $t_{0}$-tagging
of cosmics and triggering of neutrino beam events (see section~\ref{sec:fieldresp:methodology}).  When
charged particles (e.g. cosmic muons) travel through the detector in the region between the PMTs and the anode
planes, a large signal on the PMTs due to scintillation light is expected.  It has been observed that
groups of TPC wires also pick up signals in coincidence with charged particles traveling
close to the PMTs (near the anode plane).  This pick-up comes through the capacitive
coupling between the PMTs and the TPC wires and is
triggered by the large currents produced in the PMTs upon the arrival of a large amount of scintillation
light.  
This phenomenon has been observed in both the ICARUS~\cite{ICARUS} and LArIAT~\cite{LArIAT} LArTPC
experiments, and has been observed in MicroBooNE data since the beginning of data-taking.  The
characteristics of the PMT-induced TPC signals include:
\begin{itemize}
\item A range of collection plane wires that are located in front of the fired PMT
  simultaneously have a large negative raw signal corresponding to a
  positive current, which is different from the typical negative current
  induced by ionization electrons.
\item At the same time, a range of induction plane wires that are located
  in front of the fired PMT also see a coincident signal. The strength
  of this signal is suppressed compared to that of the collection plane wires due
  to the shielding of the induction plane wires by the collection plane wires.
\item These signals in the induction and collection plane wires are truly coincident
  (i.e. occurring at the same time tick). This is different from the induced
  current from ionization electrons, as ionization electrons take several
  microseconds to travel from one wire plane to the next.
\end{itemize}

\begin{figure}
\centering
\includegraphics[width=.99\textwidth]{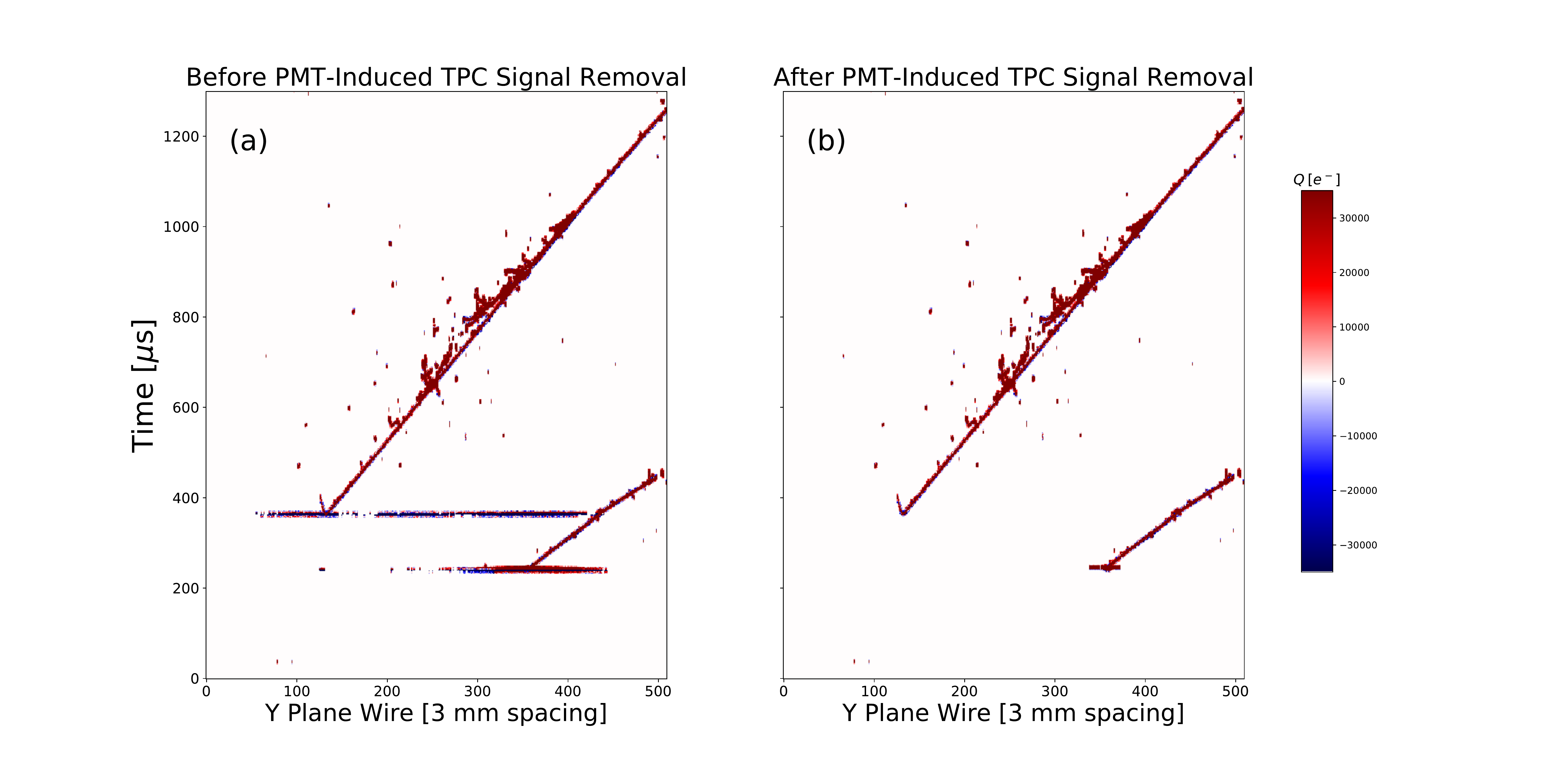}
\Put(-205,70){\fontfamily{phv}\selectfont \textbf{MicroBooNE}}
\caption{Event display of the collection plane (Y plane) before (a) and after (b)
removal of PMT pick-up noise,
after carrying out the deconvolution in both cases in units of electrons per 3~\si{\micro\second}.  The event display is
zoomed-in to a small part of the detector where PMT-induced TPC signals are observed in the event.
The horizontal blue and red streaks seen in the bottom of the left image, which are attributed to
capacitive-coupling between a PMT and the TPC wires, are largely removed via a dedicated algorithm.
} \label{fig:PMTNoiseRemovalEVD}
\end{figure}

A special algorithm is utilized to find and remove these PMT-induced TPC signals,
without compromising the real signals induced by
ionization electrons.  For collection plane wires, this is straightforward, as the
PMT-induced TPC signals behave quite differently from the normal ionization signals, with
negative instead of positive signals on the waveform.  Therefore, the PMT-induced signals
can be removed efficiently by searching for negative signals that are of high amplitude with respect
to the electronics noise level. Once the PMT-induced signal is found on the collection
plane, the induction plane waveforms at the same time tick are searched
for a group of induction plane wire signals that exhibit the signatures of PMT-induced
signals. These identified PMT-induced signals are then removed from the waveform.
Figure~\ref{fig:PMTNoiseRemovalEVD} shows a collection plane event display before and after 
the removal of the PMT pick-up noise, after carrying out the deconvolution in both cases.
The pick-up feature is largely removed via the use of the technique described above.

%% file: sigproc_data_adcbitshift_part2.tex
\begin{figure}[tb]
\centering
\begin{subfigure}{0.49\textwidth}
  \centering
  \includegraphics[width=.99\textwidth]{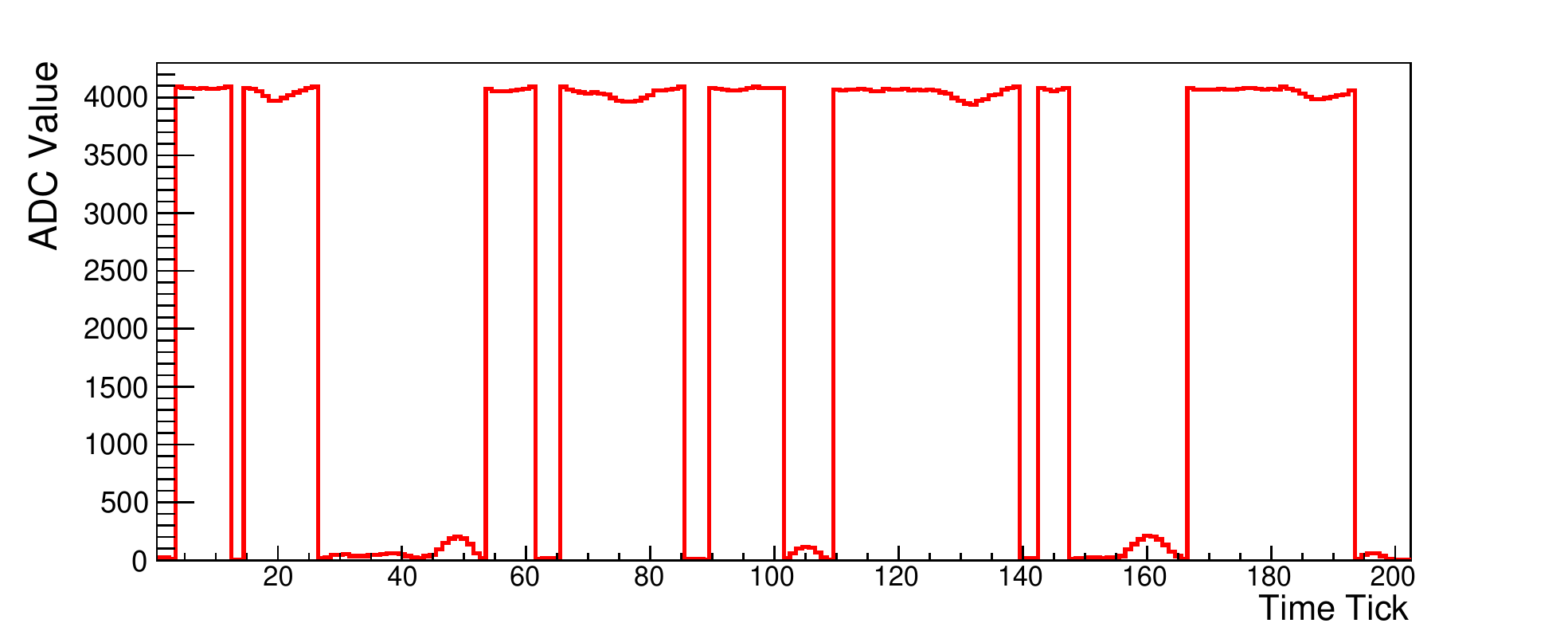}
  \caption{Example of a bit-shifted waveform.}
\end{subfigure}
\begin{subfigure}{0.49\textwidth}
  \centering
  \includegraphics[width=.99\textwidth]{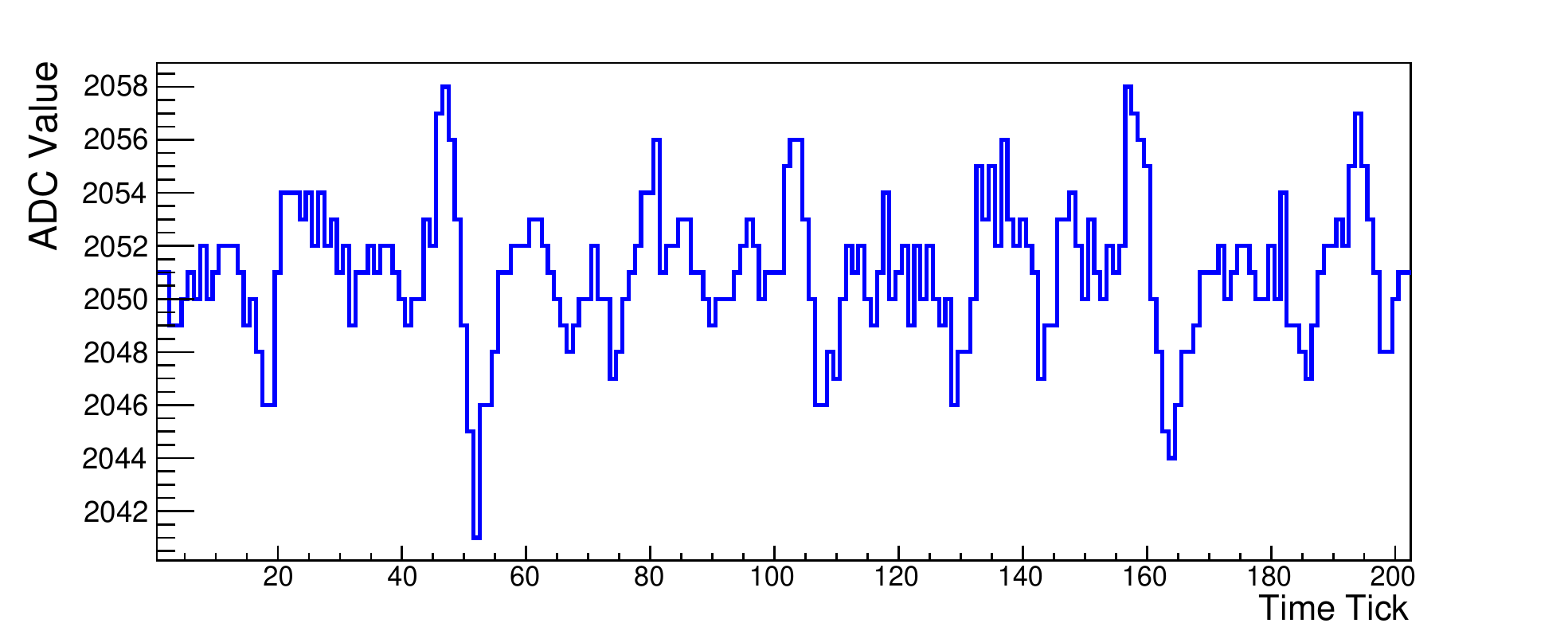}
  \caption{Example of a fixed waveform.}
\end{subfigure}
\\
\vspace{4mm}
\begin{subfigure}{0.59\textwidth}
  \centering
  \includegraphics[width=.99\textwidth]{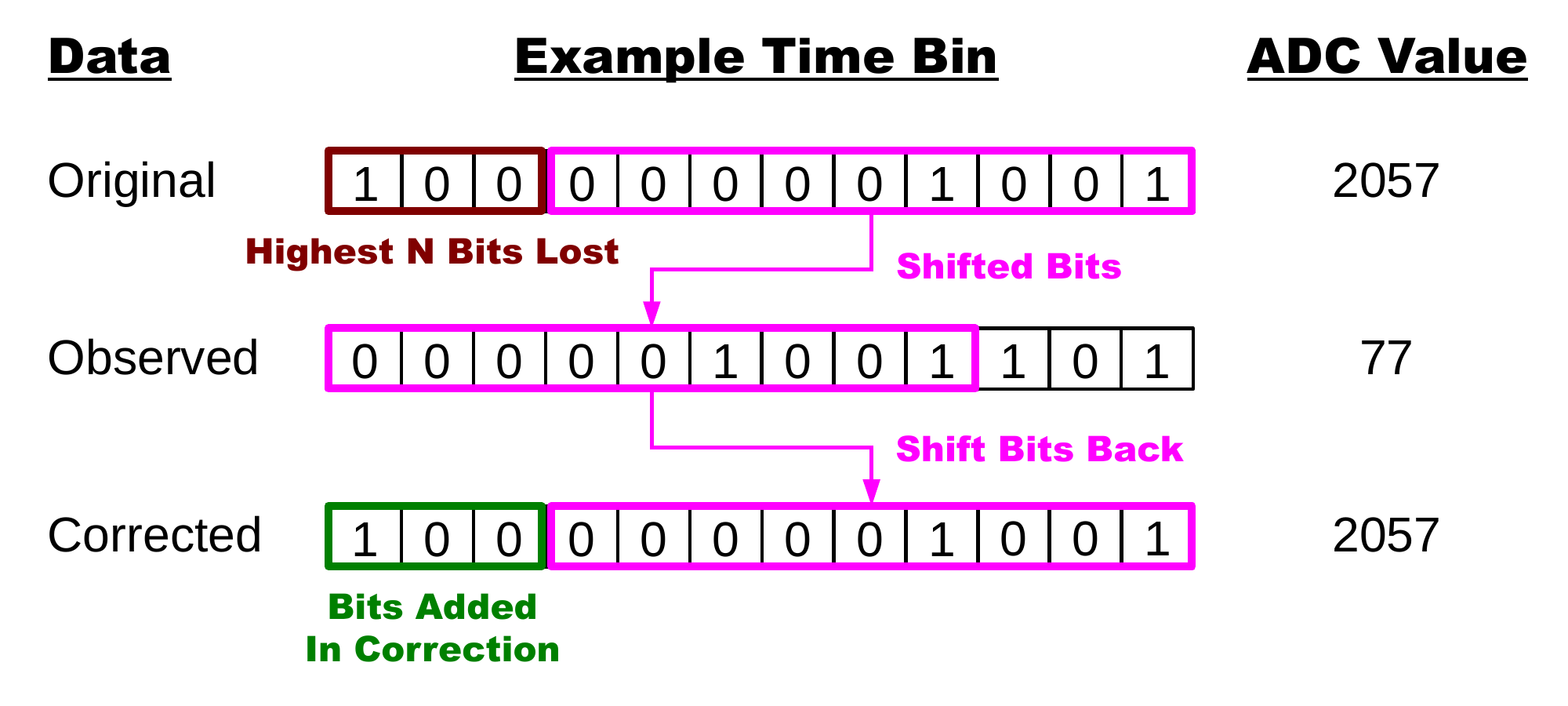}
  \caption{Illustration of correction of single ADC value.}
\end{subfigure}
\Put(-45,240){\fontfamily{phv}\selectfont \textbf{MicroBooNE}}
\caption{Visualization of the ADC bit shift issue and how it impacts ADC values in waveform data, as well as the
basic concept of the correction method utilized to recover channel waveform data affected by the ADC bit shift 
issue.  A small part of a waveform from a channel affected by the ADC bit shift
issue is shown (a).  Each ADC value on the waveform is shifted by the same number of bits, upward, leading
to the loss of the highest $N$ bits for a $N$-bit shift.  The lowest $N$ bits are replaced with
new values, as illustrated in the table at the bottom of the figure.  The corrected waveform is also shown (b).  In addition, the correction of a single ADC value is illustrated (c);  for a bit-shift of 
$N$ bits, the highest $12-N$ bits are shifted down $N$ places.  The new highest $N$ bits are selected from a collection of lowest bits associated with 
each of the time bins on the given channel's waveform before any correction is applied.  These highest $N$ bits are chosen based on their compatibility with an interpolation/extrapolation using the two time bins before and after the 
waveform time bin of interest, enforcing the fact that a channel's waveform should be relatively smooth
moving from time bin to time bin. This procedure is repeated for every channel in the event.} \label{fig:ProblemSolutionCartoon}
\end{figure}

It has been observed that periodically groups of TPC channels connected to common electronics
would experience a change of state, the main feature of which is a
multiplication of the ADC values associated with the channel waveforms by some constant
factor.  This factor was observed to be a multiple of two (e.g. two, four, eight, etc.),
a feature that is equivalent to a bit-shifting of the ADC values with a loss of higher bits.
The problem is ameliorated by power cycling the warm readout crates.  While the cause of this issue
is not yet completely understood, given that changes in the problem only occur at the beginning of
data-taking (when the ADC electronics are initialized), it is speculated that it may be due to an
occasional misconfiguration of the ADC test pattern used in the dynamic phase alignment~\cite{AlteraWhitePaper}
circuit of the field-programmable gate array (FPGA) that sets the correct order to read out the ADC
data bits for a given channel.  A visual depiction of the ADC bit-shifting issue is shown in figure~\ref{fig:ProblemSolutionCartoon}.

In order to identify problematic channels affected with the ADC bit-shifting issue, a waveform-level
algorithm is run on every channel in the noise-filtering step of event reconstruction.  This algorithm 
finds the number of bits shifted in the ADC values of a channel waveform by determining the bit that 
fluctuates the most between adjacent time bins.  This bit should be the true lowest bit as even 
small variations throughout the extent of the waveform, due to noise, may alter the value of this
bit before it is shifted in the warm electronics.
Each channel with a non-zero bit shift is categorized as a ``bad'' channel that is corrected by another
algorithm described below.  It is observed that a bad channel remains bit-shifted by the same number of
bits for the entirety of a data-taking run (up to seven hours in length), with transitions
only happening at run boundaries
(when the warm electronics are reconfigured).  After the first observation and characterization of
this problem, improvements were made to the monitoring of the detector to allow for quick response
to this issue by the MicroBooNE operations team.  However, some earlier data-taking runs at MicroBooNE
are impacted by the ADC bit shift issue and require special treatment.

A correction for each affected channel is attempted before carrying out extensive noise filtering
and signal processing of the TPC channel waveforms.  This correction is illustrated in
figure~\ref{fig:ProblemSolutionCartoon}.  As the number of bits by which the ADC values of the
channel waveform are shifted is known, the highest bits of the observed waveform can be shifted back down
to partially restore the original waveform.  After this operation, one must still determine the
highest bits of the ADC values of the channel waveform that were lost when the original waveform
was bit-shifted.  Empirically it was found that these highest bits were almost always shifted into
the lowest bits of the waveform ADC values.  In other words, the lowest bits of the observed waveforms
associated with bad channels are not random and can be corrected.  However, the lowest bits of the
waveform are not always from the same time bin that the bits were originally associated with.  In order
to pick which set of bits to assign to the highest bits
of the corrected waveform (from the collection of lowest bits associated with each of the time bins
on a given channel's waveform before any correction is applied), a smoothness criterion is employed.
This criterion, utilizing a combination of interpolation between and extrapolation across neighboring
time bins (two in each direction), makes use of the fact that a channel's waveform should be relatively
smooth moving from time bin to time bin.  An example of this correction on a subset of an affected
waveform is shown in figure~\ref{fig:ProblemSolutionCartoon}.



\begin{figure}[tb]
\centering
\includegraphics[width=.95\textwidth]{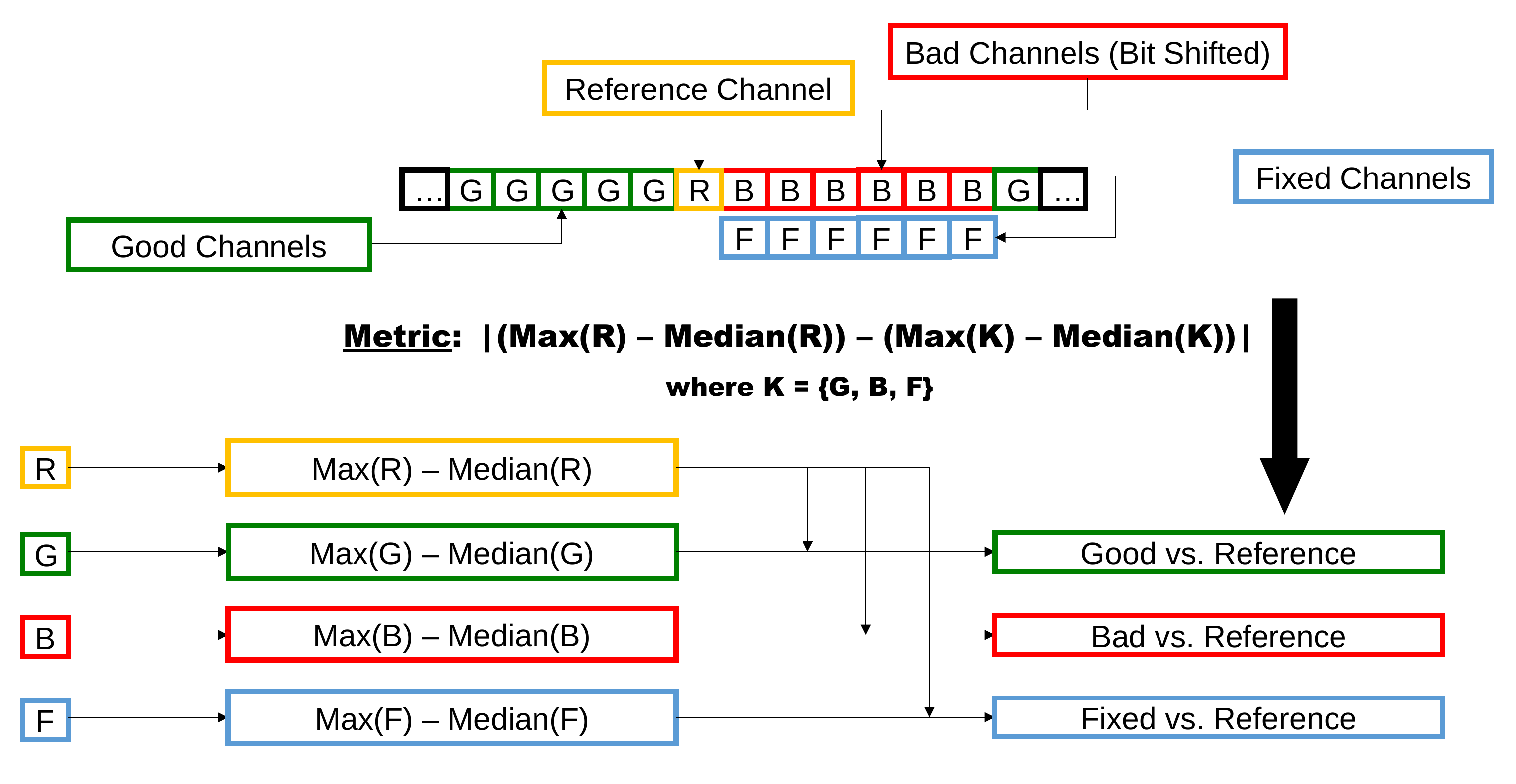}
\caption{Visualization of the construction of the ADC bit shift metric, which is utilized to validate the correction of the ADC bit shift issue.  The metric is constructed by taking the maximum ADC value on the waveform relative to the median ADC value on both the reference channel (chosen to be the closest good channel preceeding a string of bit-shifted channels) and the bad channel, taking the difference between the two.  The same metric can be constructed using a fixed channel (bad channel after the correction procedure has been carried out) or good channel instead of a bad channel.  By comparing the metric distribution constructed using the fixed (corrected) channels and the metric distribution constructed using good channels that are known to be unaffected by the ADC bit shift issue, one can validate the performance of the correction procedure using data.} \label{fig:ADCShiftCorrMetricDef}
\end{figure}

\begin{figure}[tb]
\centering
\begin{subfigure}{0.49\textwidth}
  \centering
  \includegraphics[width=.99\textwidth]{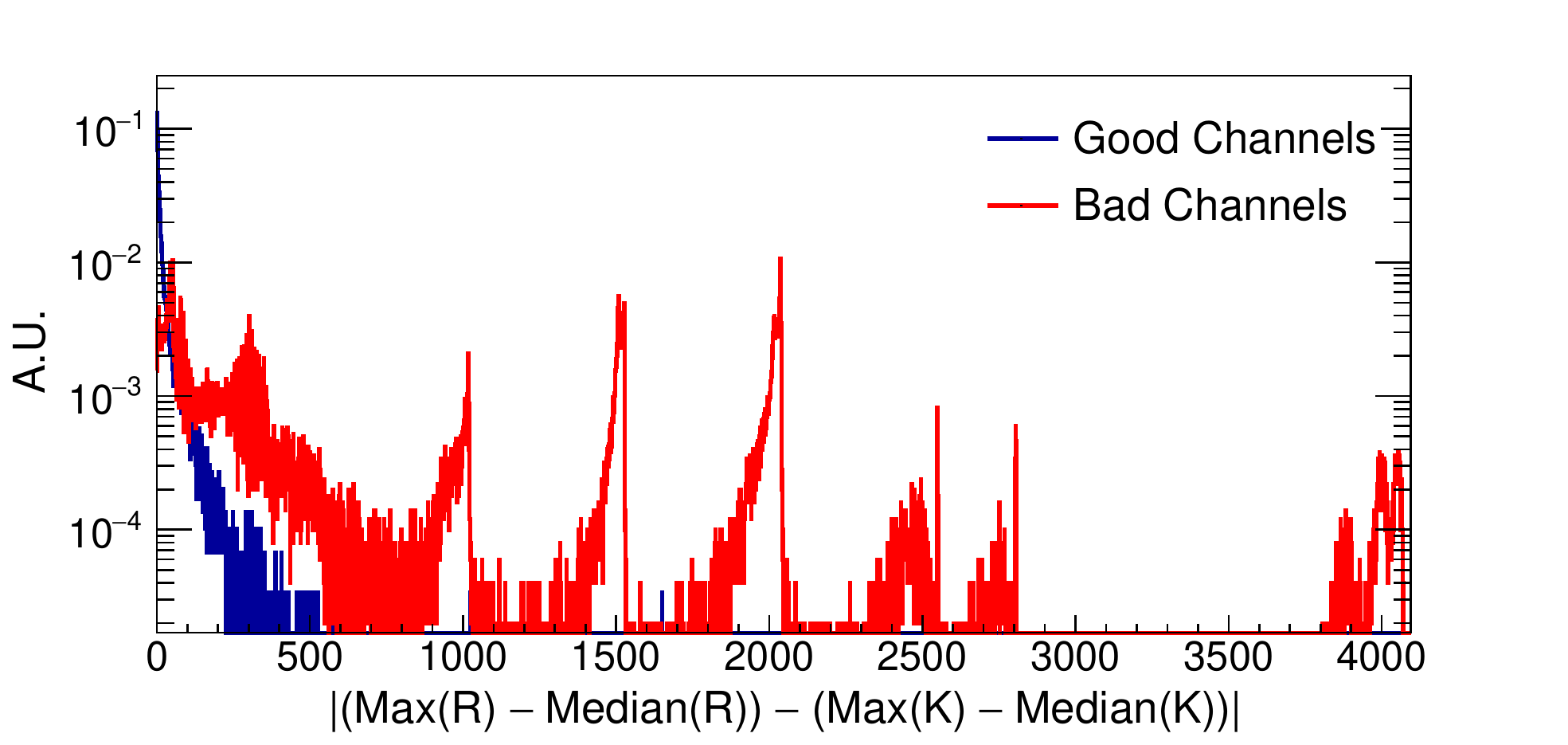}
  \caption{Good/bad channel comparison (full range).}
\end{subfigure}
\begin{subfigure}{0.49\textwidth}
  \centering
  \includegraphics[width=.99\textwidth]{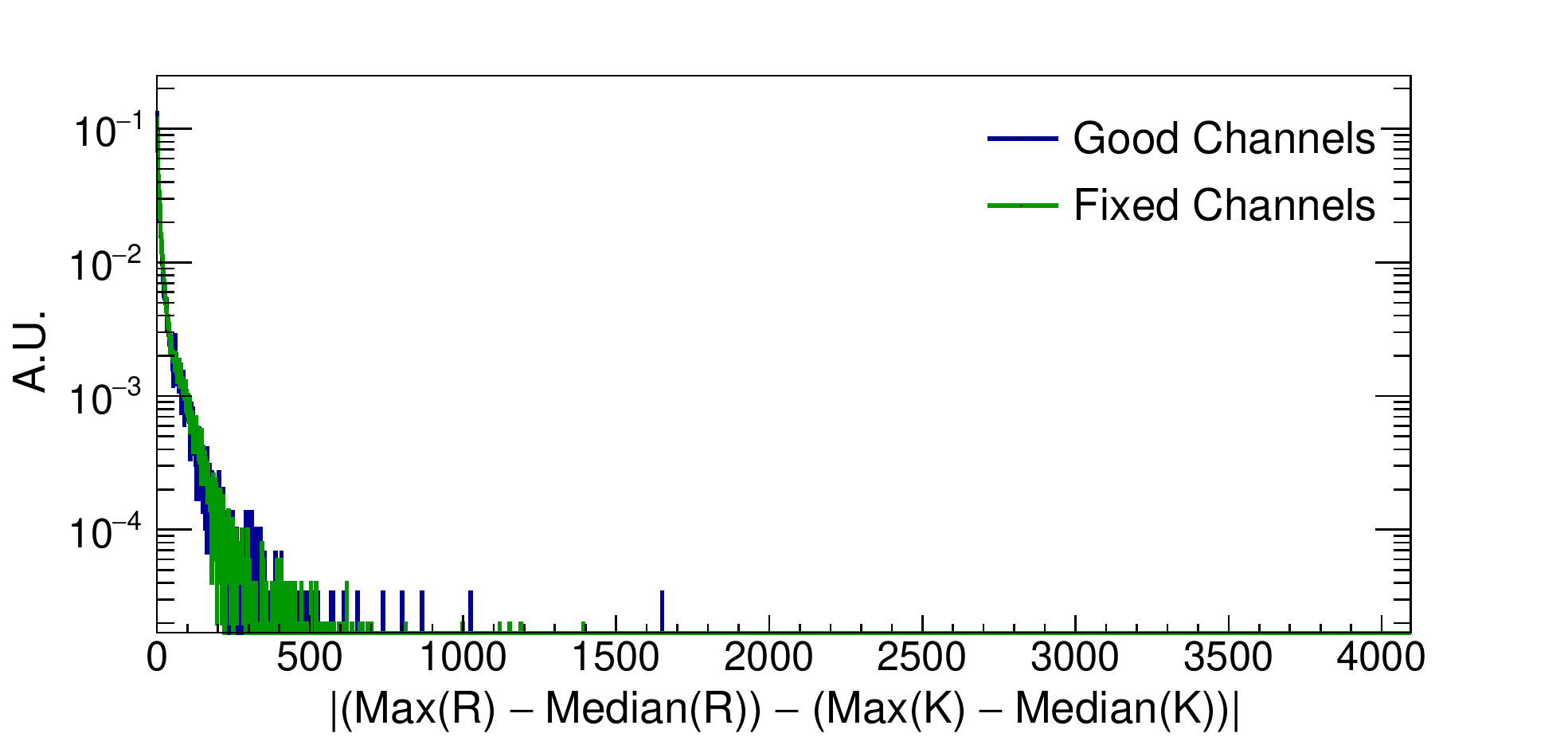}
  \caption{Good/fixed channel comparison (full range).}
\end{subfigure}
\\
\vspace{2mm}
\begin{subfigure}{0.49\textwidth}
  \centering
  \includegraphics[width=.99\textwidth]{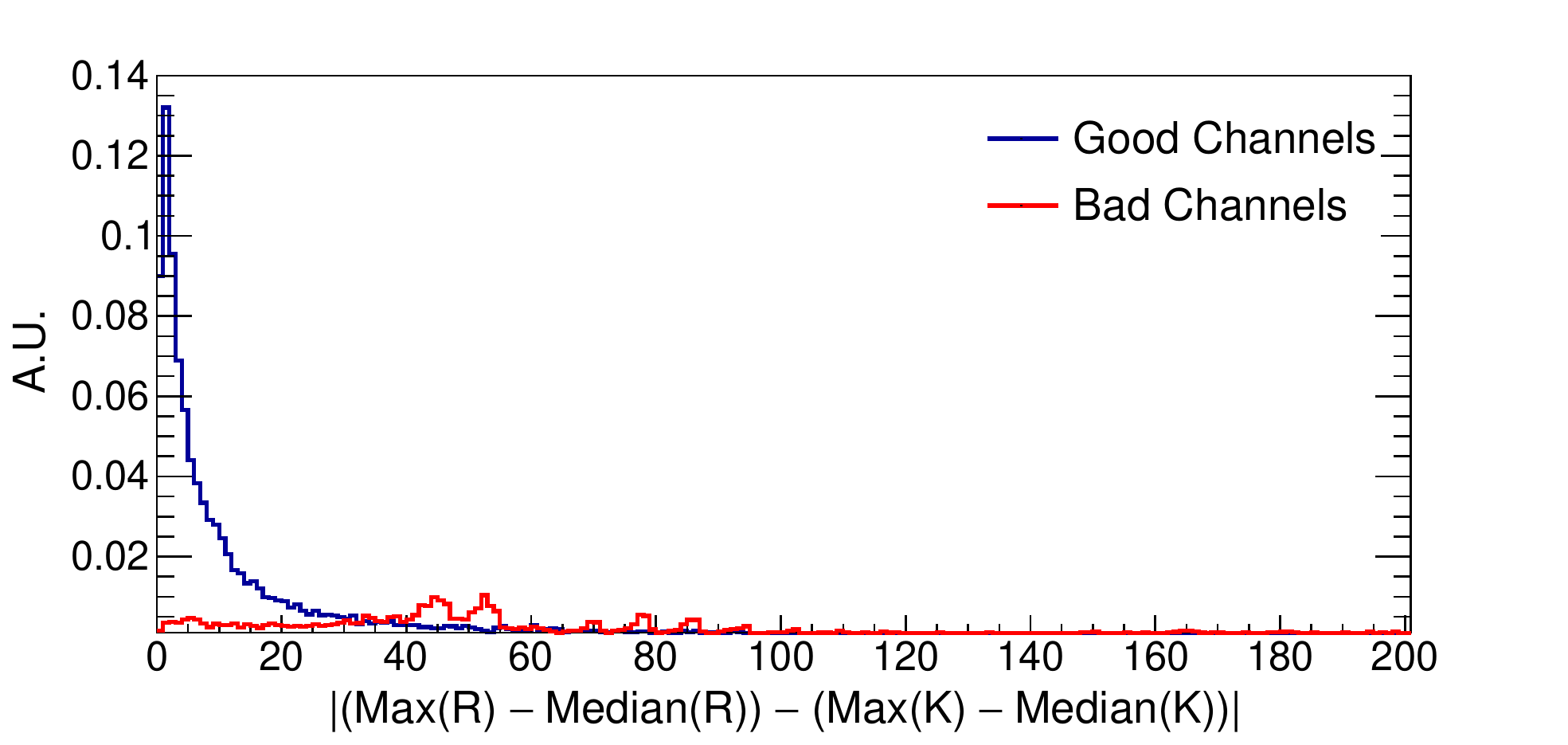}
  \caption{Good/bad channel comparison (zoom-in).}
\end{subfigure}
\begin{subfigure}{0.49\textwidth}
  \centering
  \includegraphics[width=.99\textwidth]{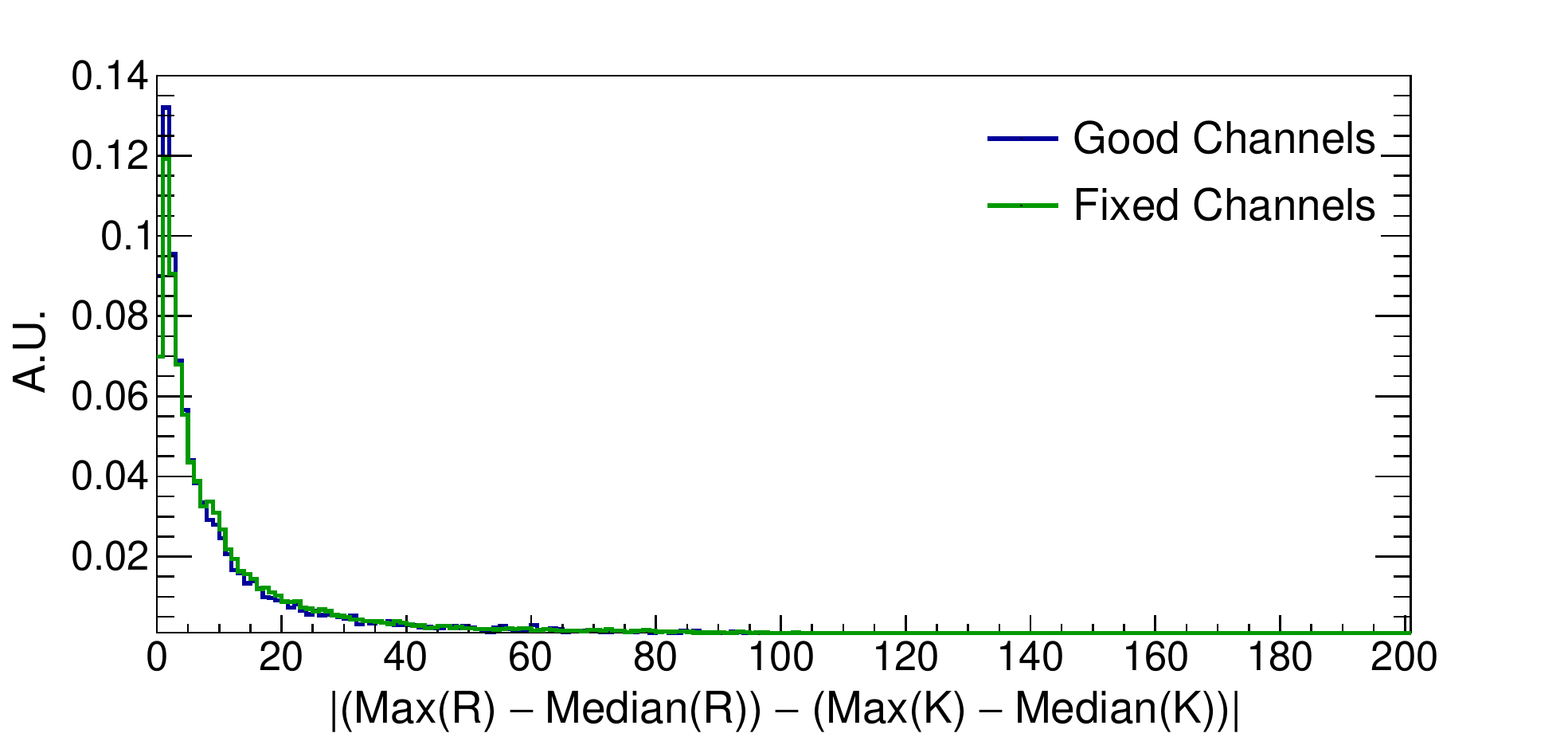}
  \caption{Good/fixed channel comparison (zoom-in).}
\end{subfigure}
\Put(-395,330){\fontfamily{phv}\selectfont \textbf{MicroBooNE}}
\caption{Results of performing the ADC bit shift correction on affected channels in MicroBooNE data.  The validation procedure outlined in figure~\ref{fig:ADCShiftCorrMetricDef} is carried out for bad, fixed, and good channels.  A comparison of the ADC bit shift metric between good and bad channels is shown in (a) and (c), while a comparison of the ADC bit shift metric between good and fixed channels is illustrated in (b) and (d).  Both full distributions of the ADC bit shift metric values, (a) and (b), and zoomed-in distributions, (c) and (d), are shown.  After the application of the correction method discussed previously, the metric distribution shows excellent agreement with respect to a ``perfect correction'' that is found in data using the comparison between good channels and a reference channel.  In the axis labels, ``R'' refers to a reference channel and ``K'' can be either ``G'' (a good channel), ``B'' (a bad channel), or ``F'' (a fixed bad channel).} \label{fig:ADCShiftCorrResults}
\end{figure}

In order to validate the correction described above, a metric that makes use of ``good'' channels (channels with waveforms unaffected by the ADC bit shift issue) physically neighboring the bad channels was developed in order to quantify how well the correction method performs on actual data.  This validation technique relies on the fact that minimum-ionizing particles (cosmic rays) will produce ionization signals that are similar in magnitude for neighboring wires.  Therefore, we expect the ``fixed'' channels (bad channels that have been corrected via the method described above) to yield measured ionization signals that are similar in amplitude to a ``reference'' channel, the closest good channel in the TPC.  The construction of this metric is described visually in figure~\ref{fig:ADCShiftCorrMetricDef}.  The metric utilizes the maximum ADC value on the waveform relative to the median ADC value on both the reference channel and the bad channel, taking the difference between the two.  This is done both before and after the correction is applied to the bad channel.  By repeating the metric calculation using a good channel instead of a bad channel, the distribution for a ``perfect correction'' is recovered.  The distribution of the ADC bit shift correction metric using bad channels, fixed channels, and good channels along with a reference channel is shown in figure~\ref{fig:ADCShiftCorrResults}.  Several hundred events from an early data-taking run that was significantly impacted by the ADC bit shift issue is utilized in this validation, including all affected channels in the event.  It is seen that the correction performs well, with the distribution of the metric using fixed channels nearly matching the distribution using good channels.

%% file: sigproc_data_PMandCBnoise_part2.tex
\begin{figure}[tb]
\centering
\includegraphics[width=.75\textwidth]{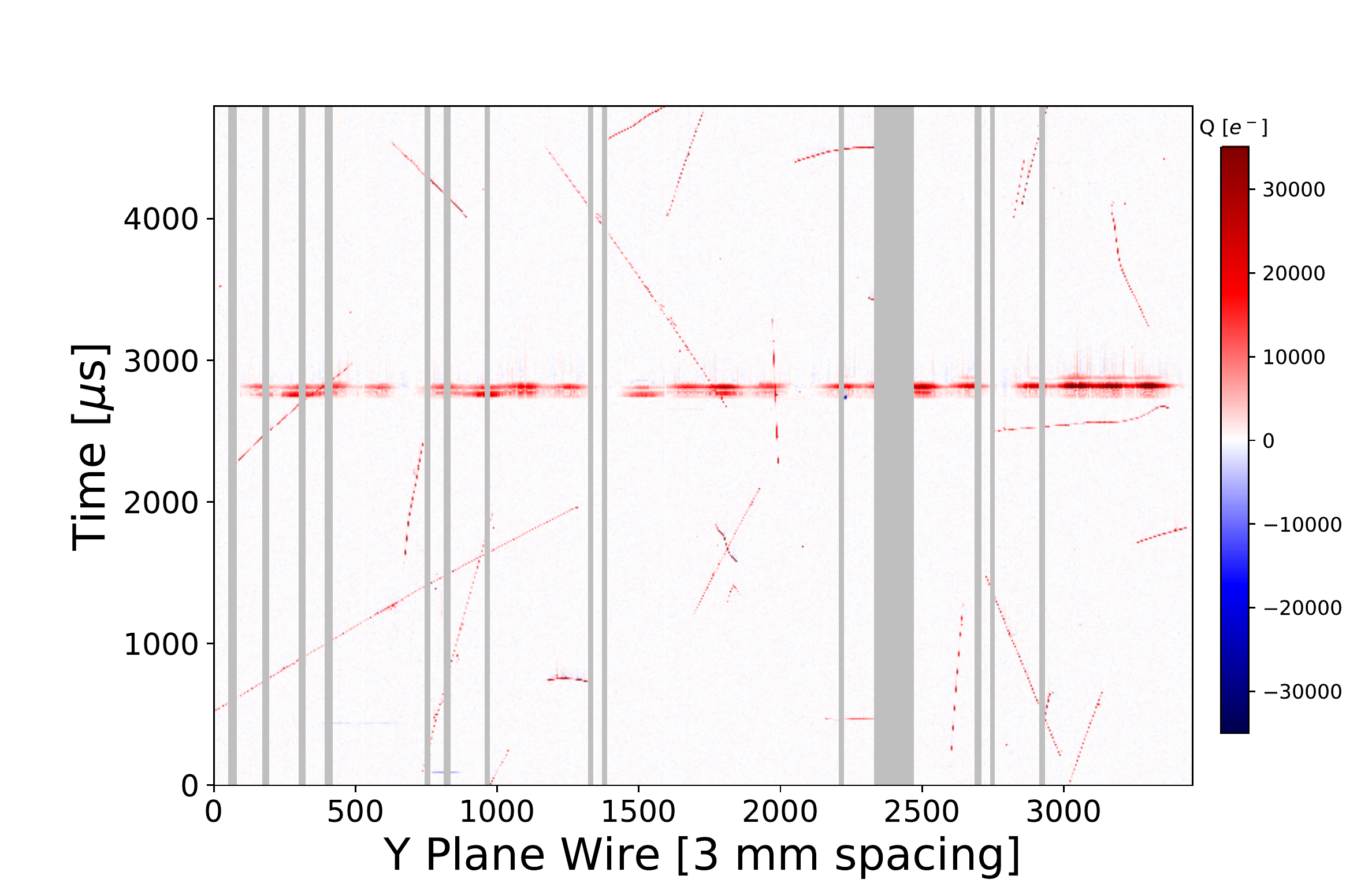}
\Put(-275,381){\fontfamily{phv}\selectfont \textbf{MicroBooNE}}
\caption{Example of an event in MicroBooNE data that is impacted by the purity-monitor-induced burst noise issue.  Shown is an event display for the waveform data associated with Y plane wires after coherent noise filtering is carried out in units of electrons per 3~\si{\micro\second}.  Noisy and unresponsive channels are shaded grey in this illustration.  One distinctive feature of the purity-monitor-induced burst noise is a strong isochronous signal across all wires that is most severe on the Y plane (as seen by the horizontal band in the display).  Another feature of this specific type of burst noise is the large amount of light observed in the PMTs in the same event (discussed later).  The noise feature is not removed from data by the coherent noise filtering procedure that is applied to raw data.} \label{fig:PMBNoiseExample}
\end{figure}
\begin{figure}[tb]
\centering
\includegraphics[width=.65\textwidth]{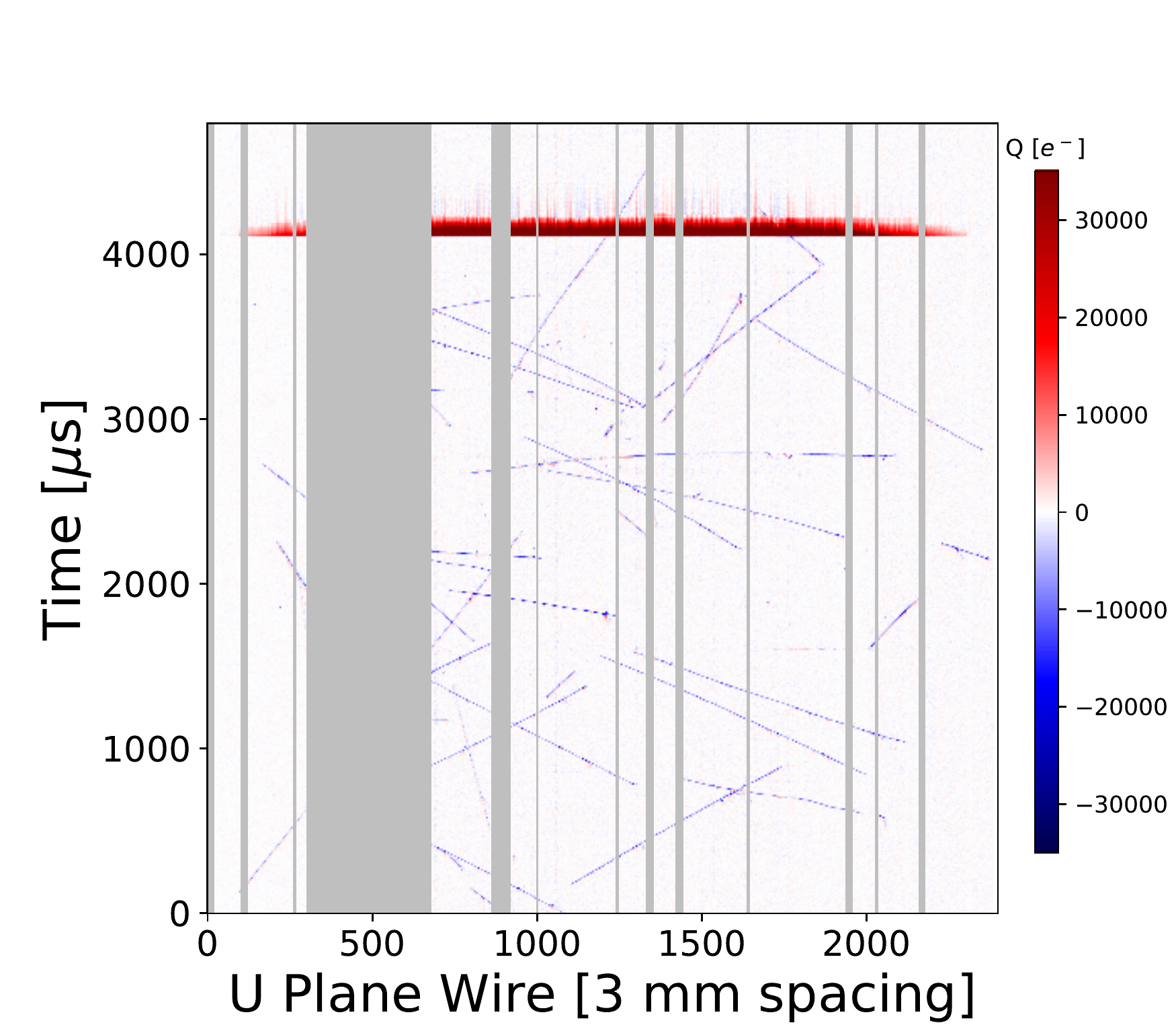}
\Put(-233,445){\fontfamily{phv}\selectfont \textbf{MicroBooNE}}
\caption{Example of an event in MicroBooNE data that is impacted by the cathode burst noise issue.  Shown is an event display for the waveform data associated with the U plane wires after coherent noise filtering is applied in units of electrons per 3~\si{\micro\second}.  Noisy and unresponsive channels are shaded grey in this illustration.  The distinctive feature of the cathode burst noise is a strong isochronous signal across all wires that is most severe on the U plane, which is seen as a band at the top of the event display.  The coherent noise filtering step does not remove this noise feature from data.} \label{fig:CBNoiseExample}
\end{figure}

Throughout data-taking with the MicroBooNE detector, a number of external noise
features were observed in TPC data.  In addition to the continuous external
noise features described in Ref.~\cite{noise_filter_paper}, there have been burst-like signals occurring
on many different TPC channels at the same measured drift time that are not associated to particle-induced
ionization.  At present this burst-like noise has been categorized into two distinct types of signals.
The first one is observed only when the liquid argon purity monitors~\cite{Acciarri:2016smi} are functioning and is
associated with a large amount of light collection in the PMTs; this noise is most intense on the collection
plane (Y plane), the wire plane closest to the PMT system.  The second one is most intense on the first induction
plane (U plane), the wire plane closest to the cathode, and is not associated with
an excess of light seen by the PMT system.  The fact that channels from all three planes share the same
electronics, but the signals are seen predominantly on the U and Y planes, respectively, for the two classes of noise,
points to a source of noise external to the electronics.  Given the locations of the U and Y wire planes in the cryostat,
and correlations with the PMT and purity monitor activity, it is currently believed that the first type of noise is associated
with the firing of the purity monitors (due to a light leak from the purity monitor flash lamps) and the second is due to
capacitive coupling between voltage fluctuations in the TPC cathode plane and the U plane wires.  For this reason, the former issue is
referred to as ``purity-monitor-induced burst noise'' and the latter issue is referred to as ``cathode burst noise''
in the discussion below.  An example of the purity-monitor-induced burst noise and cathode burst noise
is shown in figure~\ref{fig:PMBNoiseExample} and figure~\ref{fig:CBNoiseExample}, respectively.

As discussed above, the purity-monitor-induced burst noise is identifiable by both the striking signature in the TPC waveforms (most dramatic on the Y plane) and the large amount of light seen in the PMTs in the same data event.  While the TPC noise feature is isochronous, the signature seen on different wires are not completely coherent across the entire wire plane, and so this noise can not be completely removed using coherent-noise removal techniques~\cite{noise_filter_paper}.  These coherent-noise removal techniques are also designed in such a way as to prevent the removal of isochronous signals on the TPC wires due to tracks traveling nearly parallel to the wire planes, further complicating the removal of burst-like noise features.  Additionally, the excess of light seen in the PMTs due to the presence of the burst noise provides another complication as this light makes it more difficult to reconstruct the $t_{0}$ of cosmic activity in the event.  As a result, there is no simple algorithm to completely remove the burst noise from TPC and PMT data; instead, an attempt is made to filter out these events from consideration in MicroBooNE physics analyses.  This only needs to be done for data-taking runs prior to April 2016, after which the purity monitors were permanently turned off in order to mitigate this noise source.  Prior to the purity monitors being turned off, the fraction of events impacted by this type of burst noise depends on the relative firing rate of the purity monitors, leading to as many as 10\% of events impacted by the noise in a given data-taking run (lasting several hours).  However, most runs taken while the purity monitors were online see less than 0.5\% of events impacted by the purity-monitor-induced burst noise.

It was found empirically that the high amount of light produced in these events, collected by the PMTs, is enough to tag the purity-monitor-induced burst noise events.  This was determined by comparing two data-taking runs: one before the purity monitors were turned off, and one after.  In both runs, the number of ``flashes'' in the event is measured for all of the events in each of the two data-taking runs, where a ``flash'' is defined as light observed in multiple PMTs at roughly the same time (in a window of tens of nanoseconds).  In order to enhance the separation of the two populations (normal events and purity-monitor-induced burst noise events), a metric was developed (``nflash500'') which is the maximum sum of the number of flashes in the event in a 500~\si{\micro\second} window, scanning the entire event in time.  By placing a cut of nflash500~<~40, we are able to effectively remove all events from the dataset that contain the purity-monitor-induced burst noise feature.


For the cathode burst noise, there is no enhancement of light production in the event and so one must rely strictly on the TPC data in order to identify events with the noise feature present.  A common feature of this particular category of burst noise is a large signal, mostly, but not completely, coherent across all U plane channels, that is very extended in time (on the order of a millisecond).  As a result, the burst noise signal has a very low-frequency component (<~2~\si{\kHz}) that is not present in the bipolar induction plane signals from ionization electrons traveling by the U plane.  Furthermore, other coherent noises do not have power in the relevant frequency range.  The presence of this power in the low-frequency range allows us to identify and remove these events from our dataset.  A metric (``fftsum2'') for this tagging was constructed by integrating the first 2~\si{\kHz} of the power spectrum of the U plane after adding together all channel waveforms associated with that plane, coherently in time.  A cut of fftsum2~<~50000 is applied in order to mitigate the cathode burst noise events.  In February 2017, it was determined that the cause of this specific type of burst noise was due to an intermittent connection between the cathode HV feedthrough and the cathode plane.  After readjusting the cathode HV feedthrough, the noise was no longer observed in significant amounts in MicroBooNE data.

\begin{figure}[tb]
\centering
\begin{subfigure}{0.65\textwidth}
  \centering
  \includegraphics[width=.99\textwidth]{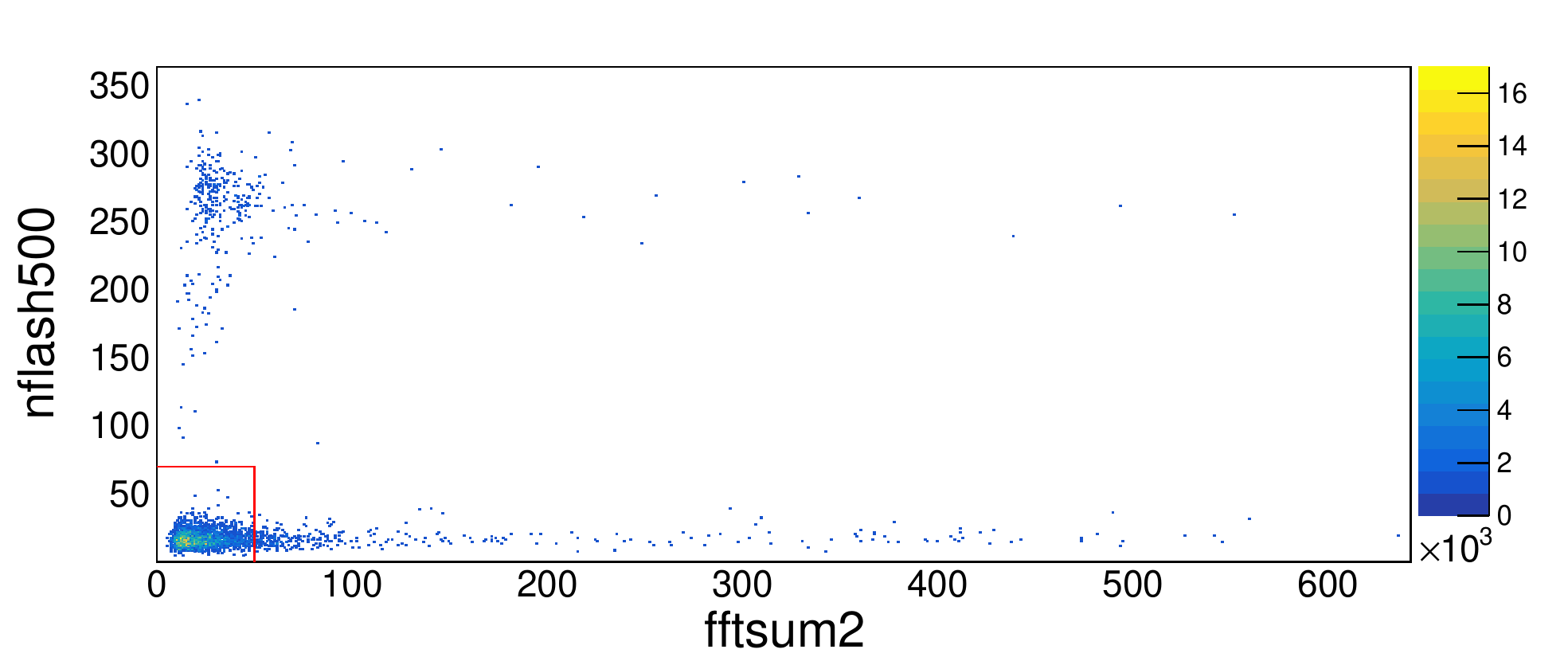}
  \caption{Values of burst noise metrics in a noisy run.}
\end{subfigure}
\\
\begin{subfigure}{0.65\textwidth}
  \centering
  \includegraphics[width=.99\textwidth]{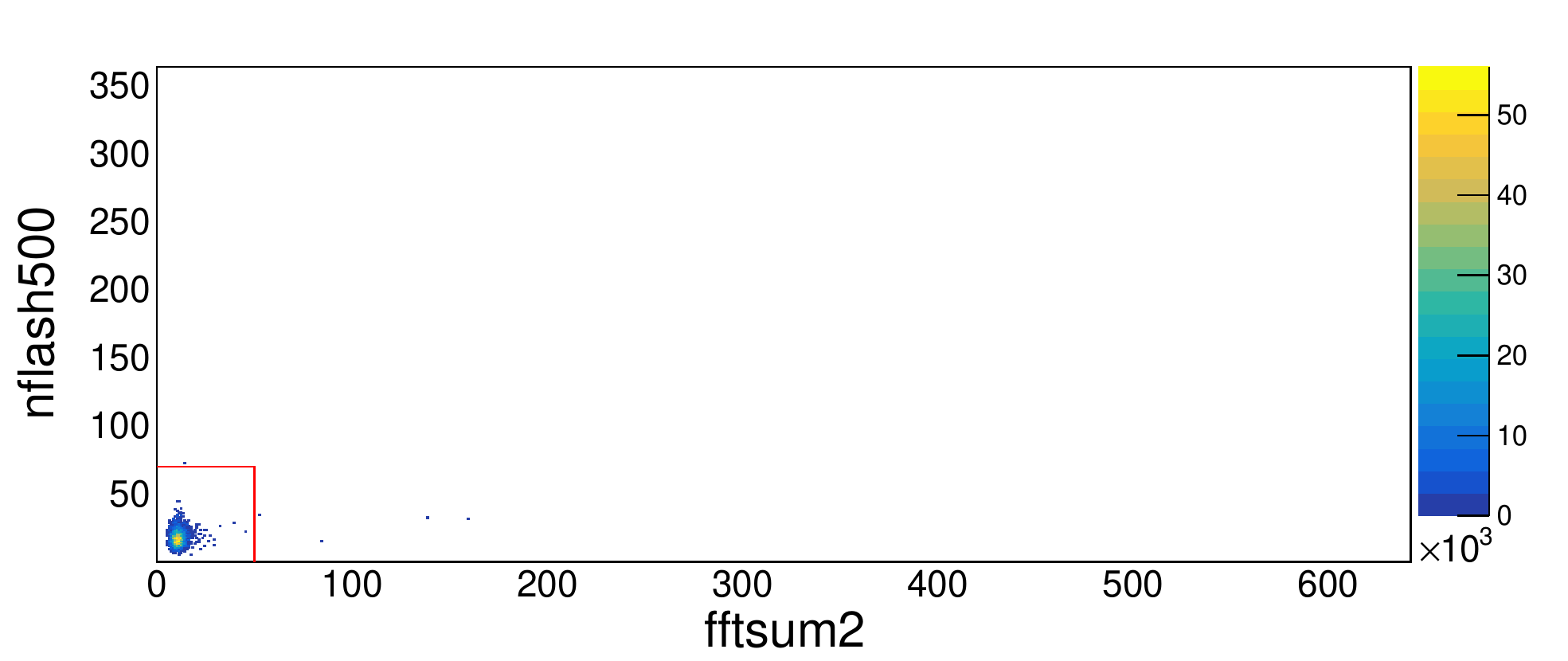}
  \caption{Values of burst noise metrics in a quiet run.}
\end{subfigure}
\Put(-105,370){\fontfamily{phv}\selectfont \textbf{MicroBooNE}}
\caption{Two-dimensional plot showing both the nflash500 and fftsum2 distributions for two data-taking runs, one that was empirically observed to have higher levels of both cathode burst and purity-monitor-induced burst events (a), and one where very little of either burst noise was observed (b).  The cut boundaries in both distributions used to remove the problematic events are shown in red.  The cuts chosen effectively remove the problematic events while leaving ``normal'' events in the sample.} \label{fig:FlashFFT2D}
\end{figure}

Two-dimensional plots showing the distribution of both nflash500 and fftsum2 for two data-taking runs (one before the purity monitors were turned off and one after they were turned off) are shown in figure~\ref{fig:FlashFFT2D}.  It was noticed in the run after the purity monitors were turned off that the cathode burst noise also reduced greatly in magnitude, though this noise feature has been seen in event data after the purity monitors were turned off as well.  Similar to the case of the purity-monitor-induced burst noise, the cathode burst noise impacts as many as 20\% of the events in some data runs, although these runs are relatively rare; in most runs, less than 0.5\% of events are significantly impacted.

The cuts on nflash500 and fftsum2 described above largely remove the purity-monitor-induced burst noise and cathode burst noise events.  However, the cathode burst noise events are not completely removed by this cut.  It is difficult to remove all cases of this noise because some of the cathode burst noise signals are very small in magnitude; because of this, it is also difficult to quantity the fraction of events that remain impacted by the cathode burst noise after application of the cuts.  Near the fftsum2 cut boundary, they are difficult to identify visually on individual waveforms, an indicator that the cut is sufficient in removing any noise that would be troublesome for downstream reconstruction of the event.  It is worth noting that a very small fraction of ``good'' events (less than $0.1\%$) are removed by the cuts.  These events were studied and determined to be associated with unusually large cosmogenic showers that can be removed in the study of lower-energy (<~2~\si{\GeV}) neutrino interaction events, such as those produced by the BNB in the MicroBooNE detector.

%% file: perf_eval_part2.tex
With the calibration of the electronics response in data, the tuning of the wire field response simulation to match 
data, the handling of the shorted wire regions in the signal processing chain for data events, and the 
cleaning or removal of certain noise events unique to data, the data collected by the MicroBooNE detector demonstrates much better agreement with simulated data.  With these conditions met, it is possible to evaluate the 
performance of the new signal processing chain (including the 2D deconvolution) using MicroBooNE data
(hereafter, ``deconvolution'' will refer to the 2D deconvolution, unless otherwise specified).

Two general metrics are used to evaluate the performance of signal processing on data events:  the
qualitative improvement in image quality after the signal deconvolution, discussed in
section~\ref{sec:results:eventdisplays}, and the quantitative improvement in cross-plane charge matching 
after the full signal processing chain, highlighted in section~\ref{sec:results:chargematching}.  
The former is simple to interpret; better image quality after deconvolution is expected to improve the pattern recognition stage of neutrino interaction reconstruction.
The latter is helpful in quantifying improvements in the charge extraction step of event reconstruction.  It is also a necessary demonstration for the application of novel LArTPC event reconstruction techniques like Wire-Cell tomographic reconstruction~\cite{wirecell},
which uses local charge correlations across two or more wire planes of a LArTPC in order to create a three-dimensional distribution of points.
Each point in the distribution has an associated charge that is estimated from the waveform data associated
with overlapping wires from different planes that share signals from the same drifting ionization charge.
This approach allows pattern recognition on three-dimensional images, as opposed to first doing pattern
recognition on the two-dimensional images from each wire plane, followed by matching across views.
This method is expected to have advantages in charge clustering and potentially neutrino
interaction vertex-finding, as it is performed in three dimensions rather than two.  In order to 
realize this reconstruction technique, it must be shown that charge can be matched across the
three wire planes using events in data, which includes the demonstration of the precise extraction
of charge information from the induction planes of a LArTPC.

%% file: perf_eval_eventdisplays_part2.tex
In order to illustrate the qualitative improvement in imaging with the updated signal processing chain, a variety of 
different events have been studied in data, making comparisons before and after the deconvolution step.  In this 
section a number of event displays are shown and discussed in order to demonstrate this improvement in terms of higher
image quality.  By high image quality, we mean that charge deposition in the detector, processed through the signal
processing chain, lead to clearly recognizable patterns that correspond to the production of physical objects in the
detector, e.g.~charged particle tracks and electromagnetic showers (from electrons or photons).

The event displays shown in this section are of localized sections of the entire TPC, without restriction to
any particular section of the detector (as was the case for the studies completed in section~\ref{sec:fieldresp}, where
the ionization charge depositions were restricted to be between 2~\si{\centi\meter} and 10~\si{\centi\meter} away from
the anode plane in order to minimize effects of diffusion). 
In all cases, the data was recorded while the neutrino beam was on under normal MicroBooNE operating conditions.
The events were selected to demonstrate the impact of applying signal processing algorithms to neutrino interaction candidates.

\begin{figure}[tb]
  \centering
  \includegraphics[width=0.8\textwidth]{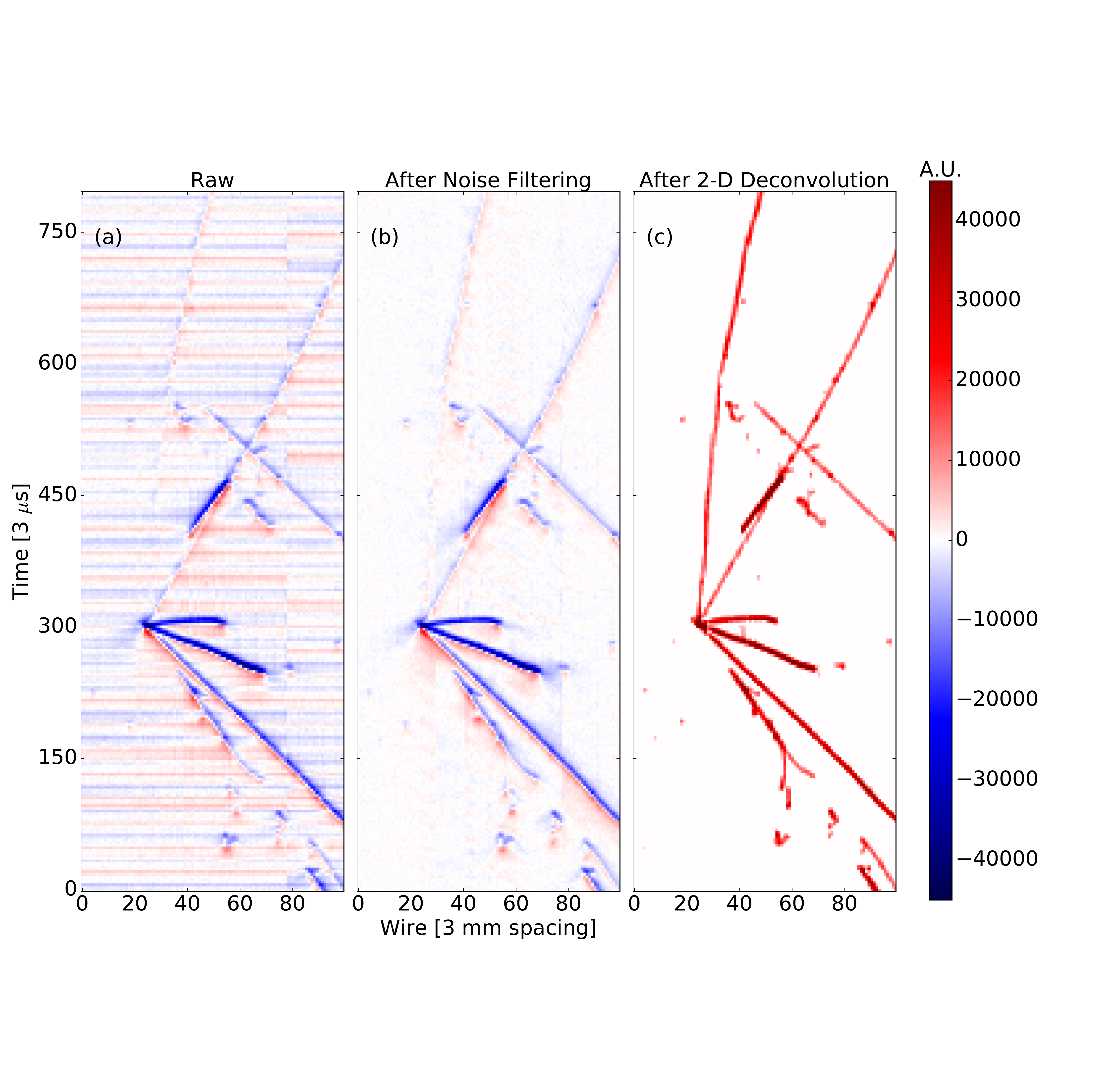}
    \Put(-315,500){\fontfamily{phv}\selectfont \textbf{MicroBooNE}}
  \vspace{-1.3cm}
  \caption{An example neutrino candidate event display from MicroBooNE data (event 41075, run 3493) showing a U plane view. (a) The raw waveform image in units of average baseline subtracted ADC scaled by 250 per 3~\si{\micro\second}. 
(b) The image after software noise-filtering in units of average baseline subtracted ADC scaled by 250 per 3~\si{\micro\second}. 
(c) The image after 2D deconvolution in units of electrons per 3~\si{\micro\second}. 
Prolonged signals associated with near-vertical tracks, such as the one at the top left of each event display window, are recovered after the deconvolution step.  Additionally, the image quality near the neutrino interaction vertex improves after the 2D deconvolution, which is expected to lead to improvements in the pattern recognition.}
  \label{fig:evd_3493_u}
\end{figure}

Figure~\ref{fig:evd_3493_u} illustrates a busy neutrino interaction candidate event with several tracks produced in the
final-state.  Shown are the raw, unprocessed image (``Raw''), the image after software noise filtering is carried
out (``After Noise Filtering'') and finally the image after the full deconvolution is completed (``After
2D Deconvolution'').  Only the U plane is shown in this comparison.  Comparing the image before and after the 
deconvolution is carried out, it can be seen that prolonged signals on the waveform are
recovered by the deconvolution step.  Furthermore, the image quality around the neutrino interaction vertex 
is sharper, as is the recovery of the electromagnetic shower activity at the bottom of the display.
The improved image quality is expected to improve neutrino reconstruction efficiencies and estimation of physics observables.

\begin{figure}[tb]
  \centering
  \includegraphics[width=0.9\textwidth]{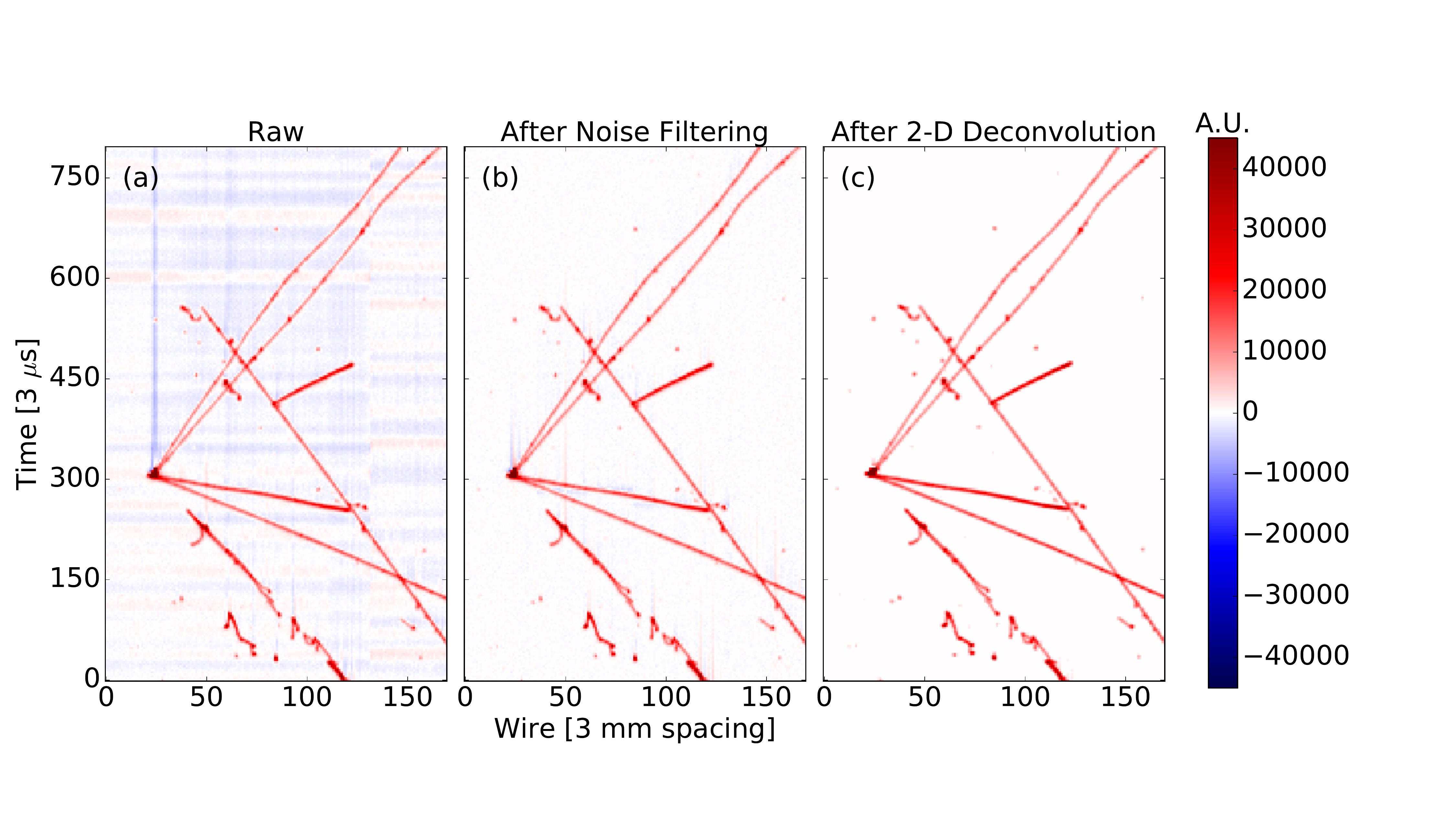}
    \Put(-355,325){\fontfamily{phv}\selectfont \textbf{MicroBooNE}}
  \vspace{-0.5cm}
  \caption{An example neutrino candidate event display from MicroBooNE data (event 41075, run 3493) showing a Y plane view. (a) The raw image in units of average baseline subtracted ADC scaled by 250 per 3~\si{\micro\second}. 
(b) The image after software noise-filtering in units of average baseline subtracted ADC scaled by 250 per 3~\si{\micro\second}.
(c) The image after the 2D deconvolution in units of electrons per 3~\si{\micro\second} are shown.  The impact from the deconvolution is minimal in the case of the Y plane, but minor improvement is observed; most notably, faint vertical lines near the neutrino interaction vertex are removed.  The improvement in this case is largely a result of the calibration of the electronics response.}
  \label{fig:evd_3493_y}
\end{figure}

Another event display, showcasing the same neutrino interaction shown in figure~\ref{fig:evd_3493_u} but 
this time in the Y plane view, is illustrated in figure~\ref{fig:evd_3493_y}.  For the collection plane, the
signal deconvolution has a much smaller impact compared to that on the induction plane views.
However, in figure~\ref{fig:evd_3493_y} one can see a small improvement in the image quality near the 
neutrino interaction vertex and close to the particle tracks after the deconvolution.  In 
particular, the faint vertical lines near the vertex (and along some of the tracks) are removed.
This is largely due to the calibration of the electronics response that is discussed in
section~\ref{sec:calib_elec_resp}.  Although the difference is not dramatic visually,
the faint vertical lines can result in fake tracks, negatively impacting the pattern
recognition.

\begin{figure}[tb]
  \centering
  \includegraphics[width=0.8\textwidth]{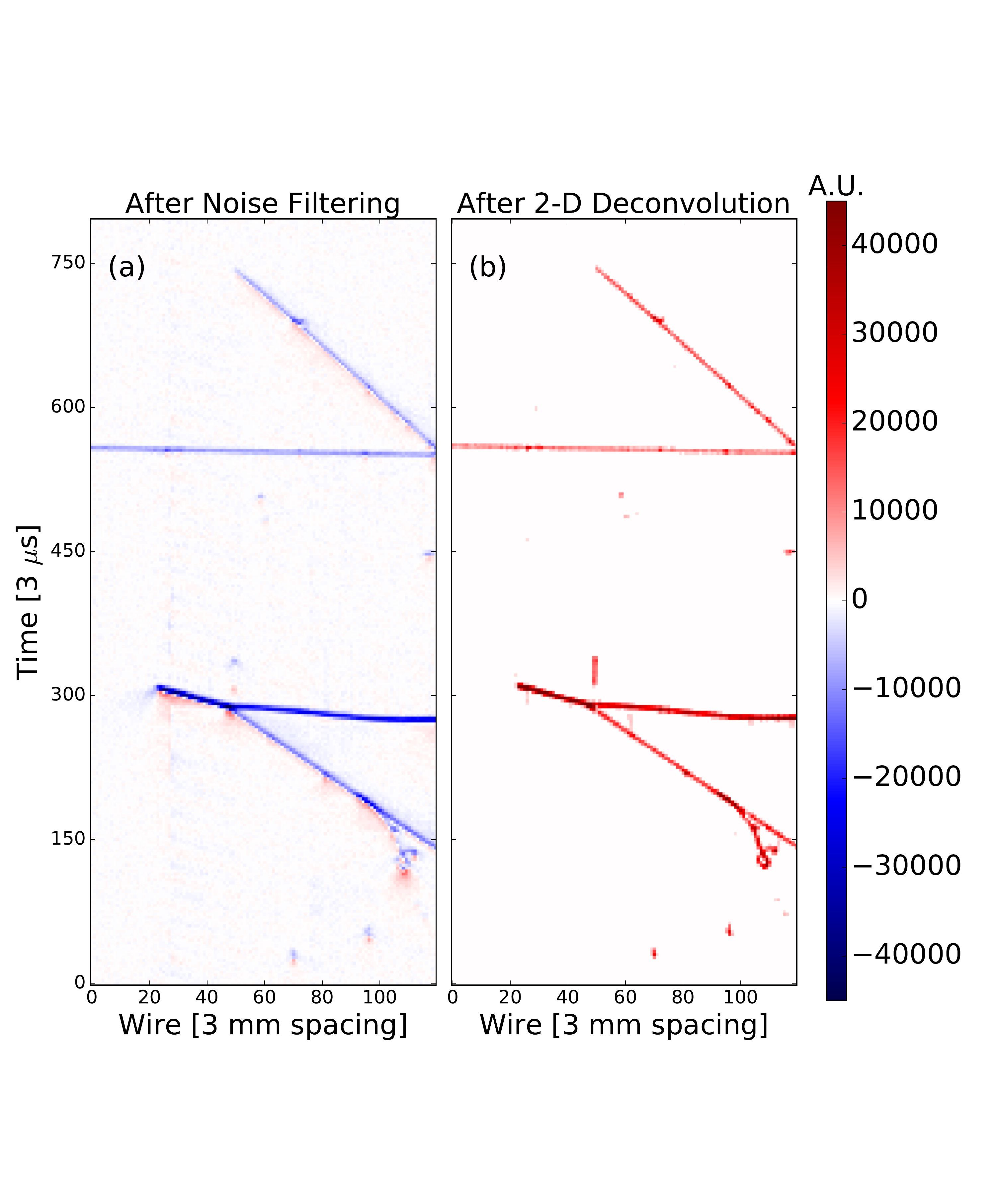}
    \Put(-310,620){\fontfamily{phv}\selectfont \textbf{MicroBooNE}}
  \vspace{-1.7cm}
  \caption{An example neutrino candidate event display from MicroBooNE data (event 53223, run 3469) showing a U plane view. 
The image after software noise-filtering in units of average baseline subtracted ADC scaled by 250 per 3~\si{\micro\second} (a) and the image after the 2D deconvolution in units of electrons per 3~\si{\micro\second} (b) are shown.  
The delta ray in the bottom right part of each event display window becomes more crisp after the deconvolution, enabling more accurate pattern recognition.  Additionally, a short track near the neutrino interaction event is recovered after the deconvolution.}
  \label{fig:evd_3469_u}
\end{figure}

Figure~\ref{fig:evd_3469_u} shows an event display of the U plane from a different event, and shows a
similar improvement after the deconvolution.  The delta ray (low-energy knock-out electron)
associated with the longer muon track at the bottom
of the display comes into focus, and the short vertical track pointing to the vertex
is recovered.  This short track may be associated with
a proton that was knocked out of an argon nucleus by a neutron associated with the neutrino 
interaction.  Accurately identifying tracks near neutrino interaction vertices is important for neutrino event reconstruction at MicroBooNE as well as at DUNE~\cite{Acciarri:2015uup}.

\begin{figure}[tb]
  \centering
  \includegraphics[width=0.8\textwidth]{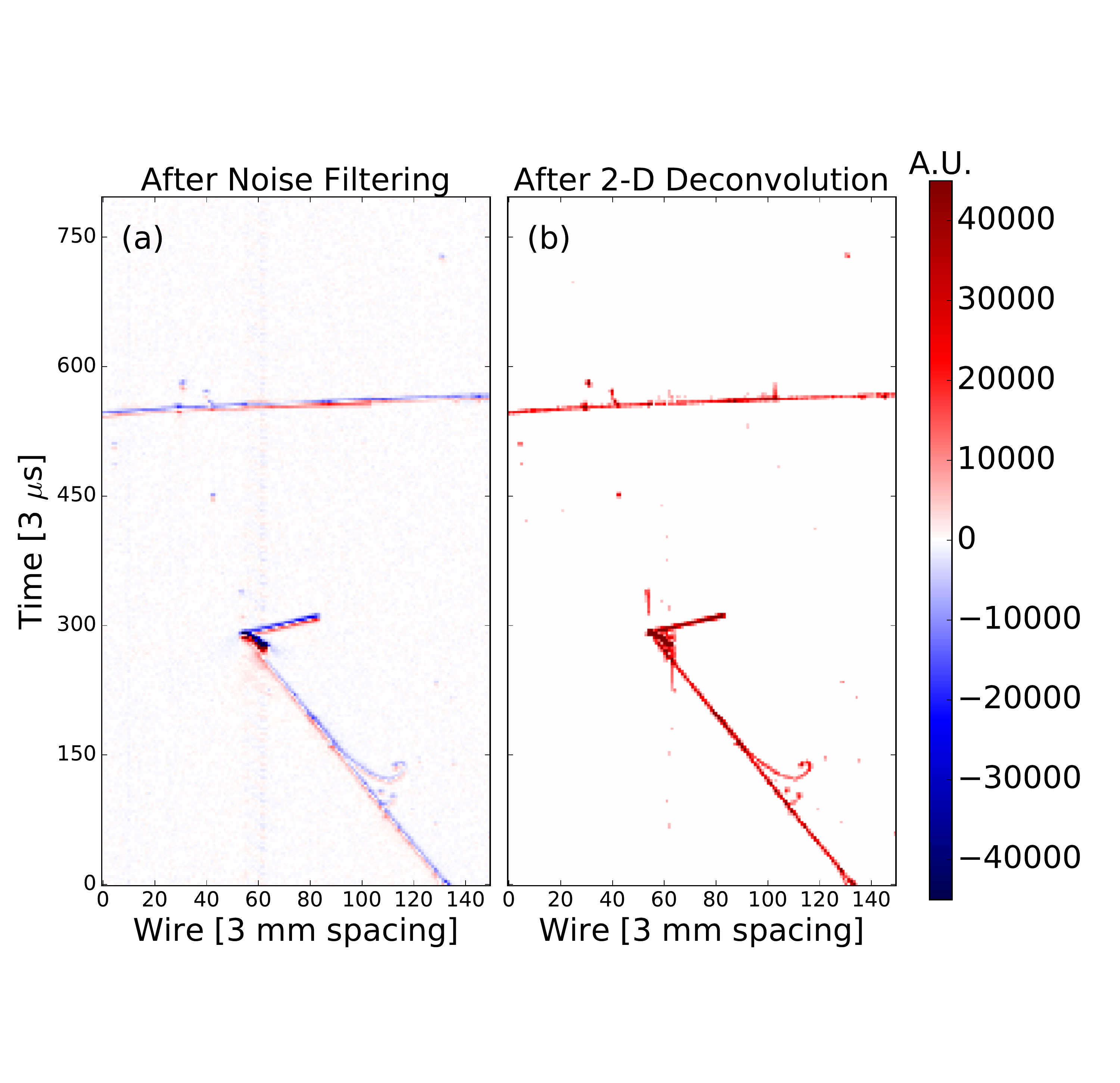}
    \Put(-310,490){\fontfamily{phv}\selectfont \textbf{MicroBooNE}}
  \vspace{-1.2cm}
  \caption{An example neutrino candidate event display from MicroBooNE data (event 53223, run 3469) showing a V plane view.  
The image after software noise-filtering in units of average baseline subtracted ADC scaled by 250 per 3~\si{\micro\second} (a) and the image after the 2D deconvolution in units of electrons per 3~\si{\micro\second} (b) are shown.  
The delta rays in the bottom right of each event display become more easily identifiable in the resultant image after the deconvolution; however, there is slight blurring of the neutrino interaction vertex due to the deconvolution step.  This is one example of where additional improvements in the signal processing chain can help improve image quality of neutrino interaction events.}
  \label{fig:evd_3469_v}
\end{figure}

Shown in figure~\ref{fig:evd_3469_v} is yet another event display corresponding to the event showcased in 
figure~\ref{fig:evd_3469_u}, this time for the V plane view.  As for the
case of the U plane for the same event, the delta rays associated with the longer muon track at the bottom of 
the display and the neutrino interaction vertex become more easily identifiable by eye in the image after the
deconvolution, and the short track near the neutrino interaction vertex is recovered.  However, one also 
notices some blurring of the event image near the neutrino interaction vertex.  This is in part due to a 
deficiency in the ROI (region of interest) finding for this particular event (see
section~\ref{sec:introduction:definitions}).  Future improvements in the ROI finding algorithm as well as a
better modeling of the wire field response will allow for further improved image quality.

%% file: perf_eval_chargematching_part2.tex
It has been shown that with the improved signal processing chain, drifting
ionization charge can be successfully extracted from the induction wire planes where
the intrinsic field response function has a bipolar shape.  In this section, it is
shown quantitatively that charge matching across the three wire planes (the unbiased estimation
of amount of ionization charge from the same drifting electron cloud when comparing signals from
the different wire planes) has been achieved in MicroBooNE data.

\begin{figure}[!htbp]
  \centering
  \includegraphics[width=0.95\textwidth]{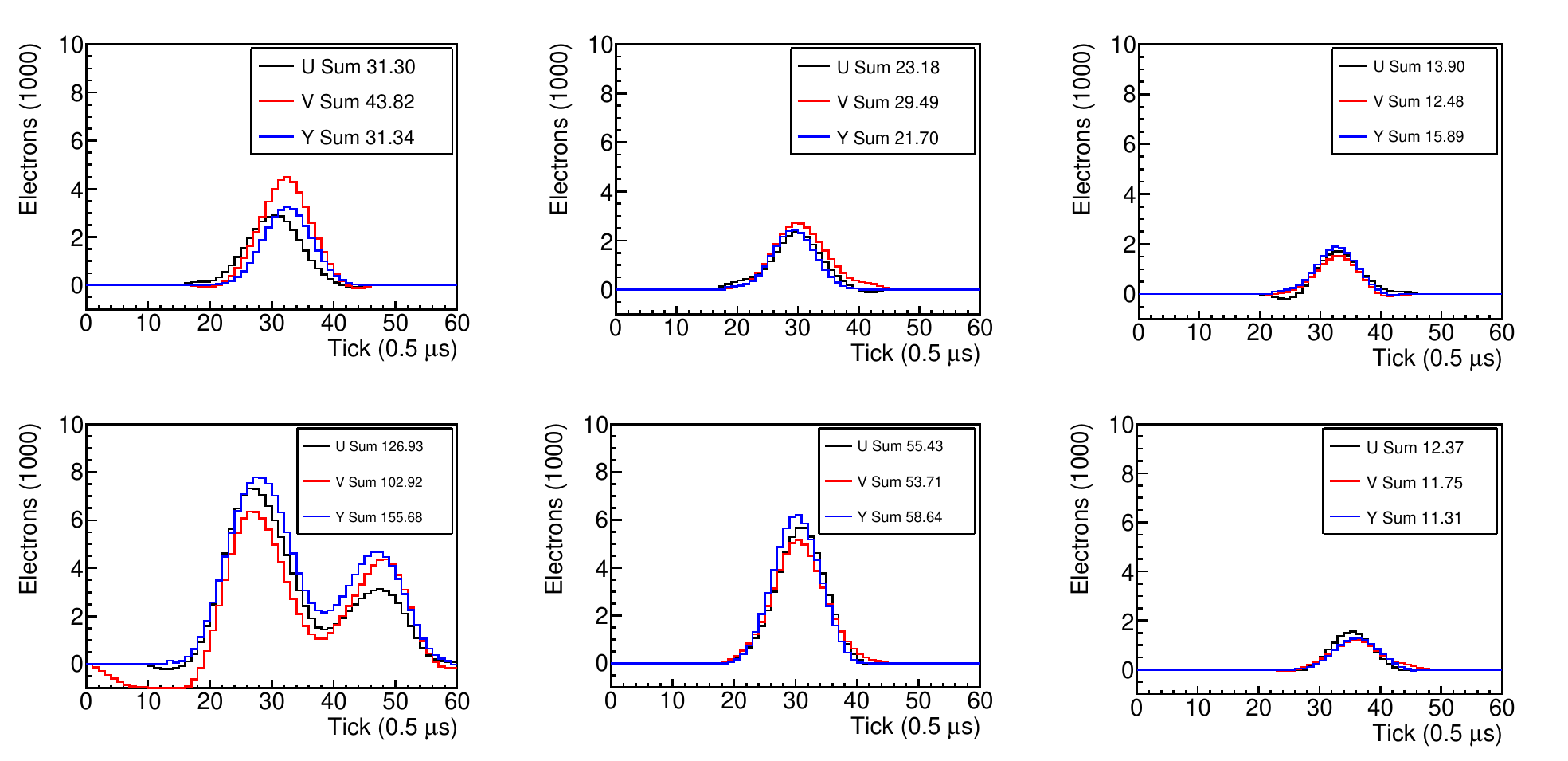}
    \Put(-88,300){\fontfamily{phv}\selectfont \textbf{MicroBooNE}}
  \caption{Individual deconvolved ionization charge distributions for a set
    of selected point sources are shown to demonstrate charge matching across the three 
    wire planes.  These distributions are obtained by summing the deconvolved waveforms associated with 
    the extent of charge deposition from each point source (the sum over the range of wires where a signal
    from the point source can be detected), independently for each plane.  The numbers in the legends show the integral of
    the charge distributions for each plane, with units in thousands of electrons.}
  \label{fig:ind_charge_match_1}
\end{figure}

Figure~\ref{fig:ind_charge_match_1} shows the extracted deconvolved ionization
charge distribution for a set of selected ``point sources'' from small, isolated ionization deposits in the TPC such
as beta decay electrons from $^{39}\mathrm{Ar}$ or other, higher-energy radiological sources in the liquid argon.  The distributions
shown are obtained by adding together all deconvolved waveforms of a given plane
associated with a point source in the detector that is identified by eye.
In general, the charge distribution as a function of time from different
wire planes are reasonably consistent with each other.  At the same time, some discrepancies do
exist in these comparisons.  Some differences can be
attributed to the inherent (or residual excess) electronics noise. 
In the bottom left image of figure~\ref{fig:ind_charge_match_1}, the V plane wires make
use of a bigger signal ROI than is necessary, leading to a reduction of reconstructed
charge in the first peak.  Future improvements to the ROI-finding algorithm, described
in section~\ref{sec:introduction:definitions}, can help ameliorate this problem.
Besides this, bias in the charge extraction can result from deviations of the local wire geometry from the 
ideal case (due to a slight sagging of the wires, for example) as well as from the use of inaccurate field response
functions in the deconvolution, among other things.  These discrepancies can be minimized with additional studies
that are beyond the scope of this article.

\begin{figure}[!htbp]
  \centering
  \includegraphics[width=0.95\textwidth]{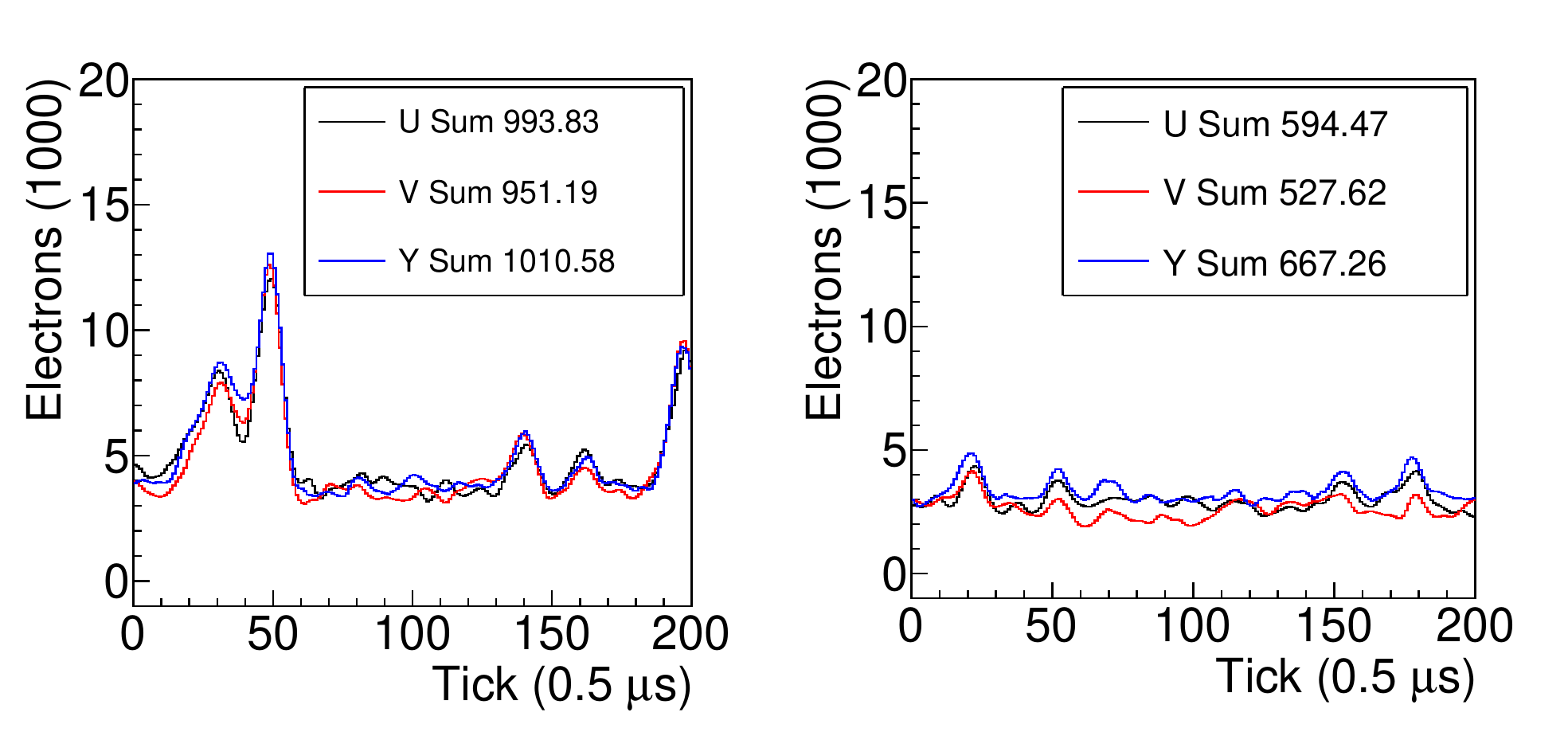}
    \Put(-100,200){\fontfamily{phv}\selectfont \textbf{MicroBooNE}}
  \caption{Individual deconvolved ionization charge distribution for two example
    line sources, associated with cosmic muons passing through the detector,
    are shown to demonstrate charge matching across the three wire planes.  These distributions are
    obtained by summing the deconvolved waveforms associated with the extent of charge deposition from
    each line source (the sum over the range of wires where a signal from the line source can be
    detected), independently for each plane.  The numbers in the legends show the integral of the charge distributions for each plane, with units in thousands of electrons.
  }
  \label{figs:ind_charge_match_2}
\end{figure}

Figure~\ref{figs:ind_charge_match_2} compares the per-plane extracted deconvolved ionization
charge distributions for two example ``line sources'' produced from cosmic muons passing through
the TPC.  As for the point source case described above, these distributions are obtained by summing 
together the deconvolved waveforms associated with each line source on a per-plane basis.  
Reasonable agreement in charge matching across the three planes is demonstrated, as in the point source case. 
The remaining differences can be explained by the inherent electronics noise, which is amplified in the
deconvolution process, as well as inaccuracies in the field response functions obtained from Garfield.  The
features in these distributions are due to large signal fluctuations in the deposited energy along the muon
track, including delta rays.

\begin{figure}[p]
\centering
\begin{subfigure}{0.49\textwidth}
  \centering
  \includegraphics[width=.86\textwidth]{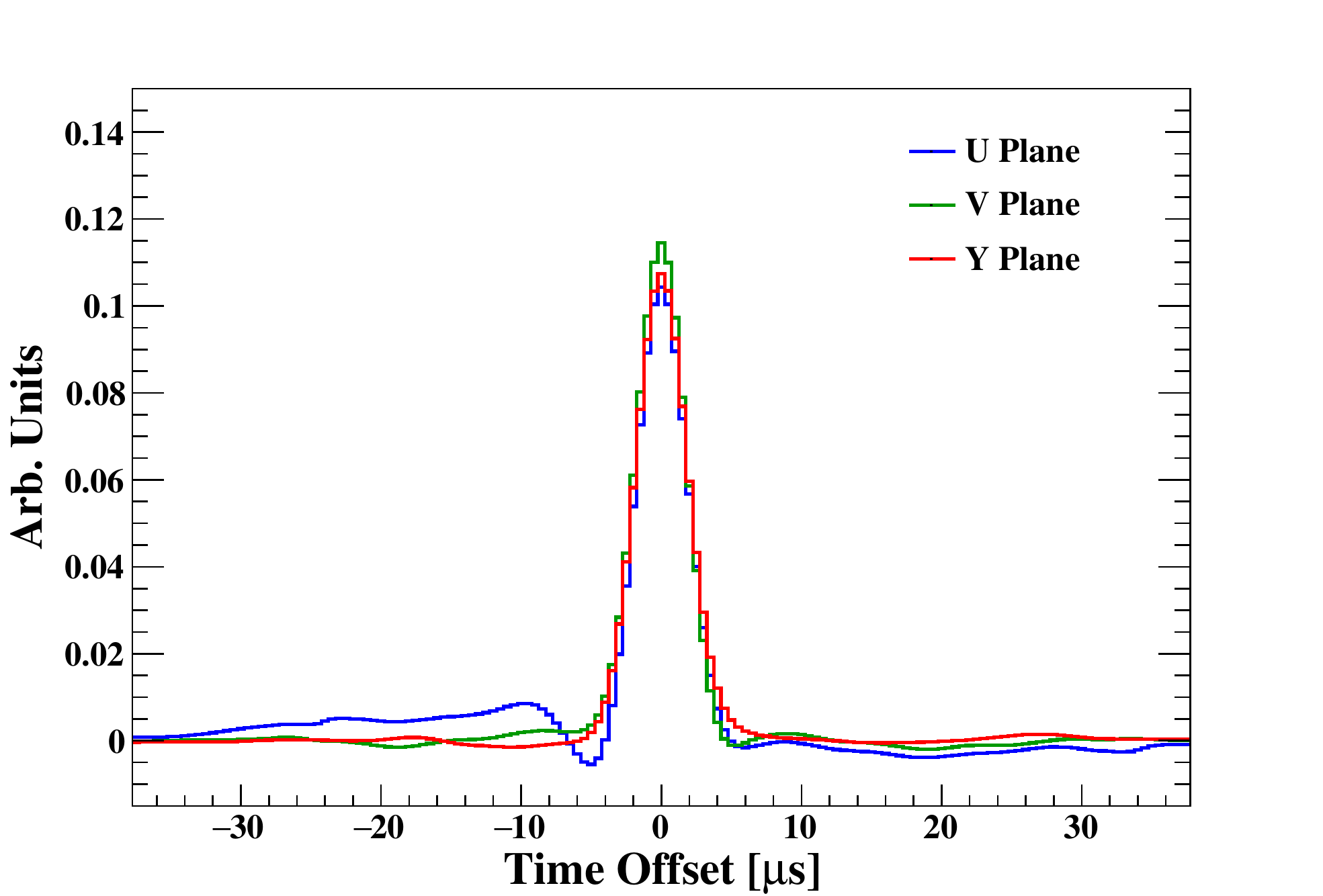}
  \Put(-155,200){\fontfamily{phv}\selectfont \textbf{MicroBooNE}}
  \caption{1D deconvolution, \ang{5}~<~$\theta_{xz}$~<~\ang{15}.}
\end{subfigure}
\begin{subfigure}{0.49\textwidth}
  \centering
  \includegraphics[width=.86\textwidth]{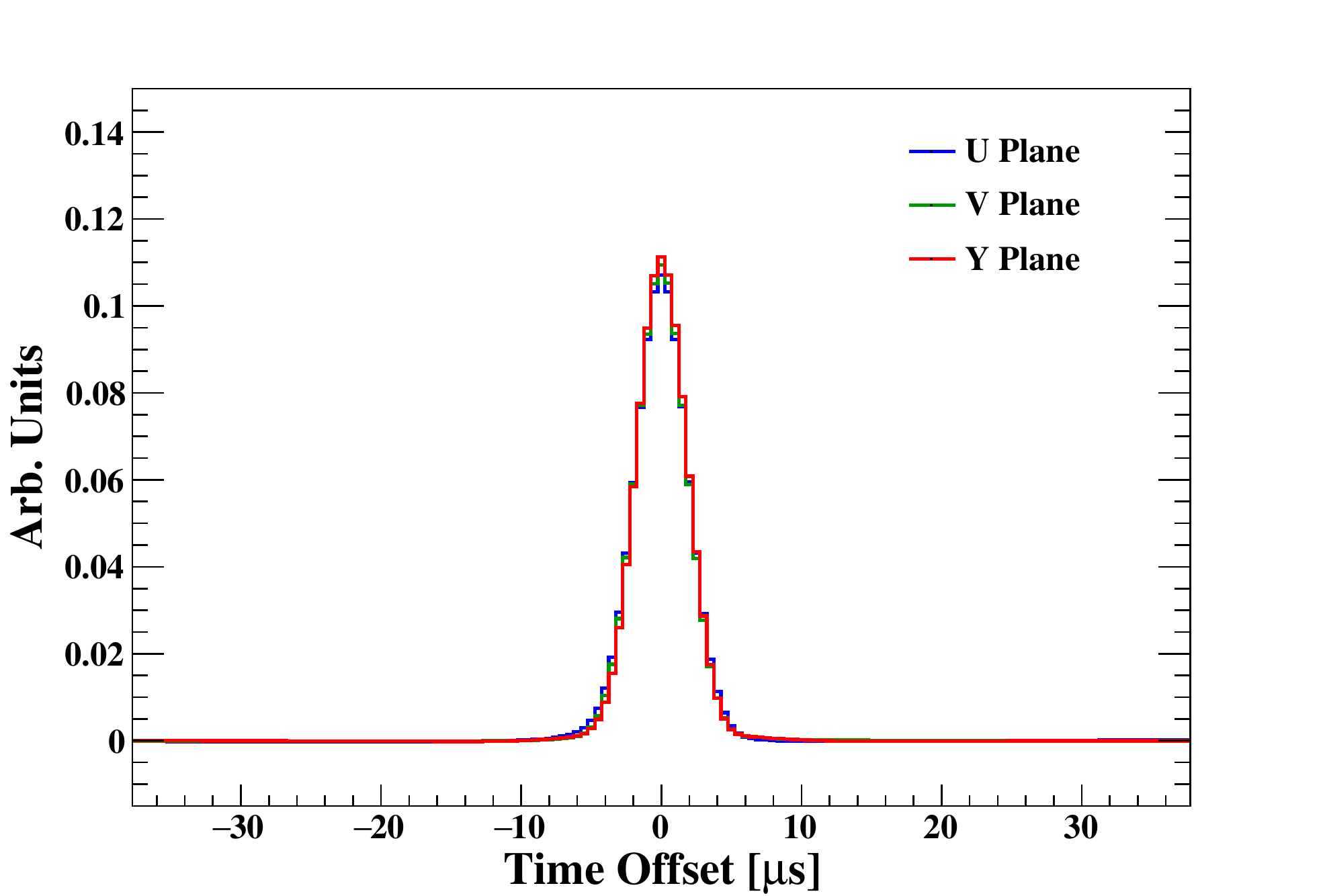}
  \caption{2D deconvolution, \ang{5}~<~$\theta_{xz}$~<~\ang{15}.}
\end{subfigure}
\begin{subfigure}{0.49\textwidth}
  \centering
  \includegraphics[width=.86\textwidth]{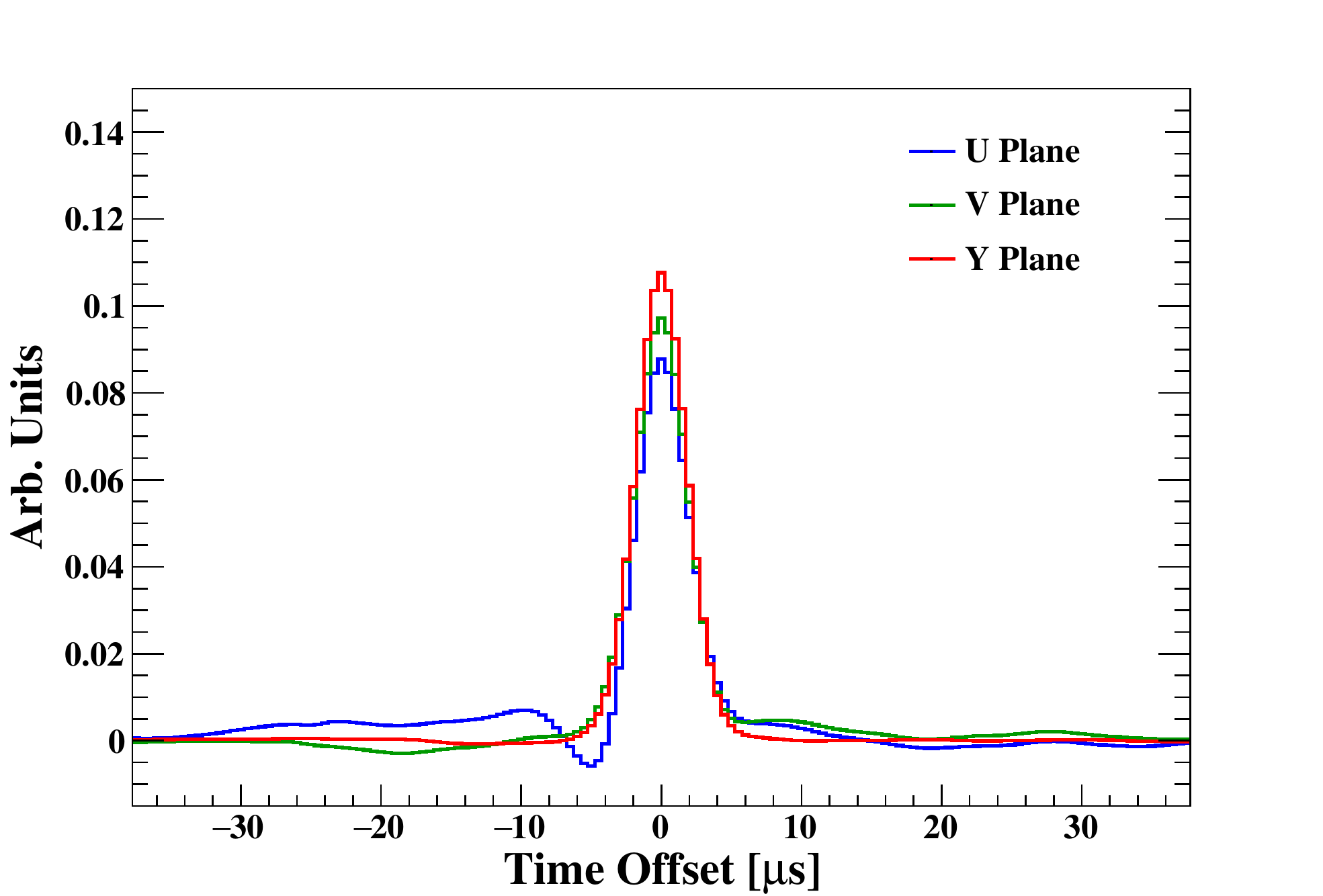}
  \caption{1D deconvolution, \ang{15}~<~$\theta_{xz}$~<~\ang{30}.}
\end{subfigure}
\begin{subfigure}{0.49\textwidth}
  \centering
  \includegraphics[width=.86\textwidth]{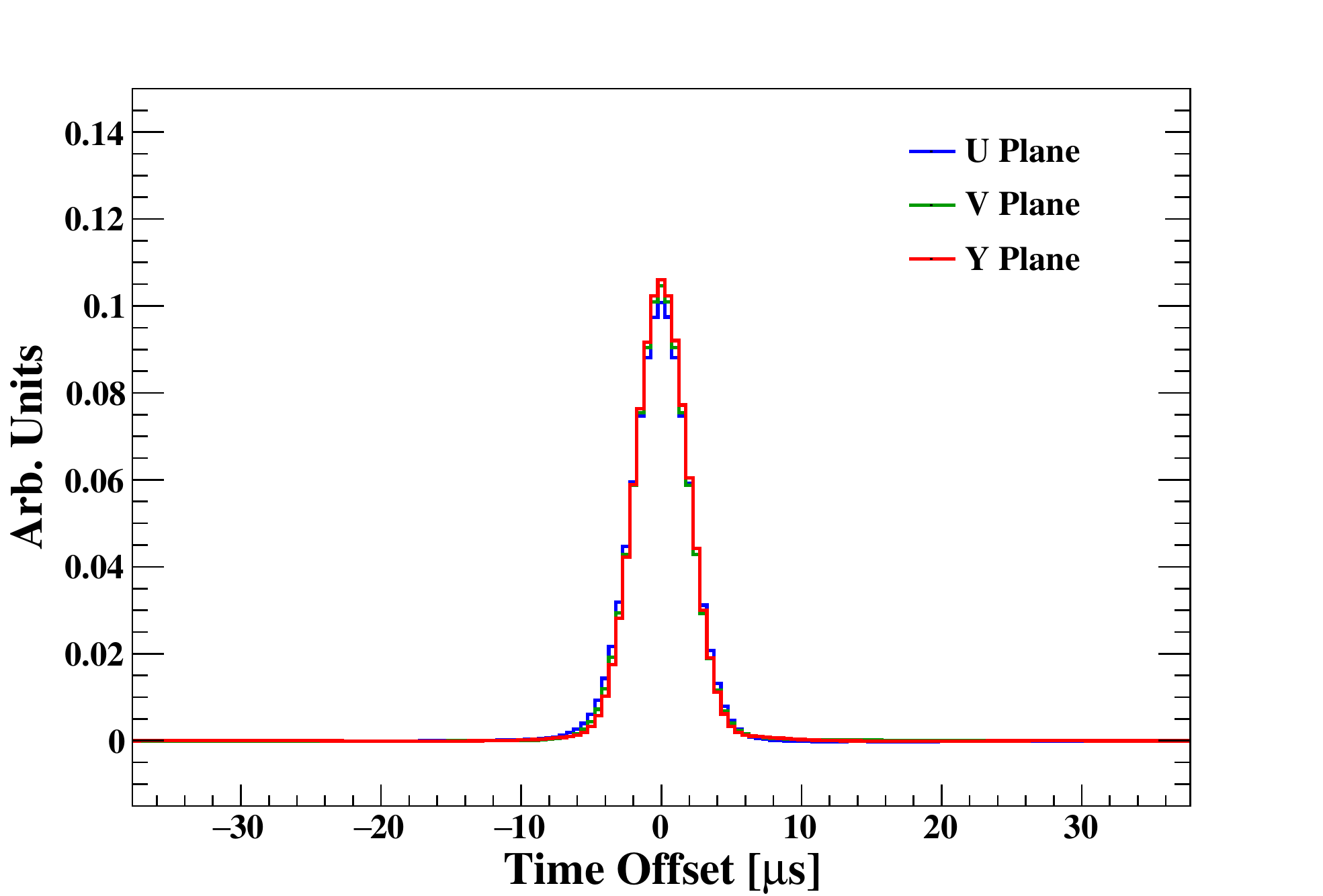}
  \caption{2D deconvolution, \ang{15}~<~$\theta_{xz}$~<~\ang{30}.}
\end{subfigure}
\begin{subfigure}{0.49\textwidth}
  \centering
  \includegraphics[width=.86\textwidth]{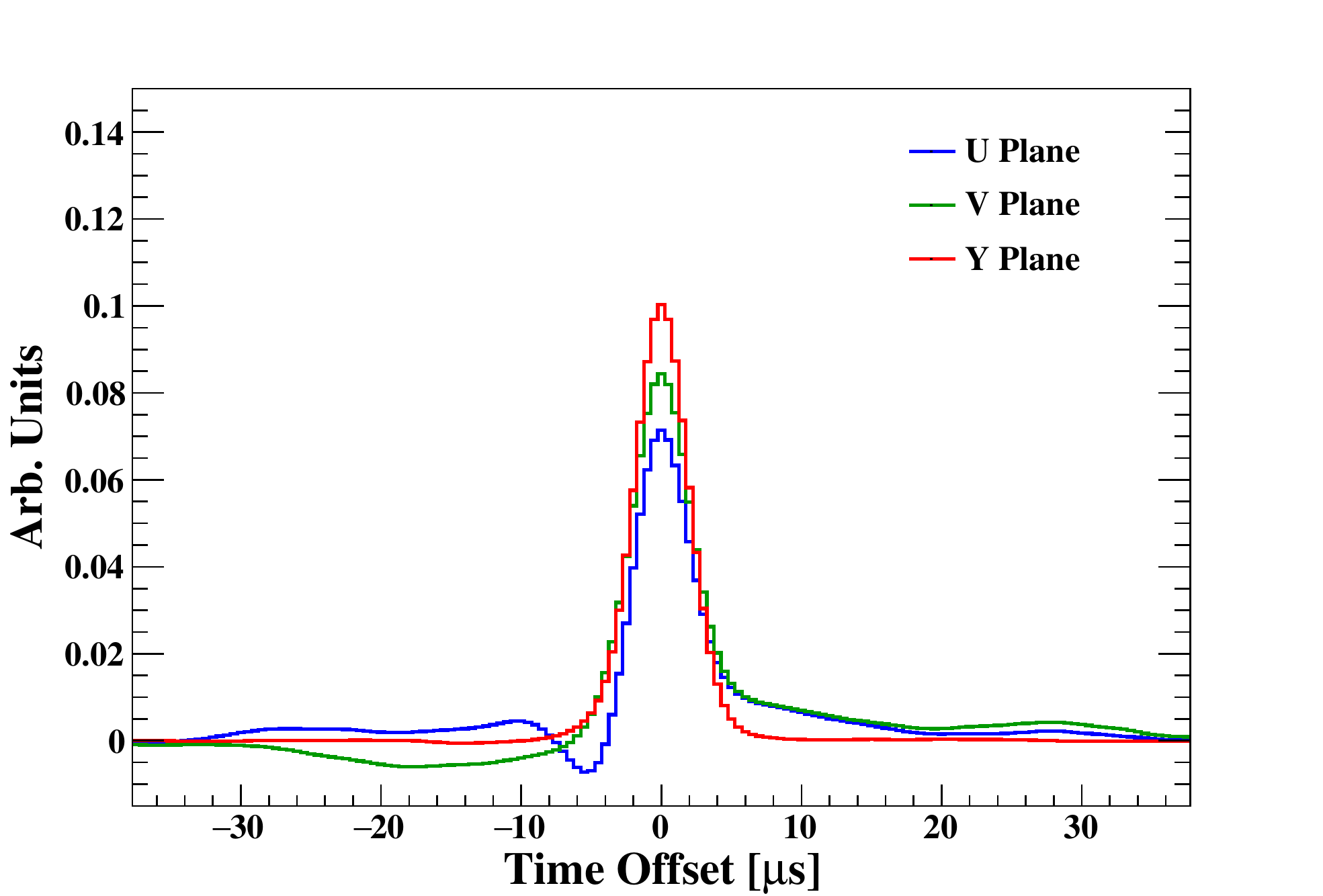}
  \caption{1D deconvolution, \ang{30}~<~$\theta_{xz}$~<~\ang{50}.}
\end{subfigure}
\begin{subfigure}{0.49\textwidth}
  \centering
  \includegraphics[width=.86\textwidth]{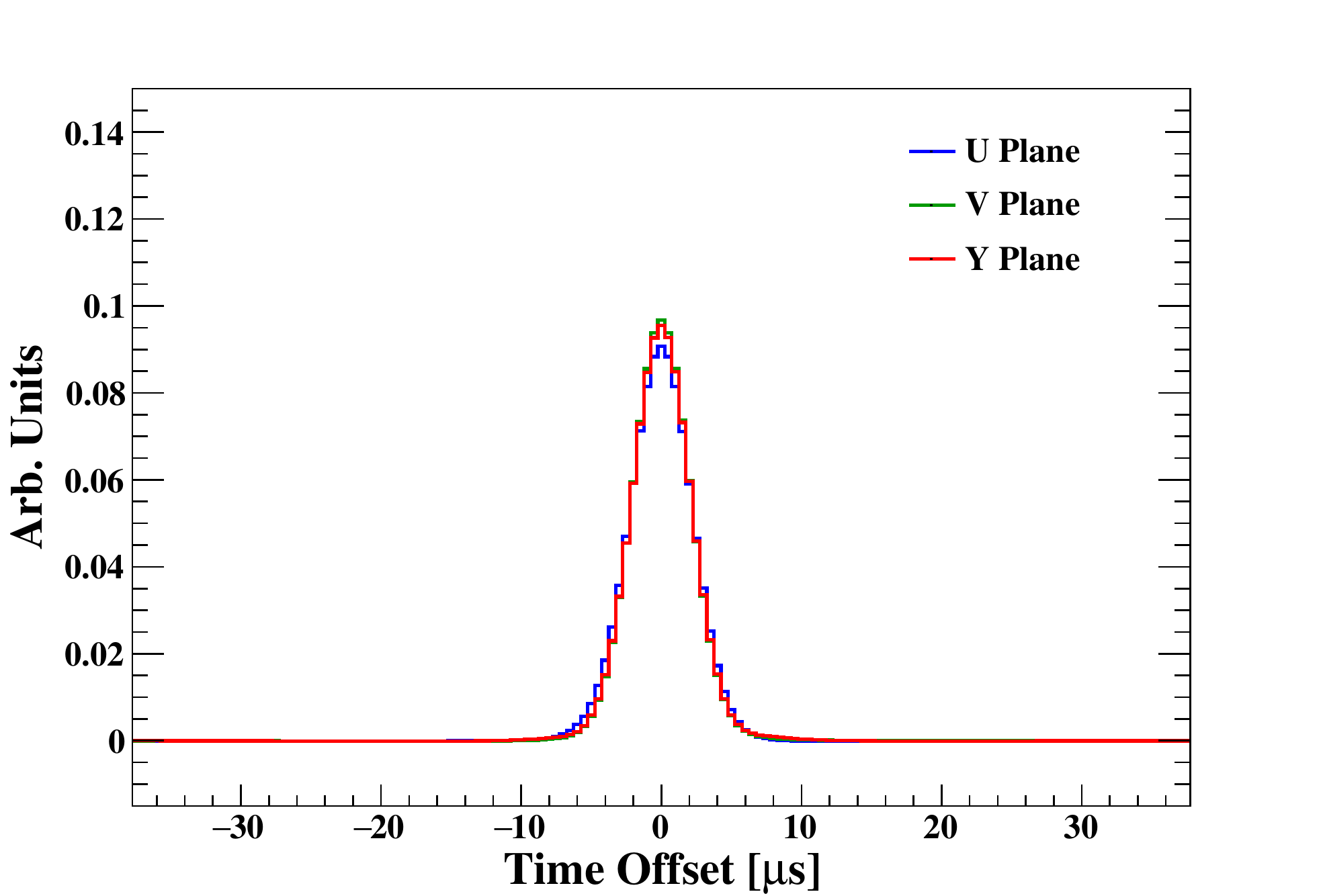}
  \caption{2D deconvolution, \ang{30}~<~$\theta_{xz}$~<~\ang{50}.}
\end{subfigure}
\begin{subfigure}{0.49\textwidth}
  \centering
  \includegraphics[width=.86\textwidth]{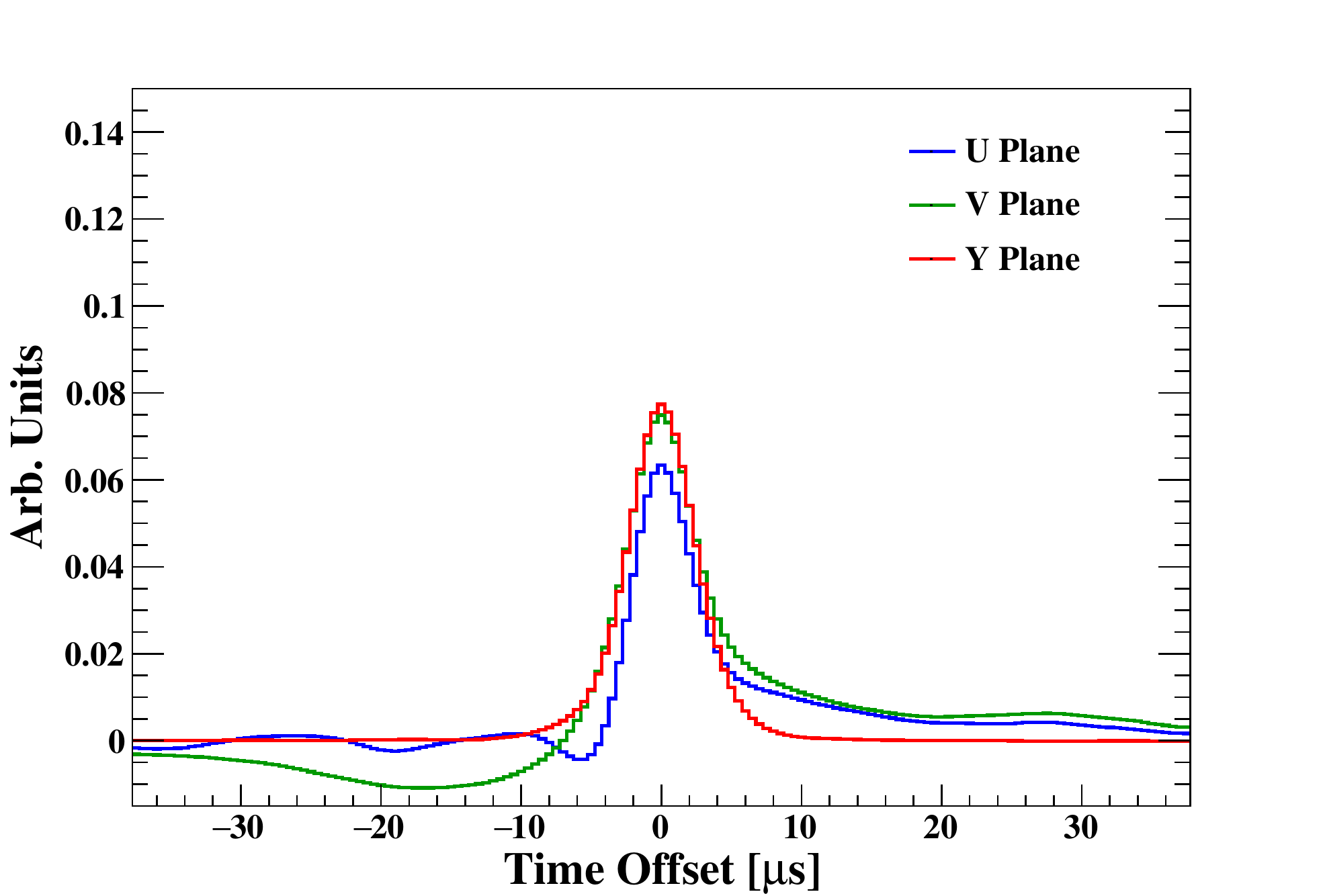}
  \caption{1D deconvolution, \ang{50}~<~$\theta_{xz}$~<~\ang{70}.}
\end{subfigure}
\begin{subfigure}{0.49\textwidth}
  \centering
  \includegraphics[width=.86\textwidth]{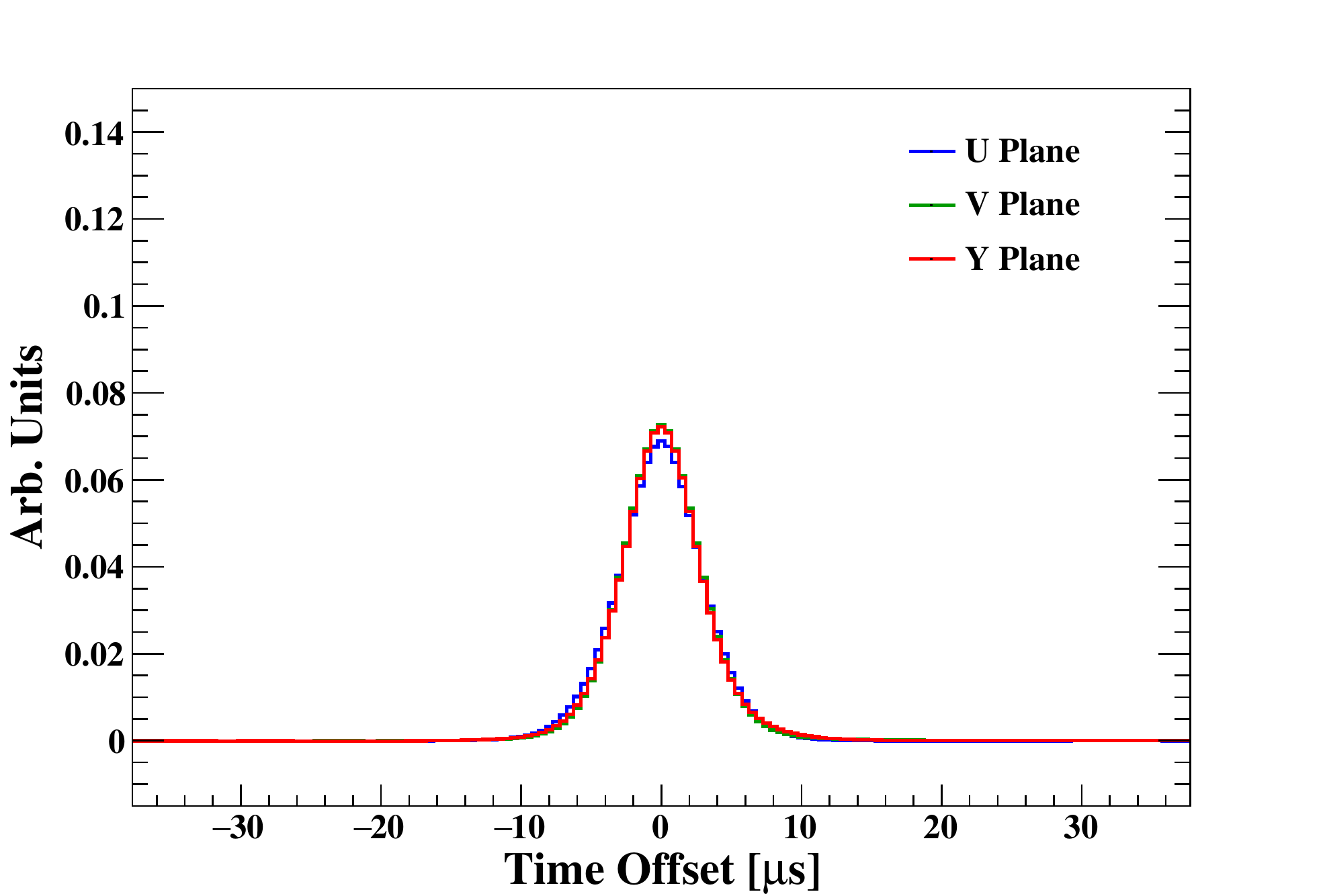}
  \caption{2D deconvolution, \ang{50}~<~$\theta_{xz}$~<~\ang{70}.}
\end{subfigure}
\caption{Comparison of average deconvolved signal shapes (both 1D and 2D deconvolution) on the U, V, and Y planes using cosmic muons near the anode plane at a variety of angles in data.  $\theta_{xz}$ is the angle of the track between the drift direction and the direction in the wire plane ($y-z$ plane) perpendicular to the wire orientation  (see figure~\ref{fig:angle}).  All distributions are normalized to unity.} \label{fig:PostDeconvShapeCompData}
\end{figure}

In addition to individual point and line sources, the average deconvolved signal shape associated 
with tracks of different angles in the detector is shown in figure~\ref{fig:PostDeconvShapeCompData}.  Results are 
shown both for the case of the 1D deconvolution and the improved 2D deconvolution in the left and right
columns of figure~\ref{fig:PostDeconvShapeCompData}, respectively.  It is clear from this comparison that the matching
of signal shapes across planes is significantly improved when moving from the 1D to the 2D 
deconvolution.  In the case of the 2D deconvolution, the average deconvolved signal shape is very close to a 
Gaussian, which is expected, because a Gaussian filter is used in the signal processing
(see section~\ref{sec:introduction:definitions}).  The improvement in moving from the 1D to the 2D
deconvolution is most noticeable for the tracks that are oriented more in the drift direction (higher angles in
figure~\ref{fig:PostDeconvShapeCompData}), as the wire field responses used in the case of the 1D deconvolution were derived from tracks nearly parallel to the anode plane.  The 2D deconvolution is able to account for
the charge induction profile of drifting charge on neighboring wires for all track angles.

\begin{figure}[!hp]
  \centering
  \begin{subfigure}{0.99\textwidth}
    \centering
    \includegraphics[width=0.99\textwidth]{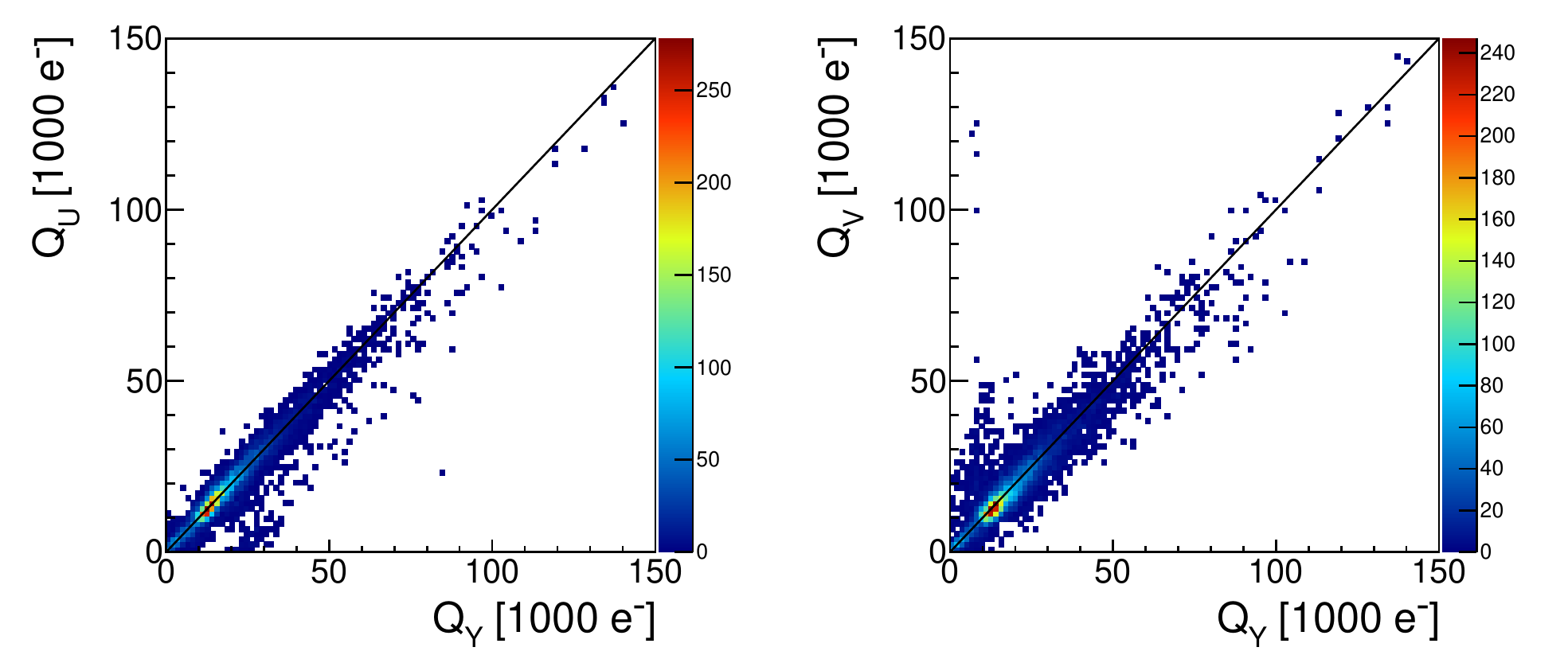}
    \Put(-370,310){\fontfamily{phv}\selectfont \textbf{MicroBooNE}}
    \caption{Scatterplots of charge extracted from induction planes versus collection plane extracted charge.}
  \end{subfigure}
  \begin{subfigure}{0.99\textwidth}
    \centering
    \includegraphics[width=0.99\textwidth]{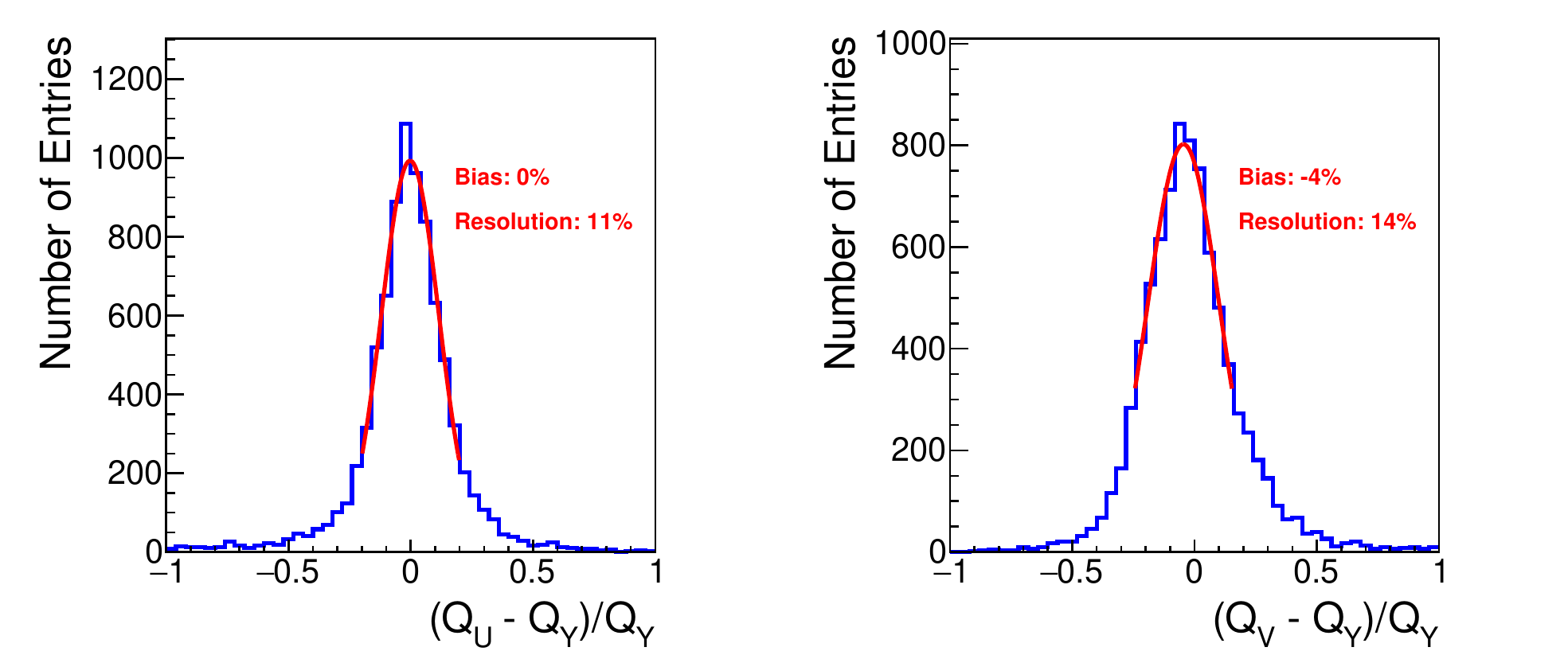}
    \Put(-370,310){\fontfamily{phv}\selectfont \textbf{MicroBooNE}}
  \caption{Projection of above scatterplots about the unity-slope line, normalized to collection plane charge.}
  \end{subfigure}
  \caption{Using a sample of cosmic muons in data, the extracted charge from the induction
  U plane and induction V plane for a ``blob''
  (matched set of wires with signal across the three planes of the LArTPC) is plotted against the
  charge from the collection Y plane (a).  The diagonal lines go through the origin and have unity slope.
  Good matching of charge between the induction planes and collection plane is observed.  The
  projection of these scatterplots along the unity-slope line are also shown for the U plane
  and V plane, normalizing to the charge observed by the Y plane in both cases (b).
  }
  \label{fig:charge_res}
\end{figure}

\afterpage{\clearpage}

\begin{figure}[!hp]
  \centering
  \begin{subfigure}{0.99\textwidth}
    \centering
    \includegraphics[width=.8\textwidth]{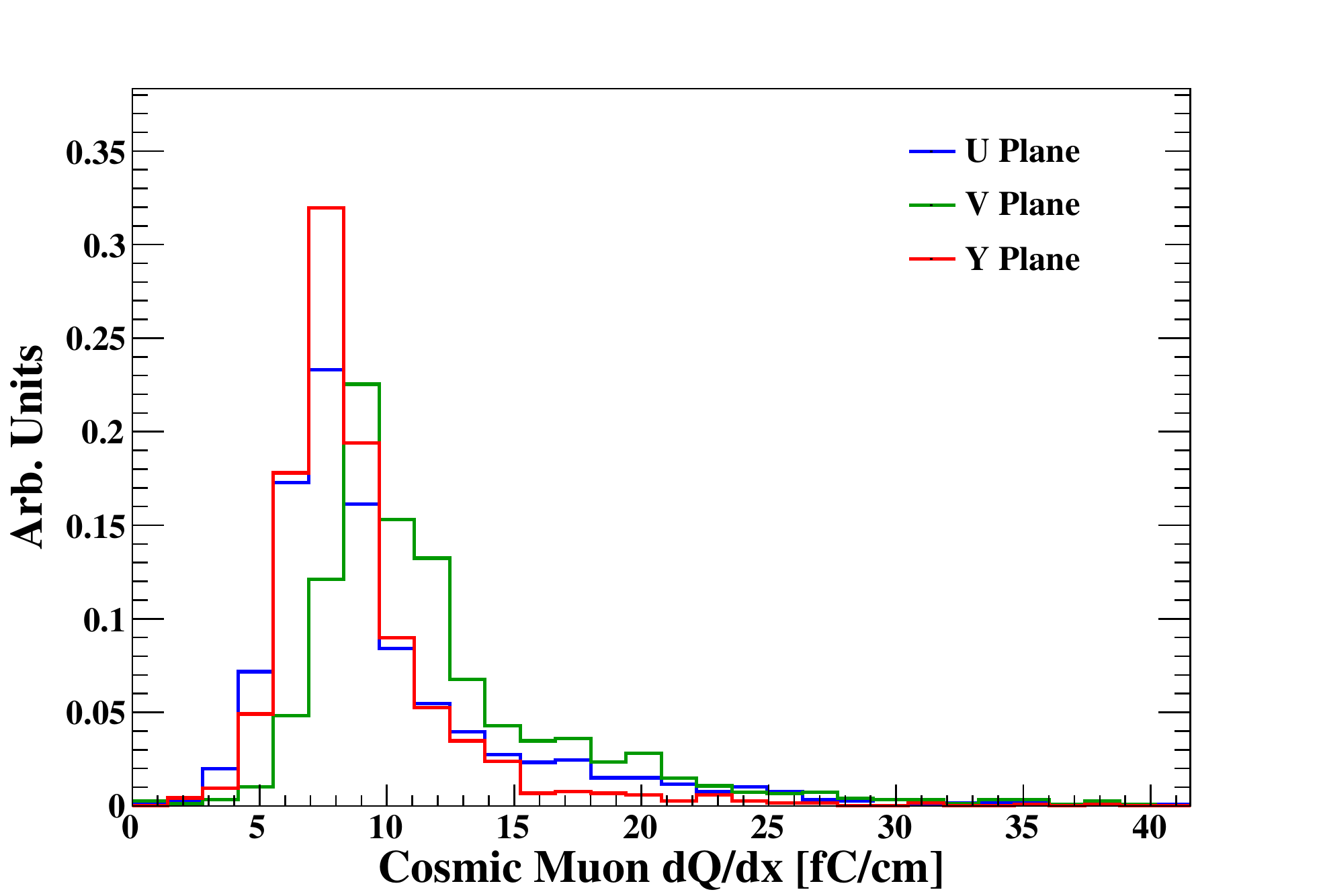}
    \caption{Cosmic muon $dQ/dx$ distribution for the case of the 1D deconvolution.}
  \end{subfigure}
  \begin{subfigure}{0.99\textwidth}
    \centering
    \includegraphics[width=.8\textwidth]{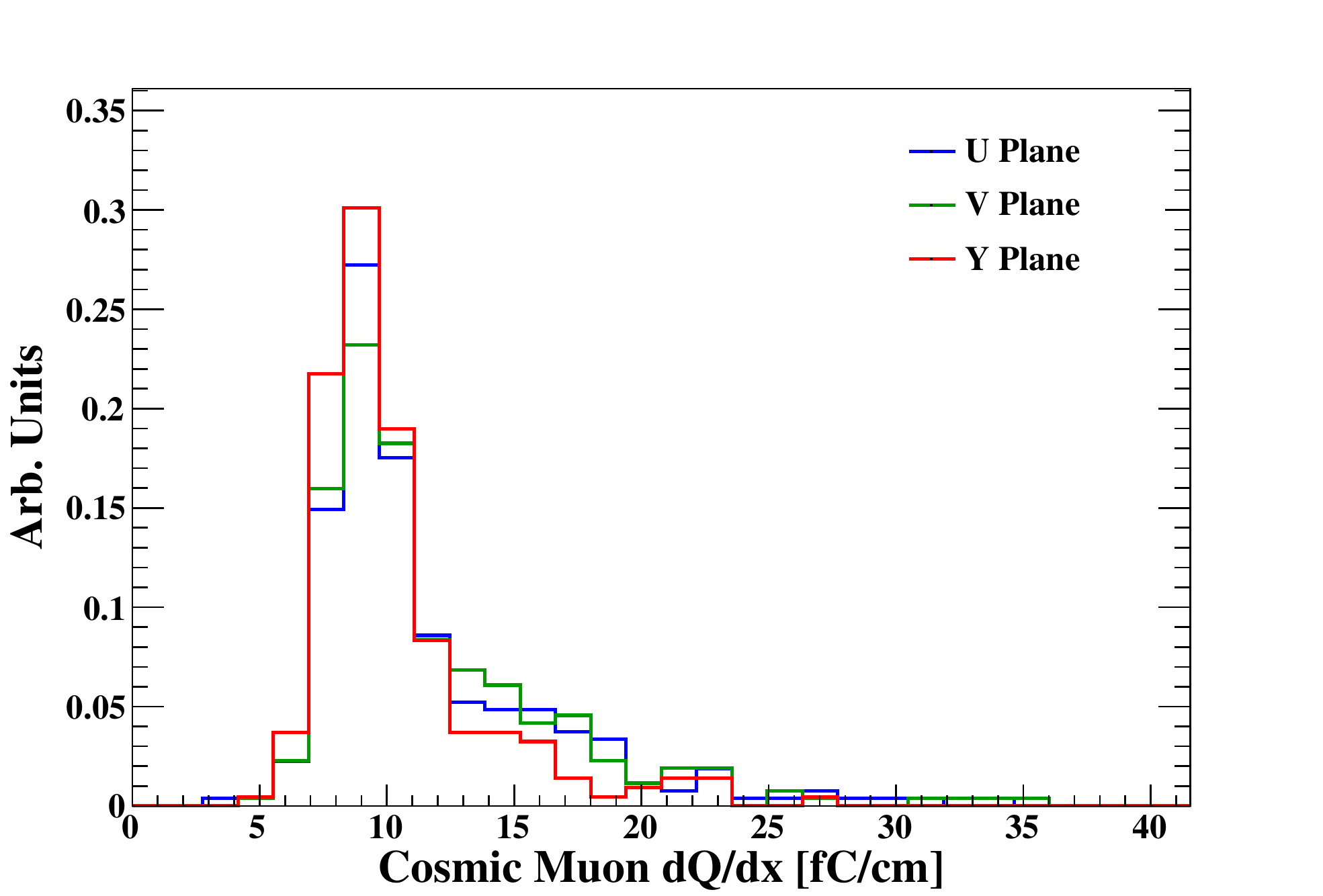}
    \caption{Cosmic muon $dQ/dx$ distribution for the case of the 2D deconvolution.}
  \end{subfigure}
  \Put(-180,565){\fontfamily{phv}\selectfont \textbf{MicroBooNE}}
  \Put(35,310){\fontfamily{phv}\selectfont \textbf{MicroBooNE}}
  \caption{Cosmic muon $dQ/dx$ as estimated from each plane (separately) using cosmic muon tracks in data.  Results are shown both for the case of the 1D (a) and the improved 2D (b) deconvolution.  
Charge estimation is done by integrating across each waveform, while the track angle used in the $dQ/dx$ calculation is obtained from the output tracks created by the Pandora pattern recognition~\cite{Acciarri:2017hat}.  Only charge deposited between 2~\si{\centi\meter} and 10~\si{\centi\meter} from the anode plane is included in the $dQ/dx$ calculation.  Tracks with a wide range of angles are used in the sample.
  } \label{fig:dQdxCompData}
\end{figure}

To further quantify the level of charge matching across the three wire planes, a more detailed 
comparison is made utilizing all of the ionization charge collected in a readout event.  First, the entire 
drift time window is divided into many different independent time slices. Each time slice consists of 
four time ticks, corresponding to 2~$\mu$\si{\second}, which is about 2.2~\si{\milli\meter} of drift in liquid
argon.  In each time slice, adjacent intraplane wires with observed signals are
grouped together to form a ``merged'' wire. These merged
wires are then compared with merged wires from the other two wire planes in order to search for 
signal activity (referred to hereafter as a ``blob'') in the transverse ($y-z$)
plane~\cite{wirecell}.  As the entire drift time window is
used, the signal activity may originate from any distance away from the anode plane (unlike in the studies of
section~\ref{sec:fieldresp}).
The top row of figure~\ref{fig:charge_res} shows the extracted charge 
from the induction wire planes (U plane on left, V plane on right) for a blob versus the extracted charge from 
the collection wire plane (Y plane) for the same blob, utilizing the 2D deconvolution for signal processing.  Good matching of charge across the wire planes is 
generally observed in these distributions.  For the U plane vs. Y plane comparison, there is a small group 
of points which has a reduced U plane charge in comparison to that extracted from the Y plane.  This is due to
prolonged signals associated with tracks nearly orthogonal to the wire planes, which are better imaged
with the 2D deconvolution, but still experience a downward bias in reconstructed charge magnitude.  Also, 
for the V plane vs. Y plane comparison, there is a small group of points which have a much 
enhanced V plane charge with respect to that extracted from the Y plane waveforms.  This is due to charge collection
on the induction wires.  Both of these occurrences are quite rare, and in general, charge reconstructed
with the induction planes are seen to be congruent with that seen by the collection plane in data.

The spread about the unity-slope lines in the top row of figure~\ref{fig:charge_res} is dominated by the charge resolution in the
induction wire planes.  This is illustrated more clearly in the bottom row of the same figure, where
the spread about the unity-slope line is profiled for both induction planes, normalizing to the charge observed by the Y plane.
By fitting the core of each distribution to a Gaussian and assuming perfect knowledge of charge
using the Y plane, it is
determined that the charge resolution of the blobs is roughly 11\% using the U plane signals and roughly 14\% using the
V plane signals.  However, this is not equivalent to the per-wire charge resolution, as the blobs used in
figure~\ref{fig:charge_res} are formed by merging signals (and noise) across neighboring wires and adjacent waveform
time bins.  The charge resolution can in principle decrease (increase) by merging more (less) information in
this way.  Furthermore, the charge resolution for the induction planes depends heavily on the signal window size.  This
varies greatly across different particle topologies observed in the TPC readout.  As a result, it is not appropriate
to interpret the above charge resolution numbers as applicable to all signals observed in data for a given wire plane.
There is also a small bias in average extracted charge between the V plane and Y plane (roughly -4\%, obtained via the
Gaussian fit), as presented in figure~\ref{fig:charge_res}.  This is likely due to an imperfect estimation of the
V plane wire field response (discussed in section~\ref{sec:fieldresp}).

While the charge resolution for the collection plane is very similar between data and
simulation, the charge resolution for the induction planes is about a factor of two worse in data than that predicted
by studies utilizing simulated events~\cite{SP1_paper}.  There are several potential reasons for this.  First, in the
full TPC simulation, the field response functions are also used in the TPC signal deconvolution.
These field response functions are based on a two-dimensional Garfield simulation, and are slightly
different from those realized by the MicroBooNE three-dimensional wire geometry.  Additionally, the local wire geometry may deviate from the ideal case, where the wire pitch 
and gap between wire planes are exactly 3~\si{\milli\meter}.  Such deviations would lead to differences 
in the field response observed in data.  Finally, the software noise filtering that is required to remove 
excess noise on TPC channels (before the hardware was modified~\cite{noise_filter_paper})
introduces some bias in the raw waveform, which can be propagated to the deconvolved waveform.
This may not be sufficiently taken into account in the simulation, in which only the inherent electronics
noise is simulated. The detailed investigation of these differences is beyond the scope of this article.

In another demonstration of the cross-plane charge matching capability of the improved signal
processing chain in MicroBooNE, the cosmic muon $dQ/dx$ distribution obtained using cosmic muon
tracks in MicroBooNE data is shown in figure~\ref{fig:dQdxCompData} for both the original 1D and
the new 2D deconvolution.  A wide variety of track angles is utilized in this sample.  In the case of the 1D 
deconvolution, the three planes do not align in terms of the estimated values of $dQ/dx$ using each plane
individually.  In the 2D case, the agreement among the planes is better, especially for the most probable
value of energy loss (MPV); the spread in the values of MPV across planes reduces from 15\% in the 1D case
to 4\% in the 2D case as determined from fits to a Landau convolved with a Gaussian.  The induction
planes experience a lower performance in terms of $dQ/dx$ resolution than the
collection plane, despite there being little bias across the three planes.  This is
in agreement with expectations from charge resolution studies carried out in Ref.~\cite{SP1_paper} using simulated
events.  Furthermore, these $dQ/dx$ distributions sample over a variety of different
residual ranges (distance from the stopping point of the particle), the only requirement being that 
the cosmic muon pierces the anode plane and only the charge between 2~\si{\centi\meter} and 
10~\si{\centi\meter} from the anode plane is included in the $dQ/dx$ measurement.  This reduces the
impact from various detector effects, such as the finite electron lifetime and the
diffusion of the drifting electrons in the liquid argon.
This leads to additional broadening, similar for all three planes, in comparison to the distribution
one might see when looking at a fixed distance from the end of a cosmic muon stopping in the detector.

In summary, the successful extraction of drifting ionization charge from induction wire planes and 
cross-plane charge matching in a LArTPC has been demonstrated with MicroBooNE data.  The reconstructed
charges in the three wire planes after deconvolution agree reasonably well for
both individual point and line charge sources as well as average signal shapes in MicroBooNE data.  Further
studies are needed in order to improve the level of agreement.  Of primary importance is to improve the
calculated field response function to better match the field response observed in data, as discussed
in section~\ref{sec:fieldresp}.

%% file: summary_part2.tex
A robust signal processing chain was developed for MicroBooNE in order to extract charge information from signals on all three TPC wire planes with minimal bias and good charge resolution~\cite{SP1_paper}.  
Calibration of the electronics chain and validation of the response simulation using data events was necessary to accomplish this.  
The conclusion from these studies is that the signal shapes in data and simulation experience much better agreement after calibration.  
A number of steps were required to achieve this result, starting with additional cleaning of data events, and identifying and removing certain relatively rare events compromised by excess noise signals.  
We have demonstrated an improved image quality and unbiased charge matching across the three wire planes of the MicroBooNE LArTPC. 
We have also presented the first quantitative evaluation of extracted charge measurements using the 2D deconvolution algorithm in LArTPC data. 
This detailed understanding of the MicroBooNE TPC response and the performance of signal processing algorithms on data is crucial for the development of physics analyses and evaluation of systematic uncertainties, as well as for the design, operation, and calibration of future LArTPC experiments.

This article has presented the precise extraction of ionization charge from induction plane wires in a single-phase LArTPC, which is expected to improve event reconstruction and provide more accurate charge measurements.    
In addition, accurate charge matching across wire planes has been demonstrated for the first time, allowing reconstruction of three-dimensional charge distributions as in Wire-Cell tomographic reconstruction~\cite{wirecell}.  This is expected to improve the reconstruction of neutrino interactions and cosmic-ray muons.